%% file: main.tex
\documentclass[10pt]{article}

\usepackage[left=2.5cm,right=2.5cm,top=2.8cm,bottom=2.8cm]{geometry}

\usepackage{amsmath}
\usepackage{booktabs}
\usepackage[caption=false]{subfig}
\usepackage{xifthen}
\usepackage{xspace}
\usepackage{listings}
\usepackage{enumitem}
\usepackage{bm} 
\usepackage{array}
\usepackage{amssymb}
\usepackage{balance}
\usepackage{url}
\usepackage{anyfontsize}
\usepackage{multirow}
\usepackage{stmaryrd}
\SetSymbolFont{stmry}{bold}{U}{stmry}{m}{n}

\usepackage{hyperref}
\usepackage{pgfplots}
\pgfplotsset{compat=1.8}
\usetikzlibrary{decorations.pathreplacing,calligraphy,backgrounds}

\usepackage{tikz} 
\usetikzlibrary{topaths,calc,shapes,decorations.pathmorphing,arrows}

\usepackage{amsmath,amsthm}
\input{macros}

\title{F-IVM: Analytics over Relational Databases under Updates}

\author{Ahmet Kara \and Milos Nikolic \and Dan Olteanu \and Haozhe Zhang}

\author{
Ahmet Kara$^1$, 
Milos Nikolic$^2$, 
Dan Olteanu$^1$, 
Haozhe Zhang$^1$ 
\\ \\
$^1$University of Zurich  
\enspace\enspace 
$^2$University of Edinburgh
}
\date{}

\begin{document}

\maketitle 
\begin{abstract}
    This article describes \DF,  a unified approach for maintaining analytics over changing relational data. We exemplify its versatility in four disciplines: processing queries with group-by aggregates and joins; learning linear regression models using the covariance matrix of the input features; building Chow-Liu trees using pairwise mutual information of the input features; and matrix chain multiplication. 

    \DF has three main ingredients: higher-order incremental view maintenance; factorized computation; and ring abstraction. \DF reduces the maintenance of a task to that of a hierarchy of simple views. Such views are functions mapping keys, which are tuples of input values, to payloads, which are elements from a ring. \DF supports efficient factorized computation over keys, payloads, and updates. It treats uniformly seemingly disparate tasks: While in the key space, all tasks require general joins and variable marginalization, in the payload space, tasks differ in the definition of the sum and product ring operations.
 
    We implemented \DF on top of DBToaster and show that it can outperform classical first-order and fully recursive higher-order incremental view maintenance by orders of magnitude while using less memory.
\end{abstract}

\paragraph{Acknowledgements}
This project has received funding from the European Union's Horizon 2020 research and innovation programme under grant agreement No 682588.

\input{introduction}
\input{systemoverview}

\input{preliminaries}
\input{factorized_ring_computation}

\input{factorized_ivm}
\input{factorized_updates}

\input{query_classes}
\input{cyclic}
\input{rings}

\input{experiments}
\input{related}

\input{conclusion}

\bibliographystyle{abbrv}
\bibliography{bibliography}

\end{document}

%% file: macros.tex
\definecolor{goodgreen}{rgb}{0.1, 0.5, 0.1}

\lstdefinelanguage{SQL}{
  basicstyle=\linespread{1.05}\ttfamily\small,
  morekeywords={SELECT, FROM, NATURAL, JOIN, GROUP, BY},
  keywordstyle=\linespread{1.05}\ttfamily\small\color{blue!45!black},
  tabsize=2,
  numbers=none,
  stringstyle=\color{white}\ttfamily,
  showspaces=false,
  showtabs=false,
  showstringspaces=false
}

\newcommand{\vecnormal}[1]{\textnormal{\bf #1}\xspace}

\newcommand{\textvec}[1]{\textnormal{\bf #1}\xspace}
\newcommand{\nop}[1]{}

\newcommand{\db}{\mathcal{D}}
\newcommand{\Dom}{\mathsf{Dom}}
\newcommand{\Codom}{\textnormal{\bf D}\xspace}

\newcommand{\bigexists}{\mathop{\lower0.5ex\hbox{\scalebox{1.5}{\ensuremath{\exists}}}}}
\newcommand{\punto}{$\hspace*{\fill}\Box$}

\newcommand{\bigO}[1]{\mathcal{O}(#1)}

\newcommand{\sch}{\mathsf{sch}}
\newcommand{\vars}{\text{vars}}
\newcommand{\tuple}[1]{(#1)}

\newcounter{CommentCounter}

\newcommand{\add}[1]{{\color{black} {#1}}}

\newcommand{\TAB}{\makebox[2.5ex][r]{}}%
\newcommand{\STAB}{\makebox[1.5ex][r]{}}%
\newcommand{\SPACE}{\makebox[0.75ex][r]{}}%

\newcommand{\LET}{\textbf{let}\xspace}%
\newcommand{\IF}{\textbf{if}\xspace}%
\newcommand{\ELSE}{\textbf{else}\xspace}%
\newcommand{\RETURN}{\textbf{return}\xspace}%
\newcommand{\MATCH}{\textbf{switch}\xspace}%
\newcommand{\VIEW}[2][]{\ifthenelse{\isempty{#1}}{\mathsf{#2}\xspace}{\mathsf{#2[\mathit{#1}]}\xspace}}
\newcommand{\VPLUS}{\uplus}
\newcommand{\VPROD}{\otimes}
\newcommand{\VPRODBIG}{\bigotimes}
\newcommand{\VSUM}{\textstyle\bigoplus}
\newcommand{\VEXISTS}[1]{\bigexists_{#1}}

\newcommand{\RING}{\textnormal{\bf D}\xspace}
\newcommand{\RINGPLUS}{+}
\newcommand{\RINGPROD}{*}
\newcommand{\RINGZERO}{\bm{0}}
\newcommand{\RINGONE}{\bm{1}}

\newcommand{\Th}{\bm{\theta}\xspace}%
\newcommand{\TR}[1]{#1^{\text{T}}}
\newcommand{\LRringC}{c\xspace}
\newcommand{\LRringS}{\bm{s}\xspace}
\newcommand{\LRringQ}{\bm{Q}\xspace}

\newcommand{\DF}{F-IVM\xspace}%
\newcommand{\DFONE}{F-IVM-ONE\xspace}%
\newcommand{\DBT}{DBT\xspace}%
\newcommand{\DBTRING}{DBT-RING\xspace}%
\newcommand{\IVM}{1-IVM\xspace}%
\newcommand{\SQLOPT}{SQL-OPT\xspace}%
\newcommand{\DFRE}{F-RE\xspace}%
\newcommand{\DBTRE}{DBT-RE\xspace}%

\newcommand{\calI}{\mathcal{I}}
\newcommand{\calR}{\mathcal{R}}

\newcommand{\calX}{\mathcal{X}}

\newcommand{\calS}{\mathcal{S}}
\newcommand{\calA}{\mathcal{A}}
\newcommand{\calB}{\mathcal{B}}

\newcounter{magicrownumbers}

\theoremstyle{plain}                  
\newtheorem{theorem}{Theorem}

\newtheorem{definition}[theorem]{Definition}
\newtheorem{remark}[theorem]{Remark}

\theoremstyle{definition}
\newtheorem{example}[theorem]{Example}

\newtheoremstyle{cited}%
{.5\baselineskip\@plus.2\baselineskip
    \@minus.2\baselineskip}
{.5\baselineskip\@plus.2\baselineskip
    \@minus.2\baselineskip}
{\itshape}
{\parindent}
{}
{.}
{.5em}
{\textsc{\thmname{#1}} \thmnote{\normalfont#3}}

%% file: introduction.tex
\section{Introduction}
\label{sec:introduction}

Supporting modern applications that rely on accurate and real-time analytics  computed over large and continuously evolving databases is a challenging data management problem~\cite{LB:SIGMOD:2015}. Special cases are the classical problems of incremental view maintenance (IVM)~\cite{Chirkova:Views:2012:FTD,DBT:VLDBJ:2014} and stream query processing~\cite{abadi2005design,madden2005tinydb}. 

Recent efforts studied the problem of computing machine learning (ML) tasks over {\em static} databases. The predominant approach loo\-sely integrates the database systems with the statistical packages \cite{MADlib:2012,Rusu:2015,MLlib:JMLR:2016,Polyzotis:SIGMOD:Tutorial:17,Kumar:SIGMOD:Tutorial:17}: First, the database system computes the input to the statistical package by joining the database relations. It then exports the join result to the statistical package for training ML models. This approach precludes real-time analytics due to the expensive export/import steps. 
Systems like Morpheus~\cite{KuNaPa15} and LMFAO~\cite{LMFAO:SIGMOD:2019} push the ML task inside the database and learn ML models over static normalized data. 
In particular, LMFAO, and its precursors F~\cite{SOC:SIGMOD:2016} and AC/DC~\cite{ANNOS:TODS:2020}, decompose the task of learning classification and regression models over arbitrary joins into factorized computation of aggregates over joins and fixpoint computation of model parameters. This factorization may significantly lower the complexity by avoiding the computation of Cartesian products lurking within joins~\cite{BKOZ:PVLDB:2013,Olteanu:FactBounds:2015:TODS}. Both the tight integration of the database computation step and of the statistical computation step as well as the factorized computation are pre-requisites for real-time analytics.

This article describes \DF\footnote{\url{https://github.com/fdbresearch/FIVM}.}, a unified approach for maintaining analytics over changing relational data.  We exemplify its versatility in four disciplines: processing queries with group-by aggregates and joins; learning linear regression models using the covariance matrix of the input features; building Chow-Liu trees using pairwise mutual information matrix of the input features; and matrix chain multiplication. 

\DF was introduced in prior work~\cite{FIVM:SIGMOD:2018}. This article revisits and extends this prior work with: a more refined analysis of \DF for the $q$-hierarchical and free-connex acyclic  queries in the presence of functional dependencies; the covariance ring over continuous and categorical features; an overview of the design of \DF; further experiments on: the covariance matrix; end-to-end linear regression models; Chow-Liu trees; $q$-hierarchical queries with eager and lazy approaches and payloads carrying the listing or the factorized representation of the query result; and path queries of increasing length on graph data to stress-test the scalability of the IVM engines.

\DF has three main ingredients: higher-order incremental view maintenance (IVM); factorized computation and data representation; and ring abstraction.

The first ingredient reduces the maintenance task to that of a hierarchy of simple views. Such views are functions mapping keys, which are tuples of input values, to payloads, which are elements from a ring. In contrast to classical (first-order) IVM, which computes changes to the query result on the fly and does not use extra views, \DF can significantly speed up the maintenance task and lower its complexity by using carefully chosen views. Yet \DF can use substantially fewer views than the fully-recursive IVM, which is used by the state-of-the-art IVM system DBToaster~\cite{DBT:VLDBJ:2014}. In our experiments, \DF outperforms first-order and higher-order IVM by up to two orders of magnitude in both runtime and memory requirements.

The second ingredient supports efficient computation and representation for keys, payloads, and updates. \DF exploits insights from query evaluation algorithms with best known complexity and optimizations that push aggregates past joins~\cite{BKOZ:PVLDB:2013,Olteanu:FactBounds:2015:TODS,FAQ:PODS:2016}. It can process bulk updates expressed as low-rank decompositions~\cite{TensorDecomp:2009,TensorDecomposition:2017} and maintain a factorized representation of query results, which is essential to achieve low complexity for free-connex acyclic and $q$-hierarchical queries.

The third ingredient allows \DF to treat uniformly seemingly disparate tasks. In the key space, all tasks require joins and variable marginalization. In the payload space, tasks differ in the ring operations. To maintain linear regression models and Chow-Liu trees under updates, \DF uses a new ring that captures the maintenance of a covariance matrix over continuous and categorical features from the input database. Furthermore, it composes rings to capture the data-dependent computation for complex analytics. Thanks to the ring abstraction, \DF is highly extensible: efficient maintenance for new analytics over relational databases is readily available as long as they come with appropriate sum and product ring operations.

\subsection{\DF by Example}
\label{ex:sql_sum_aggregate_intro}

\begin{figure}[t]
  \centering   
  \includegraphics[width=0.5\columnwidth]{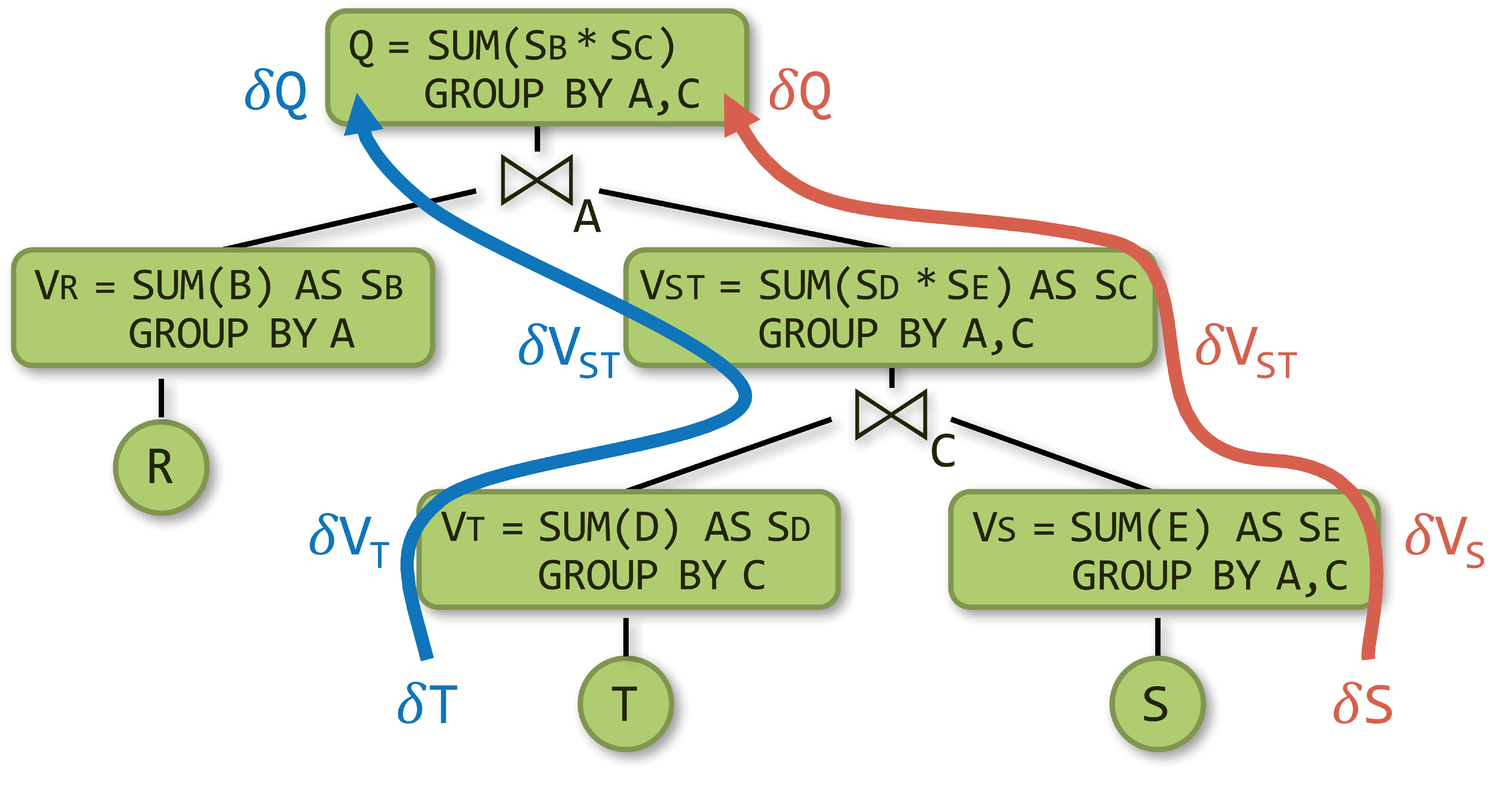}
  \caption{View tree for the query in Example~\ref{ex:sql_sum_aggregate_intro}. The propagation paths for updates to $S$ (right red) and to $T$ (left blue).}
  \label{fig:view_tree_sql}
\end{figure}

Consider the following SQL query over a database $\db$ with relations $R(A,B)$, $S(A,C,E)$, and $T(C,D)$:
\begin{lstlisting}[language=SQL, mathescape, columns=fullflexible]
  Q := SELECT A,$\;$C,$\;$SUM(B$\,$*$\,$D$\,$*$\,$E) 
       $\,$FROM   R NATURAL$\;$JOIN S NATURAL$\;$JOIN T
       $\,$GROUP$\;$BY A,$\;$C;
\end{lstlisting}
A na\"{i}ve query evaluation approach first computes the join and then the aggregate. This takes $\bigO{N^3}$ time, where $N$ is the size of $\db$.
An alternative approach exploits the distributivity of {\tt SUM} over multiplication to partially push the aggregate past joins and then combine the partial aggregates. For instance, one such partial sum over $S$ can be expressed as the view V$_\texttt{S}$:
\begin{lstlisting}[language=SQL, mathescape, columns=fullflexible]
  V$_\texttt{S}$ := SELECT A,$\;$C,$\;$SUM(E)$\;$AS$\;$S$_\texttt{E}$ 
        $\,$FROM S GROUP$\;$BY A,$\;$C;
\end{lstlisting}
In the view V$_\texttt{S}$, we identify keys, which are tuples over $(A,C)$, and payloads, which are aggregate values S$_\texttt{E}$. 
Similarly, we compute partial sums over \texttt{R} and \texttt{T} as views V$_\texttt{R}$ and V$_\texttt{T}$. These views are joined as depicted by the {\em view tree} in Figure~\ref{fig:view_tree_sql}, which is akin to a query plan with aggregates pushed past joins. This view tree computes the result of $Q$ in $\bigO{N}$ time.

Consider now the problem of learning, for each pair $(a,c)$ of $(A,C)$-values in the natural join of $R$, $S$, and $T$, a linear function $f_{a,c}$ with parameters $\theta_{0}$, $\theta_{D}$ and $\theta_{E}$ that predicts the label $B$ given features $D$ and $E$:
\begin{align*}
f(D, E) = \theta_{0} + \theta_{D} \cdot D + \theta_{E} \cdot E
\end{align*}
Our insight is that the same view tree in Figure~\ref{fig:view_tree_sql} can compute the gradient vector used for learning $f_{a,c}$, where we replace the SQL \texttt{SUM} and \texttt{*} operators.

As shown in Section~\ref{sec:application-lr}, the gradient of the square loss objective function needs the computation of three types of aggregates: the scalar $\LRringC$ that is the count aggregate \texttt{SUM(1)}; the vector $\LRringS$ of linear aggregates \texttt{SUM(i)}, for $i\in\{\texttt{B},\texttt{D},\texttt{E}\}$; and the matrix $\LRringQ$ of quadratic aggregates \texttt{SUM($i*j$)}, where $i,j\in\{\texttt{B},\texttt{D},\texttt{E}\}$. These aggregates capture the correlation between the features and the label.

We treat these aggregates as one compound aggregate $(\LRringC,\LRringS,\LRringQ)$ so we can share computation across them. This compound aggregate can be partially pushed past joins similarly to the \texttt{SUM} aggregate discussed before. Its values are carried in the key payloads of views in the view tree from Figure~\ref{fig:view_tree_sql}. For instance, the partial compound aggregate $(\LRringC_\texttt{T},\LRringS_\texttt{T},\LRringQ_\texttt{T})$ at the view V$_\texttt{T}$ computes, for each $C$-value, the count, sum, and sum of squares of the $D$-values in $T$. Similarly, the partial aggregate $(\LRringC_\texttt{S},\LRringS_\texttt{S},\LRringQ_\texttt{S})$ at the view V$_\texttt{S}$ computes, for each pair $(A,C)$, the count, sum, and sum of squares of $E$-values in $S$. In the view V$_\texttt{ST}$, which is the join of V$_\texttt{T}$ and V$_\texttt{S}$, each key $(a,c)$ is associated with the multiplication of the payloads for the keys $c$ in V$_\texttt{T}$ and $(a,c)$ in V$_\texttt{S}$. This multiplication works on compound aggregates: The scalar $\LRringC_\texttt{ST}$ is the arithmetic multiplication of $\LRringC_\texttt{T}$ and $\LRringC_\texttt{S}$; the vector of linear aggregates $\LRringS_\texttt{ST}$ is the sum of the scalar-vector products $\LRringC_\texttt{T}\LRringS_\texttt{S}$ and $\LRringC_\texttt{S}\LRringS_\texttt{T}$; finally, the matrix 
$\LRringQ_\texttt{ST}$ of quadratic aggregates is the sum of the scalar-matrix products $\LRringC_\texttt{T}\LRringQ_\texttt{S}$ and $\LRringC_\texttt{S}\LRringQ_\texttt{T}$, and of the outer products of the vectors $\LRringS_\texttt{T}$ and the transpose of $\LRringS_\texttt{S}$ and also of $\LRringS_\texttt{S}$ and the transpose of $\LRringS_\texttt{T}$. Our approach shares the computation across the aggregates: The scalar aggregates are used to scale up the linear and quadratic aggregates, while the linear aggregates are used to compute the quadratic aggregates.

We now turn to incremental view maintenance. \DF operates over view trees. Whereas for non-incre\-mental computation we only materialize the top view in the tree and the input relations, for incremental computation we may materialize additional views to speed up the maintenance task. Our approach is an instance of higher-order IVM, where an update to one relation may trigger the maintenance of several views. 

Figure~\ref{fig:view_tree_sql} shows the leaf-to-root maintenance paths under changes to $\texttt{S}$ and $\texttt{T}$. For updates $\delta{\texttt{S}}$ to $\texttt{S}$, each delta view $\delta{V_\texttt{S}}$, $\delta{V_\texttt{ST}}$, and $\delta{\texttt{Q}}$, is computed using delta rules:
%
\begin{lstlisting}[language=SQL, mathescape, columns=flexible] 
 $\delta$V$_\texttt{S}$ := $\,$SELECT A,$\;$C,$\;$SUM(E)$\;$AS$\;$S$_\texttt{E}$ 
        $\,\,$FROM $\delta$S GROUP$\;$BY A,$\;$C;
$\;\;\delta$V$_\texttt{ST}\,$:= SELECT A,$\;$C,$\;$SUM(S$_\texttt{D}$$\,$*$\,$S$_\texttt{E}$)$\;$AS$\;$S$_\texttt{C}$ 
        $\,\,$FROM V$_\texttt{T}$ NATURAL JOIN $\delta$V$_\texttt{S}$ GROUP$\;$BY A,$\;$C;
  $\delta$Q := SELECT A,$\;$C,$\;$SUM(S$_\texttt{B}$$\,$*$\,$S$_\texttt{C}$)
        $\,\,$FROM V$_\texttt{R}$ NATURAL JOIN $\delta$V$_\texttt{ST}$ GROUP$\;$BY A,$\;$C;
\end{lstlisting}
%
An update may consist of both inserts and deletes, which are encoded as keys with positive and respectively negative payloads. For the count aggregate, the payload is $1$ for an insert and   
$-1$ for a delete. For the compound aggregate, the payload is $(1, {\bf 0}_{5 \times 1}, {\bf 0}_{5 \times 5})$ for an insert and $(-1, {\bf 0}_{5 \times 1}, {\bf 0}_{5 \times 5})$ for a delete, where ${\bf 0}_{n \times m}$ is the $n$-by-$m$ matrix with all zero values.

\DF materializes and maintains views depending on the update workload. For updates to all input relations, it materializes each view in the view tree. For updates to \texttt{R} only, it materializes V$_\texttt{ST}$; for updates to \texttt{S} only, it materializes V$_\texttt{R}$ and V$_\texttt{T}$; for updates to \texttt{T} only, it materializes  V$_\texttt{R}$ and V$_\texttt{S}$. \DF takes constant time for updates to \texttt{S} and linear time for updates to \texttt{R} and \texttt{T}; these complexities are in the number of distinct keys in the views.
In contrast, the first-order IVM computes one delta query per each updated relation and without the use of extra views. It takes linear time for updates to any of the three relations for our example query. The fully-recursive higher-order IVM constructs a view tree for each delta query, so overall more views, including the view materializing the join of V$_\texttt{R}$, V$_\texttt{S}$, and $\delta{\texttt{T}}$.

\DF thus needs the same view tree and views for our query with one SUM aggregate and even for the learning task with the ten SUM aggregates. In contrast, the first-order IVM needs to compute a distinct delta query for each of these aggregates for updates to any of the three relations. DBToaster, which is the state-of-the-art fully recursive IVM, computes 31 views, ten top views and 21 auxiliary ones. Whereas \DF shares the computation across these aggregates, the other IVM approaches do not. This significantly widens the performance gap between \DF and its competitors.

%% file: systemoverview.tex
\section{Overview of the \DF System}
\label{sec:systemoverview}

\begin{figure}[t]
\centering
\includegraphics[width=0.8\textwidth]{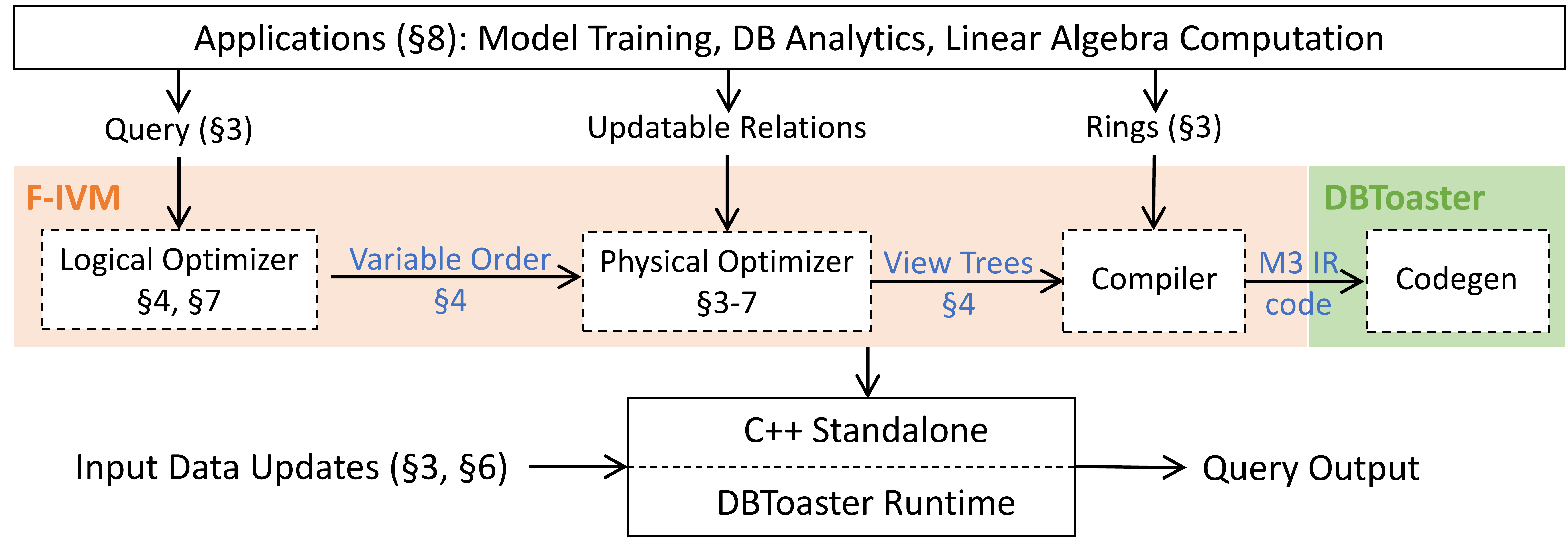}
\caption{Overview of the \DF system.}
\label{fig:system}
\end{figure}

Figure~\ref{fig:system} overviews the main components of \DF, annotated with the numbers of sections where they are discussed. Applications, e.g., database analytics,  training linear regression model and Chow-Liu trees, and linear algebra computation, rely on queries with natural joins and group-by aggregates, where each aggregate is expressed using the sum and product operations in a ring. In particular, Section~\ref{sec:applications} introduces the covariance ring over continuous and categorical features. Queries and rings serve as input to \DF, together with a stream of updates (tuple inserts and deletes) to the underlying database. Section~\ref{sec:preliminaries} details the data model, the query language supported by \DF, and the ring algebraic structure.

The logical optimizer creates a {\em variable order} for the input query (Section~\ref{sec:factorized_ring_computation}). This is akin to a query plan, albeit expressed as a partial order on the query variables as opposed to a partial order on the relations to join. Classical query evaluation uses query plans that dictate the order in which the relations are joined. \DF uses variable orders that dictate the  order in which the variables are marginalized.
For each join variable, all relations with that variable are joined. This choice is motivated by the observation that relation-at-a-time query plans is suboptimal in general, whereas the evaluation by variable orders is worst-case optimal~\cite{Ngo:SIGREC:2013}. 

Finding a good variable order for a given query is a computationally hard problem. For {\em q-hierarchical} queries~\cite{Nicole:PODS:2017}, we can efficiently find variable orders  that allow for maintenance with best guarantees in terms of update time and time to present the updated query result to the user (Section~\ref{sec:query_classes}). 
This also applies to queries, which become  q-hierarchical on databases that satisfy functional dependencies.

Given a variable order for a query, the physical optimizer creates a {\em view tree} (Section~\ref{sec:factorized_ring_computation}), which is a tree of views to support the maintenance and output enumeration of the query. Updates to base relations are propagated bottom-up in the tree, while output enumeration requires top-down access in the view tree. Depending on which base relations are updatable (dynamic) or non-updatable (static), \DF decides to materialize and maintain views in the view tree to support efficient propagation of the updates and avoid recomputation. Section~\ref{sec:factorized_IVM} discusses the view materialization problem, whereas Section~\ref{sec:factorizable_updates} discusses efficient update propagation.

Each view is accessed via indices with key-payload entries. Its primary index is a hash map over all its keys (Section~\ref{sec:preliminaries}). \DF may also need secondary and even tertiary indices, which are hash maps over different subsets of its keys. 
Such indices are updated lazily: the index updates are buffered and only executed when index access is required. 
The views for q-hierarchical queries require the primary indices to support updates that are propagated bottom-up in the view tree, and secondary indices to support output enumeration that proceeds top-down in the view tree (Section~\ref{sec:query_classes}).
\DF implements equality-based joins using in-memory hash-based join operators.
Aggregation is performed using variable marginalization. To marginalize a variable, 
\DF enumerates the entries with the same key, except for the marginalized variable, and applies the aggregation on these entries on the fly.

For a view tree and ring specification for each variable to be marginalized, the compiler outputs code in  DBToaster's intermediate representation language {\em M3}. DBToaster has its own optimizer and compiler that turns M3 code into highly optimized C++ code. This code takes the stream of input data updates, maintains the views, and enumerates the query output, relying on DBToaster's runtime library for data ingestion.

%% file: preliminaries.tex
\section{Data Model and Query Language}
\label{sec:preliminaries}

  The data model of \DF is based on relations over rings and its query language allows for natural joins and group-by aggregates over such relations.

\begin{definition}
A ring $(\RING, \RINGPLUS, \RINGPROD, \RINGZERO, \RINGONE)$ is a set $\RING$ with two closed binary operations $\RINGPLUS$ and $\RINGPROD$, the additive identity $\RINGZERO$, and the multiplicative identity $\RINGONE$ such that for all $a,b,c\in\RING$, the following 
axioms are satisfied:
\begin{enumerate}
    \item $a \RINGPLUS b = b\RINGPLUS a$.
    \item $(a \RINGPLUS b)\RINGPLUS c = a \RINGPLUS (b \RINGPLUS c)$.
    \item $\RINGZERO \RINGPLUS a = a \RINGPLUS \RINGZERO = a$.
    \item $\exists -a \in \RING: a \RINGPLUS (-a) = (-a) \RINGPLUS a = \RINGZERO$.
    \item $(a \RINGPROD b) \RINGPROD c = a \RINGPROD (b \RINGPROD c)$.
    \item $a \RINGPROD \RINGONE = \RINGONE * a = a$.
    \item $a \RINGPROD (b \RINGPLUS c) = a \RINGPROD b \RINGPLUS a \RINGPROD c$ and $(a \RINGPLUS b) \RINGPROD c = a \RINGPROD c \RINGPLUS b \RINGPROD c$.
\end{enumerate}
A semiring $(\RING, \RINGPLUS, \RINGPROD, \RINGZERO, \RINGONE)$ satisfies all of the above properties 
except the additive inverse property (Property 4) and adds the axiom 
$\RINGZERO \RINGPROD a = a \RINGPROD \RINGZERO = \RINGZERO$.
A (semi)ring for which $a \RINGPROD b = b \RINGPROD a$ is commutative. 
\punto
\end{definition}

\begin{example}
The number sets $\mathbb{Z}$, $\mathbb{Q}$, $\mathbb{R}$, and $\mathbb{C}$ with arithmetic operations $+$ and $\cdot$ and numbers $0$ and $1$ form commutative rings. The set $\mathcal{M}$ of $(n \times n)$ matrices  forms a non-commutative ring $(\mathcal{M}, \cdot, +, 0_{n,n}, I_{n})$, where $0_{n,n}$ and $I_{n}$ are the zero matrix and the identity matrix of size $(n \times n)$. The set $\mathbb{N}$ of natural numbers  is a commutative semiring but not a ring because it has no additive inverse. Further examples are the max-product semiring $(\mathbb{R}_{+}, \max, \times, 0, 1)$, the Boolean semiring $(\{ \text{true}, \text{false} \}, \lor, \land, \text{false}, \text{true})$,  and the set semiring $(2^{U}, \cup, \cap, \emptyset, U)$ of all possible subsets of a given set $U$.\punto
\end{example}


\paragraph{\textbf{Data.}}

A schema $\mathcal{S}$ is a set of variables. Let $\Dom(X)$ denote the domain of a variable $X$. A tuple $\vecnormal{t}$ over schema $\mathcal{S}$ has the domain $\Dom(\mathcal{S}) = \prod_{X \in \mathcal{S}}{\Dom(X)}$. The empty tuple $\tuple{}$ is the tuple over the empty schema.

Let $(\RING,\hspace{-0.05em} \RINGPLUS,\hspace{-0.05em} \RINGPROD,\hspace{-0.05em} \RINGZERO,\hspace{-0.05em} \RINGONE)$ be a ring. A {\em relation} $\VIEW{R}$ over schema $\mathcal{S}$ and the ring $\RING$ is a function $\VIEW{R}: \Dom(\mathcal{S}) \to \Codom$ mapping tuples over schema $\mathcal{S}$ to values in $\Codom$ such that $\VIEW{R}[\vecnormal{t}] \neq \RINGZERO$ for finitely many tuples $\vecnormal{t}$. The tuple $\vecnormal{t}$ is called a {\em key}, while its mapping $\VIEW{R}[\vecnormal{t}]$ is the {\em payload} of $\vecnormal{t}$ in $\VIEW{R}$. We use $\sch(\VIEW{R})$ to denote the schema of $\VIEW{R}$. 
The statement $\vecnormal{t} \in \VIEW{R}$ tests if $\VIEW{R}[\vecnormal{t}] \neq \RINGZERO$. The size $|\VIEW{R}|$ of $\VIEW{R}$ is the size of the set $\{ \vecnormal{t} \mid \vecnormal{t} \in \VIEW{R} \}$, which consists of all keys with non-$\RINGZERO$ payloads. 
A database $\db$ is a collection of relations over the same ring. Its size $|\db|$ is the sum of the sizes of its relations. 
This data model is in line with prior work on $K$-relations over provenance semirings~\cite{Green:2007:ProvenanceSemirings}, generalized multiset relations~\cite{Koch:Ring:2010:PODS}, and factors over semirings~\cite{FAQ:PODS:2016}.

Each relation  or materialized view $\VIEW{R}$  over schema $\mathcal{S}$ is implemented 
as a hash map or a multidimensional array that stores 
key-payload entries $(\vecnormal{t},\VIEW{R}[\vecnormal{t}])$ for each tuple 
$\vecnormal{t}$ with $\VIEW{R}[\vecnormal{t}] \neq \RINGZERO$.
The data structure can:
(1) look up, insert, and delete entries in amortized constant time, and
(2) enumerate all stored entries in $\VIEW{R}$ 
with constant \emph{delay}, i.e., the following times are constant: 
(i) the time between the start of the enumeration and outputting the first tuple, (ii) the time between outputting any two consecutive tuples, and (iii) the time between outputting the last tuple and the end of the enumeration~\cite{DurandFO07}.
For a schema $\mathcal{X} \subset \mathcal{S}$,
we use an index data structure that for any $\vecnormal{t} \in \Dom(\mathcal{X})$ can:
(4) enumerate all tuples in $\sigma_{\mathcal{X}=\vecnormal{t}} \VIEW{R}$ with constant delay,
(5) check $\vecnormal{t} \in \pi_{\mathcal{X}}\VIEW{R}$ in amortized constant time; and
(7) insert and delete index entries in amortized constant time.

We give a hash-based example data structure that supports the above operations with the stated complexities.
Consider a relation $R$ over schema $\mathcal{S}$. 
A hash table with chaining stores key-value entries of the form $(\vecnormal{t},R(\vecnormal{t}))$ for each tuple $\vecnormal{t}$ over $\mathcal{S}$ with  
$R(\vecnormal{t}) \neq \RINGZERO$. 
The entries are doubly linked to support enumeration with constant delay. 
The hash table can report the number of its entries in constant time and supports lookups, inserts, and deletes in {amortized} constant time.
%
To support index operations on a schema $\mathcal{X} \subset \mathcal{S}$, 
we create another hash table with chaining where each table entry stores an $\calX$-value 
$\vecnormal{t}$ as key and a doubly-linked list of pointers to the entries in $R$ having 
$\vecnormal{t}$ as $\mathcal{X}$-value.
Looking up an index entry given $\vecnormal{t}$ takes {amortized} constant time,
and its doubly-linked list enables enumeration of the matching entries in $R$ with constant delay. 
Inserting an index entry into the hash table additionally prepends a new pointer to the doubly-linked list for a given $\vecnormal{t}$; overall, this operation takes {amortized} constant time.
For efficient deletion of index entries, each entry in $R$ also stores back-pointers to its index entries (one back-pointer per index for $R$). 
When an entry is deleted from $R$, locating and deleting its index entries in doubly-linked lists takes constant time per index.


\paragraph{\textbf{Query Language.}}
We consider queries with natural joins and group-by aggregates:
\begin{lstlisting}[language=SQL,mathescape]
  SELECT$\;X_1,\ldots,X_f$,$\;$SUM$(g_{f+1}(X_{f+1}) * ... * g_{m}(X_{m}))$
  FROM $R_1$ NATURAL$\;$JOIN $\ldots$ NATURAL$\;$JOIN $R_n$
  GROUP$\;$BY $X_1,\ldots,X_f$
\end{lstlisting}
The group-by variables $X_1,\ldots,X_f$ are {\em free}, while the other variables $X_{f+1},\ldots,X_m$ are {\em bound}. The {\tt SUM} aggregate values are from a ring $(\RING, \RINGPLUS, \RINGPROD, \RINGZERO, \RINGONE)$. The \texttt{SUM} operator uses the addition 
$\RINGPLUS$ from $\RING$. Further aggregates can be expressed using the sum and product operations from the ring. 
A {\em lifting} function $g_{k}: \Dom(X_k) \to \RING$, for $f < k \leq m$, maps $X_k$-values to elements in $\RING$:
when marginalizing $X_k$,  we aggregate the values $g_{k}(x)$ from $\RING$ and not the values $x$ from $\Dom(X_k)$.

Instead of the verbose SQL notation, we use the following more compact encoding:
\begin{align*}
\qquad  \VIEW[X_1,\ldots,X_f]{Q} = \VSUM_{X_{f+1}} \cdots \VSUM_{X_{m}} \VPRODBIG_{i \in [n]} \VIEW[\mathcal{S}_i]{R_i}
\end{align*}
where $\VPRODBIG$ is the join operator, $\VSUM_{X_{f+1}}$ is the aggregation operator that  marginalizes over the variable $X_{f+1}$, and each relation $\VIEW{R_i}$ is a function mapping keys over schema $\mathcal{S}_i$ to payloads in $\RING$. We also need a union operator $\VPLUS$ to express updates (insert/delete) to relations.

\begin{example}
\label{ex:sql_count}
The SQL query
\begin{lstlisting}[language=SQL,mathescape,columns=flexible]
  SELECT SUM(1) FROM R NATURAL JOIN S NATURAL JOIN T 
\end{lstlisting}
over tables $R(A,B)$, $S(A,C,E)$, and $T(C,D)$ can be encoded as follows in our formalism.
The table $R$ is encoded as a relation $\VIEW{R}: \Dom(A) \times \Dom(B) \to \mathbb{Z}$ that maps tuples $(a,b)$ to their multiplicity in $R$; similarly, we encode the tables $S$ and $T$ as relations $\VIEW{S}$ and $\VIEW{T}$. We translate the SQL query into: 
  $$
    \VIEW[~]{Q} = \underset{A,B,C,D,E}{\VSUM} \VIEW[A,B]{R}  \VPROD \VIEW[A,C,E]{S} \VPROD \VIEW[C,D]{T}
  $$
  where $\underset{A,B,C,D,E}{\VSUM}$ abbreviates $\VSUM_A \cdots \VSUM_E$. The lifting functions used for marginalization map all values to $1$.
  Recall that by definition $\VIEW{R}$, $\VIEW{S}$, and $\VIEW{T}$ are finite.
  The relation $\VIEW{Q}$ maps the empty tuple $()$ to the count. 
  \punto
  \end{example}

Given a ring $(\RING, \RINGPLUS, \RINGPROD, \RINGZERO, \RINGONE)$, relations $\VIEW{R}$ and $\VIEW{S}$ over schema $\mathcal{S}_1$ and relation $\VIEW{T}$ over schema $\mathcal{S}_2$, a variable $X \in \mathcal{S}_1$, and a lifting function $g_{X}:\Dom(X) \to \RING$, we define the three operators as follows: 

\hspace{-1.9em} 
\begin{tabular}{@{~}l@{~~~~~}r@{~}r@{\;}l@{}}
\multicolumn{4}{@{~}l}{\em union:}\\
& $\forall\vecnormal{t} \in \mathsf{D}_1{:}$ & 
  $(\VIEW{R} \VPLUS \VIEW{S})[\vecnormal{t}]$ & 
  $= \VIEW{R}[\vecnormal{t}] + \VIEW{S}[\vecnormal{t}]$ \\[4pt]

\multicolumn{4}{@{~}l}{\em join:}\\
& $\forall\vecnormal{t} \in \mathsf{D}_2{:}$ & 
  $(\VIEW{S} \VPROD \VIEW{T})[\vecnormal{t}]$ & 
  $= \VIEW{S}[\pi_{\mathcal{S}_1}(\vecnormal{t})] * \VIEW{T}[\pi_{\mathcal{S}_2}(\vecnormal{t})]$\\[4pt]

\multicolumn{4}{@{~}l}{\em aggregation by marginalization:}\\[4pt]
& $\forall\vecnormal{t} \in \mathsf{D}_3{:}$ & 
  $(\VSUM_{X} \VIEW{R})[\vecnormal{t}]$ & 
  $= \textstyle\sum \,\{\, \VIEW{R}[\vecnormal{t}_1] \,\RINGPROD\, g_{X}(\pi_{\{X\}}(\vecnormal{t}_1)) \mid$
  $\vecnormal{t}_1 \,{\in}\, \mathsf{D}_1, \vecnormal{t} = \pi_{\mathcal{S}_1 \setminus \{X\}}(\vecnormal{t}_1) \}$  
\end{tabular}
\\[6pt]
where $\mathsf{D}_1 = \Dom(\mathcal{S}_1)$, $\mathsf{D}_2 = \Dom(\mathcal{S}_1 \cup \mathcal{S}_2)$, and $\mathsf{D}_3 = \Dom(\mathcal{S}_1 \setminus \{X\})$, and $\pi_{\mathcal{S}}(\vecnormal{t})$ is a tuple representing the projection of tuple $\vecnormal{t}$ on the schema $\mathcal{S}$.

\begin{example}
Consider relations over a ring $(\RING,\hspace{-0.05em} \RINGPLUS,\hspace{-0.05em} \RINGPROD,\hspace{-0.05em} \RINGZERO,\hspace{-0.05em} \RINGONE)$:
\begin{center}
  \small
  \begin{tabular}{lll}
    \begin{tabular}[t]{@{\,}l@{~}l@{~$\to$~}l@{\,}}
      $A$ & $B$ & $\VIEW{R}[A,B]$ \\\toprule
      $a_1$ & $b_1$ & $r_1$ \\
      $a_2$ & $b_1$ & $r_2$ \\
    \end{tabular}
    &
    \begin{tabular}[t]{@{\,}l@{~}l@{~}l@{~}l@{\,}}
      A & B & $\to$ & $\VIEW[A,B]{S}$ \\\toprule
      $a_2$ & $b_1$ & $\to$ & $s_1$ \\
      $a_3$ & $b_2$ & $\to$ & $s_2$ 
    \end{tabular}
    &
    \begin{tabular}[t]{@{\,}l@{~}l@{~}l@{~}l@{\,}}
      B & C & $\to$ & $\VIEW[B,C]{T}$\\\toprule
      $b_1$ & $c_1$ & $\to$ & $t_1$ \\
      $b_2$ & $c_2$ & $\to$ & $t_2$ 
    \end{tabular}
  \end{tabular}
\end{center}
The values $r_1$, $r_2$, $s_1$, $s_2$, $t_1$, $t_2$ are non-$\RINGZERO$ values from $\RING$. 
The operators $\VPLUS$, $\VPROD$, and $\oplus$ are akin to union, join, and aggregation ($g_A:\hspace{-0.1em} \Dom(A) \hspace{-0.1em}\to\hspace{-0.1em} \RING$ is the lifting for $A$):

\vspace{-6pt}
\begin{center}
  \small  \hspace*{-1em}
  \begin{tabular}{l@{\hskip 0.5em}l}
    \begin{tabular}[t]{@{\,}l@{~}l@{~}l@{~}l@{\,}}
      $A$ & $B$ & $\to$ & $(\VIEW{R} \VPLUS \VIEW{S})[A,B]$ \\\toprule
      $a_1$ & $b_1$ & $\to$ & $r_1$ \\
      $a_2$ & $b_1$ & $\to$ & $r_2+s_1$ \\
      $a_3$ & $b_2$ & $\to$ & $s_2$
    \end{tabular}
    &
    \begin{tabular}[t]{@{\,}l@{~}l@{~}l@{~}l@{~}l@{\,}}
      $A$ & $B$ & $C$ & $\to$ & $\big((\VIEW{R} \VPLUS \VIEW{S}) \VPROD \VIEW{T}\big)[A,B,C]$ \\\toprule
      $a_1$ & $b_1$ & $c_1$ & $\to$ & $r_1*t_1$ \\
      $a_2$ & $b_1$ & $c_1$ & $\to$ & $(r_2 + s_1)*t_1$ \\
      $a_3$ & $b_2$ & $c_2$ & $\to$ & $s_2*t_2$
    \end{tabular}
  \end{tabular}\\[4ex]
  \begin{tabular}[t]{@{\,}l@{~}l@{~}l@{~}l@{\,}}
      $B$ & $C$ & $\to$ & $\big(\VSUM_{A}(\VIEW{R} \VPLUS \VIEW{S}) \VPROD \VIEW{T} \big)[B,C]$ \\\toprule
      $b_1$ & $c_1$ & $\to$ & $r_1 * t_1  \,*\, g_{A}(a_1) + (r_2 + s_1) * t_1  \,*\, g_{A}(a_2)$ \\
      $b_2$ & $c_2$ & $\to$ & $s_2 * t_2 \,*\, g_{A}(a_3)$
    \end{tabular}
  \end{center}
\end{example}

\begin{example}\label{ex:sql_sum_aggregate}
Let us consider the SQL query from Section~\ref{ex:sql_sum_aggregate_intro}, which computes \texttt{SUM(R.B$\,$*$\,$T.D$\,$*$\,$S.E)} grouped by $A$, $C$. Assume that $B$, $D$, and $E$ take values from $\mathbb{Z}$.
We model the tables $R$, $S$, and $T$ as relations mapping tuples to their multiplicity, as in 
Example~\ref{ex:sql_count}. 
The variables $A$ and $C$ are free, while  $B$, $D$, and $E$ are bound. 

When marginalizing over the bound variables, we apply the same lifting function to these variables: $\forall x \in \mathbb{Z}: g_{B}(x) = g_{D}(x) = g_{E}(x) = x$. The SQL query can be expressed in our formalism as follows:
$$
  \VIEW[A,C]{Q} = \underset{B,D,E}{\VSUM}\VIEW[A,B]{R} \VPROD \VIEW[A,C,E]{S} \VPROD \VIEW[C,D]{T}
$$
The computation of the aggregate \texttt{SUM(R.B$\,$*$\,$T.D$\,$*$\,$S.E)} now happens over payloads. 
\punto
\end{example}

By using relations over rings, we avoid the intricacies of incremental computation under multiset semantics caused by the non-commutativi\-ty of inserts and deletes. We simplify delta processing by representing both inserts and deletes as tuples, with the distinction that they map to positive and respectively negative ring values. This uniform treat\-ment allows for simple delta rules for the three operators of our query language.

\bigskip

%% file: factorized_ring_computation.tex
\section{Factorized Ring Computation}
\label{sec:factorized_ring_computation}

This section introduces a framework for query evaluation  based on factorized computation and data rings. The next section extends it to incremental maintenance. 

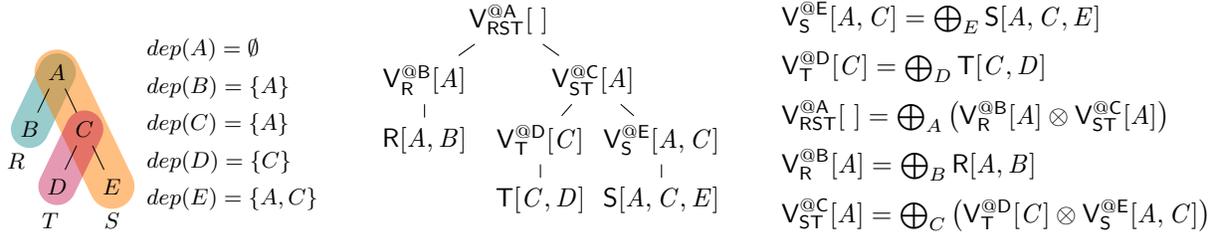
\begin{figure}[t]
\centering
\begin{minipage}[b]{0.3\linewidth}
    \scalebox{0.95}{
      \begin{tikzpicture}[xscale=0.96, yscale=0.8]
        \node at (0, 0.0) (A) {\small  $A$};
        \node at (-0.4, -1.0) (B) {\small $B$} edge[-] (A);
        \node at (0.4, -1.0) (C) {\small $C$} edge[-] (A);
        \node at (0.8, -2.0) (E) {\small $E$} edge[-] (C);
        \node at (0, -2.0) (D) {\small $D$} edge[-] (C);
        \node at (0.8, -2.6) (S) {\small $S$};
        \node at (-0.6, -1.55) (R) {\small  $R$};
        \node at (-0.1, -2.6) (T) {\small  $T$};
        \node at (2.55, -1.6) {
          \small
          \begin{tabular}{@{~~}l}
            $dep(A) = \emptyset$\\[1ex]
            $dep(B)=\{A\}$\\[1ex]
            $dep(C)=\{A\}$\\[1ex]
            $dep(D)=\{C\}$\\[1ex]
            $dep(E)=\{A,C\}$\\[8ex]
          \end{tabular}
        };
      
        \begin{pgfonlayer}{background}
          \draw[opacity=.4,fill opacity=.4,line cap=round, line join=round, line width=15pt,color=teal] (0,0.0) -- (-0.4,-1);
          \draw[opacity=.5,fill opacity=.5,line cap=round, line join=round, line width=18pt,color=orange] (0,0.0) -- (0.8,-2.0);
          \draw[opacity=.4,fill opacity=.4,line cap=round, line join=round, line width=15pt,color=purple]  (0.4,-1) -- (0, -2.0);
        \end{pgfonlayer}
      \end{tikzpicture}
      }
  \end{minipage}
  \begin{minipage}[b]{0.32\linewidth}
    \begin{tikzpicture}[xscale=0.7, yscale=0.8]
      \node at (0, 0) (A) {$\VIEW[~]{V^{@A}_{RST}}$};
      \node at (-1.6, -1) (B) {$\VIEW[A]{V^{@B}_{R}}$} edge[-] (A);
      \node at (1.6, -1) (C) {$\VIEW[A]{V^{@C}_{ST}}$} edge[-] (A);
      \node at (0.6, -2) (D) {$\VIEW[C]{V^{@D}_{T}}$} edge[-] (C);
      \node at (2.9, -2) (E) {$\VIEW[A,C]{V^{@E}_{S}}$} edge[-] (C);
      
      \node at (-1.6, -2) {$\VIEW[A,B]{R}$} edge[-] (B);
      \node at (2.9, -3) {$\VIEW[A,C,E]{S}$} edge[-] (E);
      \node at (0.6, -3) {$\VIEW[C,D]{T}$} edge[-] (D);

      \node at (0.5, -4.5) {}; 
    \end{tikzpicture}
  \end{minipage}
  \begin{minipage}[b]{0.3\linewidth}
    \begin{align*}
      &\VIEW[A,C]{V^{@E}_{S}} = \VSUM_{E} \VIEW[A,C,E]{S} \\[2pt]
      &\VIEW[C]{V^{@D}_{T}} = \VSUM_{D} \VIEW[C,D]{T} \\[2pt]
      &\VIEW[~]{V^{@A}_{RST}} = \VSUM_{A} \big(\VIEW[A]{V^{@B}_{R}} \VPROD \VIEW[A]{V^{@C}_{ST}}\big) \\[2pt]
      &\VIEW[A]{V^{@B}_{R}} = \VSUM_{B} \VIEW[A,B]{R} \\[2pt]
      &\VIEW[A]{V^{@C}_{ST}} = \VSUM_{C} \big(\VIEW[C]{V^{@D}_{T}} \VPROD \VIEW[A,C]{V^{@E}_{S}}\big)
      \\[0.5cm]
   \end{align*}  
  \end{minipage}  
  \vspace*{-0.7cm}
\caption{
(left) Variable order $\omega$ of the natural join of the relations $\VIEW[A,B]{R}$, $\VIEW[A,C,E]{S}$, and $\VIEW[C,D]{T}$; (middle) View tree over $\omega$ and $\mathcal{F} = \emptyset$;
(right)  View definitions. 
}
\label{fig:example_payloads}
\end{figure}


\paragraph{\textbf{Variable Orders.}}
Classical query evaluation makes use of  query plans that dictate the order in which the relations are joined. We use a different evaluation approach based on variable orders that dictate the  order in which we marginalize each join variable. This approach may require to join several relations at a time if they have the same variable. Our choice is motivated by the complexity of the evaluation problem for join queries: standard (relation-at-a-time) query plans are provably suboptimal, whereas the evaluation by variable orders can be worst-case optimal~\cite{Ngo:SIGREC:2013}.

Given a join query $Q$, a variable $X$ {\em depends} on a variable $Y$ if both are in the schema of a relation in $Q$.

\begin{definition}[adapted from \cite{Olteanu:FactBounds:2015:TODS}]\label{def:vo}
A {\em variable order} $\omega$ for a join query $Q$ is a pair $(F,dep)$, where $F$ is a rooted forest with one node per variable in $Q$, and {\em dep} is a function mapping each variable $X$ to a set of variables in $F$. It satisfies the following constraints:
\begin{itemize}
\item For each relation in $Q$, all of its variables lie along a root-to-leaf path in $F$.
\item For each variable $X$, $dep(X)$ is the subset of its ancestors in $F$ on which the variables in the subtree rooted at $X$ depend.
\end{itemize}
\end{definition}

\begin{example}
Consider the query
from Example~\ref{ex:sql_count} that joins the relations $\VIEW{R}[A,B]$, $\VIEW[A,C,E]{S}$, and 
$\VIEW[C,D]{T}$.   
Figure~\ref{fig:example_payloads}  gives a variable order (top left) for
the query.
  Variable $D$ has ancestors $A$ and $C$, yet it only depends on $C$ since $C$ and $D$ appear in the same relation $\VIEW[]{T}$ and $D$ does not occur in any relation together with $A$. Thus, $dep(D)=\{C\}$. Given $C$, the variables $D$ and $E$ are independent of each other.
\punto
\end{example}
For a query $Q$ with free variables, a variable order is \emph{free-top} if no bound variable is an ancestor of a free variable~\cite{KNOZ20}. Variable orders are a different syntax~\cite{Olteanu:FactBounds:2015:TODS} for hypertree decompositions~\cite{Gottlob99}. They are more natural for algorithms that proceed one variable at a time.

\textbf{View Trees.} Our framework relies on a variable order $\omega$ for the input query $Q$ to describe the structure of the computation and indicate which variable marginalizations are pushed past joins. Based on $\omega$, we construct a tree of views that represent \DF's data structure to support query maintenance and enumeration.

\begin{figure}[t]
\centering
\setlength{\tabcolsep}{3pt}
\renewcommand{\arraystretch}{1.2}
\begin{tabular}{@{}c@{}c@{~~~}l}
  \toprule
  \multicolumn{3}{c}{$\tau$(\text{variable order} $\omega$, \text{free variables} $\mathcal{F}$) : view tree} \\
  \midrule
  \multicolumn{3}{l}{\MATCH $\omega$:} \\
  \midrule 
  \phantom{a} & $\VIEW{R}$\hspace*{1em} & \RETURN $\VIEW[\mathit{\sch(\VIEW{R})}]{R}$ \\
  \cmidrule{2-3} \\[-6pt] 
  &
  \begin{minipage}[b]{1.8cm}
    \begin{tikzpicture}[xscale=0.4, yscale=1]
      \node at (0,-2)  (n4) {$X$};
      \node at (-1,-3)  (n1) {$\omega_1$} edge[-] (n4);
      \node at (0,-3)  (n2) {$\ldots$};
      \node at (1,-3)  (n3) {$\omega_k$} edge[-] (n4);
      \node at (0,-4.5) {~};
    \end{tikzpicture}
    \vspace{1.5cm}
  \end{minipage}
  &
  \begin{minipage}[b]{6.3cm}
\LET $T_i = \tau(\omega_i, \mathcal{F}), \ \forall i\in[k] $\\[0.5ex]  
  \LET $\VIEW[\mathit{keys_i}]{V^{@\omega_i}_{rels_i}} = \text{ root of } T_i, \ \forall i\in[k] $\\[0.5ex]
    \LET $\mathit{keys}=\mathit{dep}(X) \cup (\mathcal{F} \cap \vars(\omega))$ \\[0.5ex]
    \LET $\mathsf{rels}=\bigcup_{i\in[k]}\mathsf{rels}_i$\\[0.5ex]
    \IF $X \notin \mathcal{F}$ \\[0.5ex]
    \TAB $\VIEW[\mathit{keys}]{V^{@X}_{rels}}= \VSUM_{X} \VPRODBIG_{i \in [k]} \VIEW[\mathit{keys_i}]{V^{@\omega_i}_{rels_i}}$\\[0.5ex]
    \ELSE \\[0.5ex]
    \TAB $\VIEW[\mathit{keys}]{V^{@X}_{rels}}=\VPRODBIG_{i \in [k]} \VIEW[\mathit{keys_i}]{V^{@\omega_i}_{rels_i}}$\\[0.5ex]
    \RETURN $
				\left\{
				\begin{array}{@{~~}c@{~~}}
					\tikz {
						\node at (1.2,-1)  (n4) {$\VIEW[\mathit{keys}]{V^{@X}_{rels}}$};
						\node at (0.8,-1.75)  (n1) {$T_1$} edge[-] (n4);
						\node at (1.25,-1.75)  (n2) {$\ldots$};
						\node at (1.8,-1.75)  (n3) {$T_k$} edge[-] (n4);
					}
				\end{array}  \right.$
  \end{minipage}
  \\
  \bottomrule
\end{tabular}
\caption{Creating a view tree $\tau(\omega, \mathcal{F})$ for a variable order $\omega$ and a set of free variables $\mathcal{F}$.}
\label{fig:static_view_tree_algo}
\end{figure}
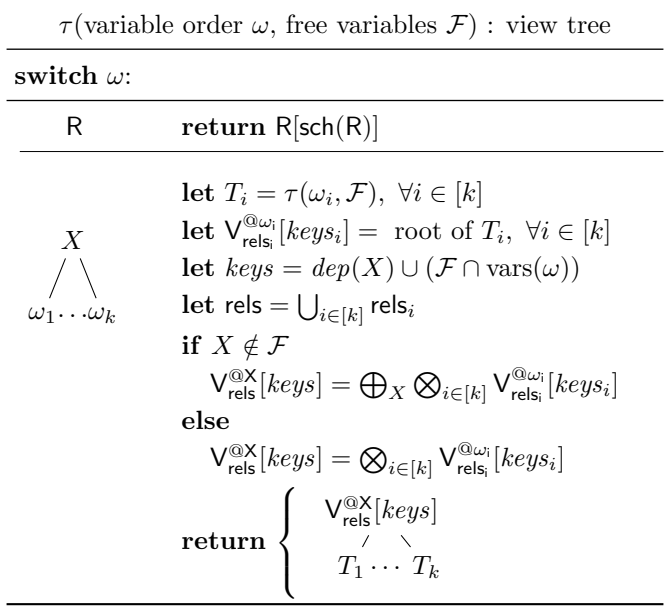

Figure~\ref{fig:static_view_tree_algo} gives a function $\tau$ that constructs a view tree $\tau$ for a variable order $\omega$ and the set $\mathcal{F}$ of free variables of the query $Q$. Without loss of generality, we assume that $\omega$ is a single rooted tree. Otherwise, we apply $\tau$  to each tree in $\omega$ to obtain a set of view trees.
For simplicity, we assume that $\omega$ was first extended with relations as children under their lowest variable. 

The function  $\tau$ maps the variable order to a view tree of the same tree structure, yet with each variable $X$ replaced by a view $\VIEW[keys]{V^{@X}_{rels}}$. This notation states that the view $\VIEW{V}$ is (recursively) defined over the input relations {\sf rels}, has free variables $keys$, and it corresponds to the variable $X$ in $\omega$; in case of a view for an input relation $\VIEW{R}$, we use the simplified notation $\VIEW[\sch(\VIEW{R})]{R}$.

The base case (leaf in the extended variable order) is that of an input relation: We construct a view that is the relation itself. At a variable $X$ (inner node), we distinguish two cases: If $X$ is a bound variable, then we construct a view that marginalizes out $X$ in the natural join of the views that are children of the current view; we thus first join on $X$, then apply the lifting function for $X$ on its values, and aggregate $X$ away. If $X$ is a free variable, however, then we retain it in the view schema without applying the lifting function to its values. The schema of the view consists of $dep(X)$ and the free variables in the subtree of $\omega$ rooted at $X$.

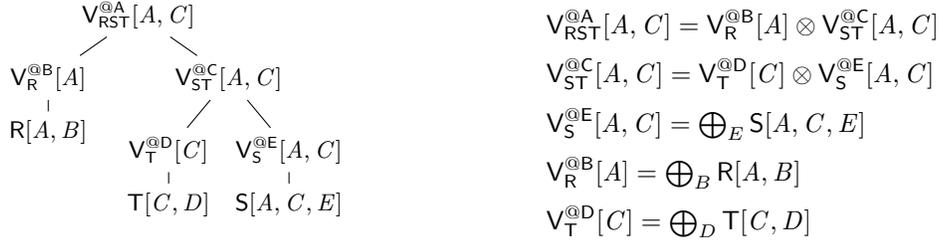
\begin{figure}[t]
  \centering
  \begin{tikzpicture}
        \node at (0, 0.5) (x) {
    \begin{minipage}[b]{0.2\linewidth}
      \small
      \hspace{-0.25cm}
      \begin{tikzpicture}[xscale=0.8, yscale=0.24]  
        \node at (0, -1.5) (A) {$\VIEW[A,C]{V^{@A}_{RST}}$};
        \node at (-1.5, -5) (B) {$\VIEW[A]{V^{@B}_{R}}$} edge[-] (A);
        \node at (1.5, -5) (C) {$\VIEW[A,C]{V^{@C}_{ST}}$} edge[-] (A);
        \node at (0.5, -9) (D) {$\VIEW[C]{V^{@D}_{T}}$} edge[-] (C);
        \node at (2.5, -9) (E) {$\VIEW[A,C]{V^{@E}_{S}}$} edge[-] (C);
        
        \node at (-1.5, -8) {$\VIEW[A,B]{R}$} edge[-] (B);
        \node at (2.5, -12) {$\VIEW[A,C,E]{S}$} edge[-] (E);
        \node at (0.5, -12) {$\VIEW[C,D]{T}$} edge[-] (D);
  \end{tikzpicture}
    \end{minipage}
    };

        \node at (8.2, 0.5) (x) {
     \begin{minipage}[b]{0.3\linewidth}       
        \begin{align*}
          &\VIEW[A,C]{V^{@A}_{RST}} = \VIEW[A]{V^{@B}_{R}} \VPROD \VIEW[A,C]{V^{@C}_{ST}} \\[2pt]
          &\VIEW[A,C]{V^{@C}_{ST}} = \VIEW[C]{V^{@D}_{T}} \VPROD \VIEW[A,C]{V^{@E}_{S}} \\[2pt]
          &\VIEW[A,C]{V^{@E}_{S}} = \VSUM_{E} \VIEW[A,C,E]{S} \\[2pt]
          &\VIEW[A]{V^{@B}_{R}} =  \VSUM_{B}\VIEW[A,B]{R} \\[2pt]
          &\VIEW[C]{V^{@D}_{T}} =  \VSUM_{D}\VIEW[C,D]{T}
        \end{align*}
    \end{minipage}
};
\end{tikzpicture}    
  \caption{
  (left) View tree over the variable order $\omega$ in Figure~\ref{fig:example_payloads} and 
  $\mathcal{F} = \{A,C\}$; (right) View definitions.
  }
  \label{fig:view tree_free_vars}
\end{figure}

\begin{figure}[t]
  \begin{minipage}[b]{\linewidth}
    \centering
    \small
    \begin{tikzpicture}[xscale=0.75, yscale=0.35]         
    
      \node at (2, 0) {
        \begin{tabular}{@{\,}l@{\,}  @{\,}c@{\,}c@{\,}l@{\,}}
          & $()$ & $\rightarrow$ & \ $\VIEW[\;]{V^{@A}_{RST}}$ \\[0.5ex]\toprule
          & $()$ & $\rightarrow$ & 
          $\VIEW{V^{@C}_{ST}}[a_1] \RINGPROD \VIEW{V^{@B}_{R}}[a_1] \RINGPROD g_A(a_1) \RINGPLUS \VIEW{V^{@C}_{ST}}[a_2] \RINGPROD \VIEW{V^{@B}_{R}}[a_2] \RINGPROD g_A(a_2) $           
          \\\bottomrule
        \end{tabular}
      };   

      \node[anchor=north west] at (-7, -2.5) {
        \begin{tabular}{@{\,}l@{\,} @{\,}c@{\,}c@{\,}l@{\,}}
          & $A$ & $\to$ & $\VIEW[A]{V^{@B}_{R}}$ \\[0.5ex]\toprule
          & $a_1$ & $\rightarrow$ & $p_{1} \RINGPROD  g_B(b_1)  + p_{2} \RINGPROD  g_B(b_2)$ \\
          & $a_2$ & $\rightarrow$ & $p_{3} \RINGPROD  g_B(b_3)$ \\
          & $a_3$ & $\rightarrow$ & $p_{4} \RINGPROD  g_B(b_4)$\\\bottomrule
        \end{tabular}
      };
      
      \node[anchor=north west] at (-7, -9) {
      \begin{tabular}{@{\,}l@{~~}l@{~$\to$~}l@{\,}}
        $A$ & $B$ & $\VIEW{R}[A,B]$\\[0.5ex]\toprule
        $a_1$ & $b_1$ & $p_1$ \\
        $a_1$ & $b_2$ & $p_2$\\  
        $a_2$ & $b_3$ & $p_3$\\
        $a_3$ & $b_4$ & $p_4$\\\bottomrule
        \end{tabular}
      };

      \node [anchor=north west] at (1, -2.5) {
        \begin{tabular}{@{\,}l@{\,} @{\,}c@{\,}c@{\,}l@{\,}}
          & $A$ & $\rightarrow$ & $\VIEW[A]{V^{@C}_{ST}}$ \\[0.5ex]\toprule
          & $a_1$ & $\rightarrow$ & 
          $\VIEW{V^{@E}_{S}}[a_1,c_1] \RINGPROD \VIEW{V^{@D}_{T}}[c_1] \RINGPROD g_C(c_1) + 
          \VIEW{V^{@E}_{S}}[a_1,c_2] \RINGPROD \VIEW{V^{@D}_{T}}[c_2] \RINGPROD g_C(c_2)$ \\[0.05cm]  
          & $a_2$ & $\rightarrow$ & $\VIEW{V^{@E}_{S}}[a_2,c_2] \RINGPROD \VIEW{V^{@D}_{T}}[c_2] \RINGPROD g_C(c_2)$ \\
          \bottomrule 
        \end{tabular}
      };

      \node [anchor=north west] at (1, -8) {
        \begin{tabular}{@{\,}l@{\,} @{\,}c@{\,}c@{\,}l@{\,}}
          & $C$ & $\to$ & $\VIEW[C]{V^{@D}_{T}}$ \\[0.5ex]\toprule
          & $c_1$ & $\rightarrow$ & $ p_9 \RINGPROD g_D(d_1)$\\
          & $c_2$ & $\rightarrow$ & $p_{10} \RINGPROD  g_D(d_2)  + p_{11} \RINGPROD  g_D(d_3)$ \\
          & $c_3$ & $\rightarrow$ & $p_{12} \RINGPROD  g_D(d_4)$ \\\bottomrule
        \end{tabular}
      };

      \node [anchor=north west] at (8, -8) {
        \begin{tabular}{@{\,}l@{\,} @{\,}c@{\,}c@{\,}c@{\,}l@{\,}}
          & $A$ & $C$ & $\to$ & $\VIEW[A,C]{V^{@E}_{S}}$ \\[0.5ex]\toprule
          & $a_1$ & $c_1$ & $\rightarrow$ & $p_{5} \RINGPROD  g_E(e_1)  + p_{6} \RINGPROD  g_E(e_2)$ \\
          & $a_1$ & $c_2$ & $\rightarrow$ & $p_{7} \RINGPROD  g_E(e_3)$  \\
          & $a_2$ & $c_2$ & $\rightarrow$ & $p_{8} \RINGPROD  g_E(e_4) $ \\\bottomrule
        \end{tabular}
      };
      
      \node [anchor=north west] at (1, -14.5) {
       \begin{tabular}{@{\,}l@{~~}l@{~$\to$~}l@{\,}}
        $C$ & $D$ & $\VIEW{T}[C,D]$ \\[0.5ex]\toprule
        $c_1$ & $d_1$ & $p_9$\\
        $c_2$ & $d_2$ & $p_{10}$\\
        $c_2$ & $d_3$ & $p_{11}$\\
        $c_3$ & $d_4$ & $p_{12}$\\\bottomrule
      \end{tabular}
      };

      \node [anchor=north west] at (8, -14.5) {
      \begin{tabular}{@{\,}l@{~~}l@{~~}l@{~$\to$~}l@{\,}}
        $A$ & $C$ & $E$ & $\VIEW{S}[A,C,E]$ \\[0.5ex]\toprule
        $a_1$ & $c_1$ & $e_1$ & $p_5$\\
        $a_1$ & $c_1$ & $e_2$ & $p_6$\\
        $a_1$ & $c_2$ & $e_3$ & $p_7$\\
        $a_2$ & $c_2$ & $e_4$ & $p_8$\\\bottomrule
      \end{tabular}
      };
    \end{tikzpicture}
  \end{minipage}
\caption{
Contents of the views in the view tree from Figure~\ref{fig:example_payloads} in case the relations 
$\VIEW{R}$,
$\VIEW{S}$, and $\VIEW{T}$ are over a ring 
$(\RING, \RINGPLUS, \RINGPROD, \RINGZERO, \RINGONE)$ with $p_i \in \RING$ for
$i \in [12]$.}
\label{fig:count}
\end{figure}

\begin{example}
\label{ex:views_count}
Figure~\ref{fig:example_payloads} shows the view tree constructed by the function $\tau$ from 
Figure~\ref{fig:static_view_tree_algo} over the variable order $\omega$ and the empty set of free variables.
Figure~\ref{fig:view tree_free_vars} depicts 
  the view tree constructed over the
  same variable order but for 
the set  $\mathcal{F} = \{A,C\}$ of free variables.
  
Figure~\ref{fig:count} gives the  
  contents of the views in the view tree from Figure~\ref{fig:example_payloads}, where 
$\VIEW{R}$,
$\VIEW{S}$, and $\VIEW{T}$ are relations over a ring $\RING$ with payloads $p_i \in \RING$ for
$i \in [12]$. 
Assume that $\RING$ is the $\mathbb{Z}$ ring, 
each tuple in these relations 
is mapped to $1$,
i.e.,  $p_i = 1$ for $i \in [12]$, and 
the lifting functions map all
values to $1$.
Then, the view tree 
computes the {\tt COUNT} query 
from Example~\ref{ex:sql_count}
and the root view $\VIEW{V_{RST}^{@A}}$ maps the empty tuple to 
  the overall count $10$,
which is the number 
of tuples in the natural join of $\VIEW{R}$, $\VIEW{S}$, and 
$\VIEW{T}$.
\punto
\end{example}

By default, the function $\tau$ in Figure~\ref{fig:static_view_tree_algo} constructs one view per variable in the variable order $\omega$. A wide relation (with many variables) leads to long branches in $\omega$ with variables that are only local to this relation. This is, for instance, the case of our retailer dataset used in 
Section~\ref{sec:experiments}. Such long branches create long chains of views, where each view marginalizes one bound variable over its child view in the chain. For practical reasons, we compose such long chains into a single view that marginalizes several variables at a time.

%% file: factorized_IVM.tex
\section{Factorized Higher-Order IVM}
\label{sec:factorized_IVM}

We introduce incremental view maintenance in our factorized ring computation framework. 
Unlike evaluation, the incremental maintenance of the query result may require the materialization and maintenance of views. 
An update to a relation $\VIEW{R}$ triggers changes in all views from the leaf $\VIEW{R}$ to the root of the view tree.

\textbf{Updates.} The insertion (deletion) of a tuple $\textvec{t}$ into (from) a relation $\VIEW{R}$ is expressed as a delta relation $\delta\VIEW{R}$ that maps $\textvec{t}$ to $\RINGONE$ (and respectively $-\RINGONE$). In general, $\delta\VIEW{R}$ can be a relation, thus a collection of tuples mapped to payloads. The updated relation is then the union of the old relation and the delta relation: $\VIEW{R} := \VIEW{R} \VPLUS \delta\VIEW{R}$.

\textbf{Delta Views.} For each view $\VIEW{V}$ affected by an update, a {\em delta view} $\VIEW{\delta{V}}$ defines the change in the view content. In case the view $\VIEW{V}$ represents a relation $\VIEW{R}$, then $\VIEW{\delta{V}}=\VIEW{\delta{R}}$ if there are updates to $\VIEW{R}$ and  $\VIEW{\delta{V}}=\emptyset$ otherwise. If the view is defined using operators on other views, $\delta\VIEW{V}$ is derived using the following delta rules:
\begin{align*}
  \quad
  \delta{(\VIEW{V_1} \VPLUS \VIEW{V_2})} &= \delta{\VIEW{V_1}} \VPLUS \delta{\VIEW{V_2}} \\
  \delta{(\VIEW{V_1} \VPROD \VIEW{V_2})} &= (\delta{\VIEW{V_1}} \VPROD \VIEW{V_2}) \VPLUS (\VIEW{V_1} \VPROD\delta{\VIEW{V_2}}) \VPLUS (\delta{\VIEW{V_1}} \VPROD \delta{\VIEW{V_2}})\\
  \delta{(\VSUM_{X}\VIEW{V})} &= \VSUM_{X}\delta{\VIEW{V}}
\end{align*}

The correctness of the rules follows from the associativity of $\VPLUS$ and the distributivity of $\VPROD$ over $\VPLUS$; $\VSUM_{X}$ is equivalent to the repeated application of $\VPLUS$ for the possible values of $X$. The derived delta views are subject to standard simplifications: If $\VIEW{V}$ is not defined over the updated relation $\VIEW{R}$, then its delta view $\VIEW{\delta{V}}$ is empty, and then we propagate this information using the identities $\emptyset \VPLUS \VIEW{V} = \VIEW{V} \VPLUS \emptyset = \VIEW{V}$ and $\emptyset \VPROD \VIEW{V} = \VIEW{V} \VPROD \emptyset = \emptyset$.

\medskip

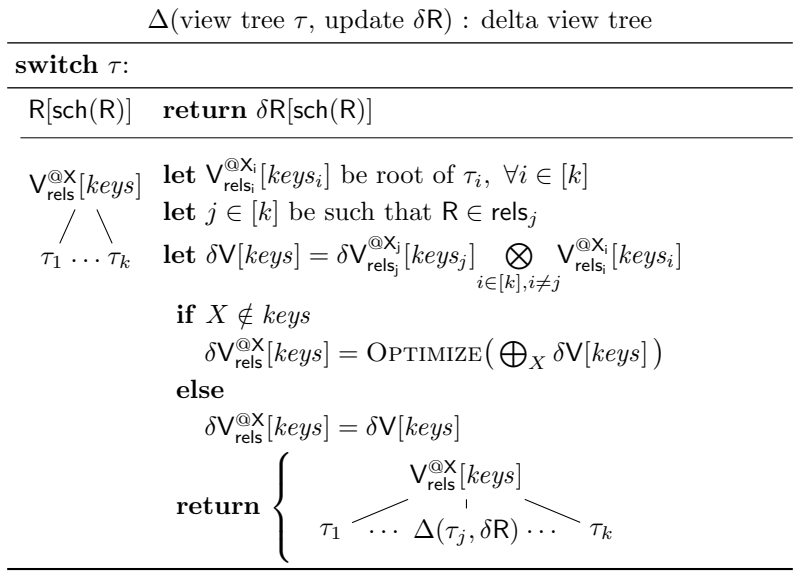
\begin{figure}[t]
\centering
\setlength{\tabcolsep}{3pt}
\begin{tabular}{@{}c@{}c@{~~~~}l@{}}
    \toprule
    \multicolumn{3}{c}{$\Delta$(\text{view tree} $\tau$, \text{update} $\VIEW{\delta{R}}$) : delta view tree}\\
    \midrule
    \multicolumn{3}{l}{\MATCH $\tau$:}\\
    \midrule
    \phantom{a} & $\VIEW{R}[\mathit{\sch(\VIEW{R})}]$ & \hspace*{-0.5em}\RETURN $\VIEW{\delta{R}}[\mathit{\sch(\VIEW{R})}]$\\[2pt]
    \cmidrule{2-3} \\[-6pt]
    &    
    \begin{minipage}[t]{1.6cm}
        \begin{tikzpicture}[xscale=0.45, yscale=1]
            \node at (0,-2)  (n4) {$\VIEW[{\it keys}]{V^{@X}_{rels}}$};
            \node at (-1,-3)  (n1) {$\tau_1$} edge[-] (n4);
            \node at (0,-3)  (n2) {$\ldots$};
            \node at (1,-3)  (n3) {$\tau_k$} edge[-] (n4);
        \end{tikzpicture}        
    \end{minipage}
    &
    \begin{minipage}[t]{8.2cm}
\vspace{-4.5em}
 \hspace*{-0.5em}\LET $\VIEW[{\it keys_i}]{V^{@X_i}_{rels_i}}\text{ be root of } \tau_i, \ \forall i\in[k]$\\[0.5ex]
\hspace*{-0.5em}\LET $j\in[k] \text{ be such that } \VIEW{R}\in\mathsf{rels}_j$ \\[0.5ex]    
\hspace*{-0.5em}\LET $\delta\VIEW[\mathit{keys}]{V}  = \delta\VIEW[{\it keys_j}]{V_{rels_j}^{@X_j}} \underset{i\in[k],i\neq j}{\VPRODBIG}\!\VIEW[{\it keys_i}]{V^{@X_i}_{rels_i}}$\\[0.5ex]
\IF $X \notin {\it keys}$ \\[0.5ex]   
\TAB $\delta\VIEW[\mathit{keys}]{V^{@X}_{rels}}  = \textsc{Optimize}\big( \VSUM_{X} \delta\VIEW[{\it keys}]{V} \, \big)$\\[0.5ex]
\ELSE \\[0.5ex]   
\TAB $\delta\VIEW[\mathit{keys}]{V^{@X}_{rels}}  = \delta\VIEW[{\it keys}]{V}$\\[0.5ex]
        \RETURN $
				\left\{
				\begin{array}{@{~~}c@{~~}}
					\tikz {
				\node at (1,-1)  (n4) {$\VIEW[\mathit{keys}]{V^{@X}_{rels}}$};
				\node at (-0.8,-1.75)  (n1) {$\tau_1$} edge[-] (n4);
				\node at (-0.1,-1.75)  (n2) {$\ldots$};
				\node at (1,-1.75)  (n3) {$\Delta(\tau_j, \delta \VIEW{R})$} edge[-] (n4);
				\node at (2,-1.75)  (n2) {$\ldots$};
				\node at (2.8,-1.75)  (n3) {$\tau_k$} edge[-] (n4);
					}
				\end{array}  \right.$
    \end{minipage}
    \\
    \bottomrule
\end{tabular}
\caption{Creating a delta view tree $\Delta(\tau,\VIEW{\delta{R}})$ for a view tree $\tau$ to process an update $\VIEW{\delta{R}}$ to relation $\VIEW{R}$. }
\label{fig:dynamic_view_tree_algo}
\end{figure}

\textbf{Delta Trees.} Under updates to one relation, a view tree becomes a delta tree where the affected views become delta views. The function $\Delta$ in Figure~\ref{fig:dynamic_view_tree_algo} 
replaces the views along the path from the updated relation to the root with delta views.
The {\sc Optimize} method rewrites delta view expressions to exploit factorized updates by avoiding the materialization of Cartesian products
and pushing marginalization past joins (see Section~\ref{sec:factorizable_updates}).

\begin{example}
\label{ex:delta_view_tree}
Consider again the query from Example~\ref{ex:sql_count}, its view tree in Figure~\ref{fig:example_payloads}, and the same relations over the $\mathbb{Z}$ ring 
and the lifting functions that map all values to $1$ as in Example~\ref{ex:views_count}. 
An update $\VIEW{\delta{T}}[C,D]=\{ \tuple{c_1,d_1} \to -1, \tuple{c_2, d_2} \to 3 \}$ triggers 
delta computation at each view from the leaf $\VIEW{T}$ to the root of the view tree:
\begin{align*}
  \qquad
  \delta\VIEW[C]{V^{@D}_{T}} &= \VSUM_{D} \delta\VIEW[C,D]{T} \\
  \delta\VIEW[A]{V^{@C}_{ST}} &= \VSUM_{C} \delta\VIEW[C]{V^{@D}_{T}} \VPROD \VIEW[A,C]{V^{@E}_{S}} \\
  \delta\VIEW[~]{V^{@A}_{RST}} &= \VSUM_{A} \VIEW[A]{V^{@B}_{R}} \VPROD \delta\VIEW[A]{V^{@C}_{ST}}
\end{align*}

Given that $\VIEW{V^{@E}_{S}} =$ 
$\{(a_1,c_1) \to 2,$ $(a_1,c_2) \to 1,$ $(a_2,c_2) \to 1\}$ 
and $\VIEW{V^{@B}_{R}} =$ 
$\{a_1 \to 2,$ $a_2 \to 1,$ $a_3 \to 1\}$,
we obtain
 $\delta\VIEW{V^{@D}_{T}}[C] = \{c_1 \to -1,$ $c_2 \to 3\}$, 
 $\delta\VIEW{V^{@C}_{ST}}[A] = \{a_1 \to 1,$ $a_2 \to 3\}$, and 
 $\delta\VIEW{V^{@A}_{RST}} = \{() \to 5\}$. 
  
A single-tuple update to $\VIEW{T}$ fixes the values for $C$ and $D$. Computing $\delta\VIEW{V^{@D}_{T}}$ then takes constant time. The delta view $\delta\VIEW{V^{@C}_{ST}}$ iterates over all possible $A$-values for a fixed $C$-value, which takes linear time; $\delta\VIEW{V^{@A}_{RST}}$ incurs the same linear-time cost. A single-tuple update to $\VIEW{R}$ or $\VIEW{S}$
fixes all variables on a leaf-to-root path in the delta view tree, giving a constant view maintenance cost.
\punto
\end{example}

In contrast to classical (first-order) IVM that only requires maintenance of the query result~\cite{Chirkova:Views:2012:FTD}, our approach is higher-order IVM as updates may trigger ma\-intenance of several interrelated views. The fully-recur\-sive IVM scheme of DBToaster~\cite{Koch:Ring:2010:PODS,DBT:VLDBJ:2014} creates one materialization hierarchy per relation in the query, whereas we use one view tree for all relations. This view tree relies on variable orders to decompose the query into views and factorize its computation and maintenance.

\textbf{Which Views to Materialize and Maintain?} 
The answer to this question depends on which relations may change. 
The set of the updatable relations determines the possible delta propagation paths in a view tree, and these paths may use materialized views.

\begin{figure}[t]
\centering
\setlength{\tabcolsep}{3pt}
%
\begin{tabular}[t]{@{}c@{}c@{~~~}l}
    \toprule
    \multicolumn{3}{c}{$\mu$(\text{view tree} $\tau$, \text{updatable relations} $\mathcal{U}$) : view set}\\
    \midrule
    \multicolumn{3}{l}{\MATCH $\tau$:}\\
    \midrule 
    \phantom{ab}
    &
    \hspace{-4mm}
    \begin{minipage}[t]{1.5cm}
        \begin{tikzpicture}[xscale=0.45, yscale=1]
            \node at (0,-2)  (n4) {$\mathit{root}$};
            \node at (-1,-3)  (n1) {$\tau_1$} edge[-] (n4);
            \node at (0,-3)  (n2) {$\ldots$};
            \node at (1,-3)  (n3) {$\tau_k$} edge[-] (n4);
        \end{tikzpicture}
    \end{minipage}
    &
    \begin{minipage}[t]{8.5cm}
        \vspace{-4em}
        $\mathit{children} = \{ \VIEW{V_i} \text{ is root of } \tau_i \}_{i\in[k]}$\\[0.5ex]
        $\mathit{m\_root} = \IF\SPACE (\mathit{root} \,\text{ has no parent})\SPACE \{ \VIEW{\mathit{root}} \} \SPACE\ELSE\SPACE \emptyset$\\[0.5ex]
        $\mathit{m\_children} = 
          \{ \VIEW{V_i} \mid \VIEW{V_i}, \VIEW{V_j} \in \mathit{children}, $\\[0.5ex]
        $\TAB\TAB\TAB\TAB\TAB\TAB\TAB\SPACE \VIEW{V_i} \neq \VIEW{V_j}, \mathsf{rels}(\VIEW{V_j}) \cap \mathcal{U} \neq \emptyset \}$\\[0.5ex]
        $\RETURN\SPACE \mathit{m\_root} \,\cup\, \mathit{m\_children} \,\cup\, \bigcup_{i\in[k]} \mu(\tau_i, \mathcal{U})$
    \end{minipage}\\
    \bottomrule
\end{tabular}
\caption{Deciding which views in a view tree $\tau$ to materialize in order to support updates to a set of relations $\mathcal{U}$. The notation $\mathsf{rels}(\VIEW{V_j})$ denotes the relations under the view $\VIEW{V_j}$ in $\tau$.}
\label{fig:materialization_view_tree_algo}
\end{figure}

Propagating changes along a leaf-to-root path is co\-mputationally most effective if each delta view joins with sibling views that are already materialized. 
Figure~\ref{fig:materialization_view_tree_algo} gives an algorithm that reflects this idea: Given a view tree $\tau$ and a set of updatable relations $\mathcal{U}$, the algorithm traverses the tree top-down to discover the views that need to be materialized. The root of the view tree $\tau$ is always stored as it represents the query result. Every other view $\VIEW{V_i}$ is stored only if there exists a sibling view $\VIEW{V_j}$ defined over an updatable relation.

\begin{example}
We continue with our query from Example~\ref{ex:delta_view_tree}. For updates to $\VIEW{T}$ only, i.e., $\mathcal{U} = \{ \VIEW{T} \}$, we store the root $\VIEW{V^{@A}_{RST}}$ and the views $\VIEW{V^{@E}_{S}}$ and $\VIEW{V^{@B}_{R}}$ used to compute the deltas $\VIEW{\delta{V^{@C}_{ST}}}$ and $\VIEW{\delta{V^{@A}_{RST}}}$. Only the root view is affected: 
  $\VIEW[~]{V^{@A}_{RST}} = \VIEW[~]{V^{@A}_{RST}} \VPLUS \VIEW[~]{\delta{V^{@A}_{RST}}}$.
It is not necessary to maintain other views. To also support updates to $\VIEW{R}$ and $\VIEW{S}$,  we  need to materialize $\VIEW{V^{@C}_{ST}}$ and $\VIEW{V^{@D}_{T}}$. 
If no updates are supported, then only the root view is stored. 
\punto
\end{example}

For queries with free variables, several views in their (delta) view trees may be identical: This can happen when all variables in their keys are free and thus cannot be marginalized. For instance, a variable order $\omega$ for the query from Example~\ref{ex:sql_sum_aggregate} may have the variables $A$ and $C$ above all other variables, in which case their views are the same in the view tree for $\omega$. We then store only the top view out of these identical views.

\textbf{IVM Triggers.} For each updatable relation $\VIEW{R}$, \DF constructs a trigger procedure that takes as input an update $\VIEW{\delta{R}}$ and implements the maintenance schema of the corresponding delta view tree. This procedure also maintains all materialized views needed for the given update workload. 

A bulk of updates to several relations is handled as a sequence of updates, one per relation. Update sequences can also happen when updating a relation $\VIEW{R}$ that occurs several times in the query. The instances representing the same relation are at different leaves in the delta tree and lead to changes along multiple leaf-to-root paths. 

%% file: factorized_updates.tex
\section{Factorizable Updates}
\label{sec:factorizable_updates}

Our focus so far has been on supporting updates represented by delta relations. We next consider an alternative approach that decomposes a delta relation into a union of factorizable relations. The cumulative size of the decomposed relations can be much less than the size of the original delta relation. Also, the complexity of propagating a factorized update can be much lower than that of its unfactorized (listing) representation, since the factorization makes explicit the independence between query variables and enables optimizations of delta propagation such as pushing marginalization past joins. Besides the factorized view computation, this is the second instance where \DF exploits factorization.

Factorizable updates arise in many domains such as linear algebra and machine learning. Section~\ref{sec:applications} demonstrates how our framework can be used for the incremental evaluation of matrix chain multiplication, recovering prior work on this~\cite{NEK:SIGMOD:2014}.  Matrix chain computation can be phrased in our language of joins and aggregates, where matrices are binary relations. Changes to one row/column in an input matrix may be expressed as a product of two vectors. In general, an arbitrary update matrix can be decomposed into a sum of rank-$1$ matrices, each of them  expressible as products of vectors, using low-rank tensor decomposition methods~\cite{TensorDecomp:2009,TensorDecomposition:2017}.

\begin{example}
Arbitrary relations can be decomposed in\-to a union of factorizable relations. The relation $\VIEW[A,B]{R}$ $= \{(a_i,b_j) \to 1\mid i\in[n],j\in[m]\}$ can be decomposed as $\VIEW[A]{R_1}\VPROD\VIEW[B]{R_2}$, where $\VIEW[A]{R_1}=\{(a_i) \to 1\mid i\in[n]\}$ and $\VIEW[B]{R_2}=\{(b_j) \to 1\mid j\in[m]\}$. We thus reduced a relation of size $nm$ to two relations of cumulative size $n+m$. If $\VIEW{R}$ were a delta relation, the delta views on top of it would now be expressed over $\VIEW[A]{R_1}\VPROD\VIEW[B]{R_2}$ and their computation can be factorized as done for queries in Section~\ref{sec:factorized_ring_computation}. Product decomposition of relations can be done in linearithmic time in both the number of variables and the size of the relation~\cite{WSD:2008}. 

Consider now $\VIEW[A,B]{R'}$ $=$ $\VIEW[A,B]{R}$ $\VPLUS$ $\{(a_{n+1},b_j) \to 1\mid j\in[m-1]\}$ with $\VIEW{R}$ as above. We can decompose each of the two terms in $\VIEW{R'}$ similarly to $\VIEW{R}$, yielding overall $n+2m$ values instead of $nm+m-1$. A different decomposition with $n+m+3$ values is given by a factorizable over-approximation of $\VIEW{R'}$ compensated by a small product with negative payload: $\{(a_i)\to 1\mid i\in[n+1]\}\VPROD\{(b_j)\to 1\mid j\in[m]\}\VPLUS\{(a_{n+1})\to 1\}\VPROD\{(b_m)\to -1\}$.\punto
\end{example}

The {\sc Optimize} method used in the delta view tree algorithm in Figure~\ref{fig:dynamic_view_tree_algo} exploits the distributivity of join $\VPROD$ over marginalization $\VSUM_{X}$ to push the latter past the former and down to the views with variable $X$. This optimization is reminiscent of pushing aggregates past joins in databases and variable elimination in probabilistic graphical models~\cite{FAQ:PODS:2016}. In case the delta views express Cartesian products, then they are not materialized but instead kept factorized.

\begin{example}
\label{ex:factorized-update}
Consider the query $\VIEW{Q}$ from Example~\ref{ex:delta_view_tree} 
and its view tree in Figure~\ref{fig:example_payloads}.
In the delta view tree derived for updates to $\VIEW{S}$, the top-level delta is computed as:
\begin{align*}
\VIEW[~]{\delta{V^{@A}_{RST}}} = \VSUM_{A} \VIEW[A]{V^{@B}_{R}} \VPROD 
\big(&  \VSUM_{C} \VIEW[C]{V^{@D}_{T}} \VPROD \\
& \underbrace{\hspace{3em}\underbrace{\VSUM_{E} \VIEW[A,C,E]{\delta{S}}}_{\VIEW[A,C]{\delta{V^{@E}_{S}}}}\big)}_{\VIEW[A]{\delta{V^{@C}_{ST}}}}
\end{align*}
A single-tuple update $\VIEW{\delta{S}}$ binds variables $A$, $C$, and $E$, and computing $\VIEW{\delta{V^{@A}_{RST}}}$ requires $\bigO{1}$ lookups in $\VIEW{V^{@D}_{T}}$ and $\VIEW{V^{@B}_{R}}$. An arbitrary-sized update $\VIEW{\delta{S}}$ can then be processed in $\bigO{|\VIEW{\delta{S}}|}$ time.

Assume now that $\VIEW{\delta{S}}$ is factorizable as $\VIEW[A,C,E]{\delta{S}} = \VIEW[A]{\delta{S_{A}}} \VPROD \VIEW[C]{\delta{S_{C}}} \VPROD \VIEW[E]{\delta{S_{E}}}$. In the construction of the delta view tree, the {\sc Optimize} method exploits this factorization to push the marginalization past joins at each variable; for example, the delta at $E$ becomes:
\begin{align*}
\VIEW[A,C]{\delta{V^{@E}_{S}}} &= \VSUM_{E} \VIEW[A]{\delta{S_{A}}} \VPROD \VIEW[C]{\delta{S_{C}}} \VPROD \VIEW[E]{\delta{S_{E}}} \\
&= \VIEW[A]{\delta{S_{A}}} \VPROD \VIEW[C]{\delta{S_{C}}} \VPROD \VSUM_{E} \VIEW[E]{\delta{S_{E}}}
\end{align*}
We also transform the top-level delta into a product of three views:
\begin{align*}
\VIEW[~]{\delta{V^{@A}_{RST}}} = 
&\big( \VSUM_{A} \VIEW[A]{V^{@B}_{R}} \VPROD \VIEW[A]{\delta{S_{A}}} \big) \VPROD
\big( \VSUM_{C} \VIEW[C]{V^{@D}_{T}} \VPROD \VIEW[C]{\delta{S_{C}}} \big) \VPROD 
 \big( \VSUM_{E} \VIEW[E]{\delta{S_{E}}} \big)
\end{align*}
The computation time for this delta is proportional to the sizes of the three views representing the update:
$\bigO{\min(|\VIEW{V^{@B}_{R}}|, {|\VIEW{\delta{S_{A}}}|}) + \min(|\VIEW{V^{@D}_{T}}|, |\VIEW{\delta{S_{C}}}|) + |\VIEW{\delta{S_{E}}}|}$.
\punto
\end{example}

%% file: query_classes.tex
\section{\DF for Special Query Classes}

\label{sec:query_classes}

This section shows how \DF maintains {\em free-connex ($\alpha$-)acyclic} queries~\cite{DynYannakakis:SIGMOD:2017} and {\em $q$-hierarchical} queries~\cite{Nicole:PODS:2017}. The analysis for these queries is refined into: (i) the preprocessing phase, where the view tree is constructed; (ii) the enumeration phase, where we present the query result one tuple at a time; and (iii) the update phase, where we update the view tree. The following data complexity\footnote{The {\em data complexity} is a function of the database size.} claims assume that the ring operations require constant time, otherwise the complexity results stated in this section have an extra multiplying factor to account for the complexity of the ring operations.

\begin{theorem}\label{th:special-cases}
  Let a query $\VIEW[]{Q}$ and a database of size $N$.

  \DF can maintain $\VIEW[]{Q}$ with $O(N)$ preprocessing, $O(1)$ enumeration delay, and $O(N)$ single-tuple update in case $\VIEW[]{Q}$ is free-connex acyclic.
  
  \DF can maintain $\VIEW[]{Q}$ with $O(N)$ preprocessing, $O(1)$ enumeration delay, and $O(1)$ single-tuple update in case $\VIEW[]{Q}$ is $q$-hierarchical.
\end{theorem}

Section~\ref{sec:cyclic_queries} discusses an important extension of our view tree construction to better support cyclic queries.


\subsection{Free-Connex Acyclic Queries}
\label{sec:free-connex}

We first introduce the class of free-connex acyclic que\-ries and then explain how \DF maintains them.

\begin{definition}[\cite{Yannakakis81,BraultPhD13}]
  A {\em join tree} for a query is a tree, where each node is a relation and if any two nodes have variables in common, then all nodes along the path between them also have these variables. 

  A query is \emph{($\alpha$-)acyclic} if it admits a join tree.
  %
  A query is \emph{free-connex acyclic} if it is acyclic and remains acyclic after adding a new relation whose schema consists of the free variables of the query. 
\end{definition}

\begin{example}
\label{ex:acyclic}
Consider the query 
$\VIEW[A,B,C]{Q} = \VSUM_{D}\VSUM_{E}$ $\VIEW[A,B]{R} \VPROD \VIEW[A,C,E]{S} \VPROD \VIEW[C,D]{T}$. 
  A possible join tree for $\VIEW[]{Q}$ is 
  $\VIEW[A,B]{R} - \VIEW[A,C,E]{S} - \VIEW[C,D]{T}$, where 
  ``$-$" denotes the parent-child relationship.
  Hence, $\VIEW[]{Q}$ is acyclic. 
  
Consider the triangle query 
$\VIEW[~]{Q_{\vartriangle}} = \VSUM_{A}\VSUM_{B}\VSUM_{C}$ $\VIEW[A,B]{R} \VPROD \VIEW[B,C]{S} \VPROD \VIEW[A,C]{T}$. A possible tree built from the relations  
of $\VIEW[]{Q_{\vartriangle}}$ is
$\VIEW[A,B]{R} - \VIEW[B,C]{S} - \VIEW[A,C]{T}$.    
The variable $A$ occurs in the first and last relations but not in the middle relation; thus, this tree is not a join 
tree for $\VIEW[]{Q_{\vartriangle}}$. One can show that any 
tree built from the relations of $\VIEW[]{Q_{\vartriangle}}$ is not a join tree. 
Hence, $\VIEW[]{Q_{\vartriangle}}$ is not acyclic.

The tree $\VIEW[A,B]{R} - \VIEW[A,B,C]{U} - \VIEW[A,C,E]{S} - \VIEW[C,D]{T}$
is a join tree of $\VIEW[]{Q}$ extended with the relation 
$U$ whose schema consists of the free variables of $\VIEW[]{Q}$.
Hence, $\VIEW[]{Q}$ is free-connex acyclic. 
Consider now the variant $\VIEW[]{Q'}$ 
of $\VIEW[]{Q}$ where only the variables $B$ and $C$ are free.
Adding a fresh relation $U'$ with schema $(B,C)$ to 
$\VIEW[]{Q'}$ turns it into a cyclic query $\VIEW[]{Q''}$
that does not admit a join tree. 
\punto
\end{example}

\begin{figure}[t]
\centering
\setlength{\tabcolsep}{3pt}
\begin{tabular}{@{}c@{}c@{~~~}l}
  \toprule
  \multicolumn{3}{c}{$\nu$ (\text{free-top variable order} $\omega$) : view tree} \\
  \midrule
  \multicolumn{3}{l}{\MATCH $\omega$:} \\
  \midrule 
  \phantom{a} & $\VIEW{R}$\hspace*{2.5em} & \RETURN $\VIEW[\mathit{\sch(\VIEW{R})}]{R}$ \\
  \cmidrule{2-3} \\[-6pt] 
  &
  \begin{minipage}[b]{2.5cm}
    \begin{tikzpicture}[xscale=0.4, yscale=1]
      \node at (0,-2)  (n4) {$X$};
      \node at (-1,-3)  (n1) {$\omega_1$} edge[-] (n4);
      \node at (0,-3)  (n2) {$\ldots$};
      \node at (1,-3)  (n3) {$\omega_k$} edge[-] (n4);
      \node at (0,-4.5) {~};
    \end{tikzpicture}
    \vspace{3.4cm}
  \end{minipage}
  &
\begin{minipage}[b]{6.5cm}
\LET $T_i  = \tau(\omega_i), \ \forall i\in[k] $\\[0.5ex]
\LET $\VIEW[\mathit{keys_i}]{V^{@\omega_i}_{rels_i}} = \text{ root of } T_i, \ \forall i\in[k] $\\[0.5ex]
\LET $\mathit{keys}=\{X\} \cup \mathit{dep}(X)$ \\[0.5ex]
\LET $\mathsf{rels}=\bigcup_{i\in[k]}\mathsf{rels}_i$\\[0.5ex]
\LET $\VIEW[\mathit{keys}]{H^{@X}_{rels}}= \VPRODBIG_{i \in [k]} \VIEW[\mathit{keys_i}]
{V^{@\omega_i}_{rels_i}}$\\[0.5ex]
\LET $\VIEW[\mathit{keys}\setminus \{X\}]{V^{@X}_{rels}}= \VSUM_{X} \VIEW[\mathit{keys}]{H^{@X}_{rels}}$\\[0.5ex]
\IF $X$ has more than one child ($k\geq 2$) \\[0.5ex]
  \TAB \IF $X$ has no sibling \\[0.5ex]
    \TAB\TAB\TAB \RETURN $
				\left\{
				\begin{array}{@{~~}c@{~~}}
					\tikz {
						\node at (1.4,-1)  (n4) {$\VIEW[\mathit{keys}]{H^{@X}_{rels}}$};
						\node at (0.8,-1.75)  (n1) {$T_1$} edge[-] (n4);
						\node at (1.25,-1.75)  (n2) {$\ldots$};
						\node at (1.8,-1.75)  (n3) {$T_k$} edge[-] (n4);
					}
				\end{array}  \right.$  \\[0.5ex]
    \TAB\ELSE \\[0.5ex]
    \TAB\TAB\TAB\RETURN $
				\left\{
				\begin{array}{@{~~}c@{~~}}
					\tikz {
						\node at (1.2,-0.25)  (n) {$\VIEW[\mathit{keys}\setminus \{X\}]{V^{@X}_{rels}}$};
						\node at (1.2,-1)  (n4) {$\VIEW[\mathit{keys}]{H^{@X}_{rels}}$} edge[-] (n);
						\node at (0.8,-1.75)  (n1) {$T_1$} edge[-] (n4);
						\node at (1.25,-1.75)  (n2) {$\ldots$};
						\node at (1.8,-1.75)  (n3) {$T_k$} edge[-] (n4);
					}
				\end{array}  \right.$ \\[0.5ex]
\ELSE \\[0.5ex]        
  \TAB \IF $X$ has no sibling \\[0.5ex]
    \TAB\TAB\TAB \RETURN $T_1$ \\[0.5ex]
  \TAB\ELSE \\[0.5ex]
    \TAB\TAB\TAB\RETURN $
				\left\{
				\begin{array}{@{~~}c@{~~}}
					\tikz {
						\node at (1.2,-0.25)  (n) {$\VIEW[\mathit{keys}\setminus \{X\}]{V^{@X}_{rels}}$};
						\node at (1.2,-1)  (n1) {$T_1$} edge[-] (n);
					}
				\end{array}  \right.$      
  \end{minipage}
  \\
  \bottomrule
\end{tabular}
\caption{Creating a view tree for a free-top variable order.}
\label{fig:static_view_tree_algo_free-connex}
\end{figure}
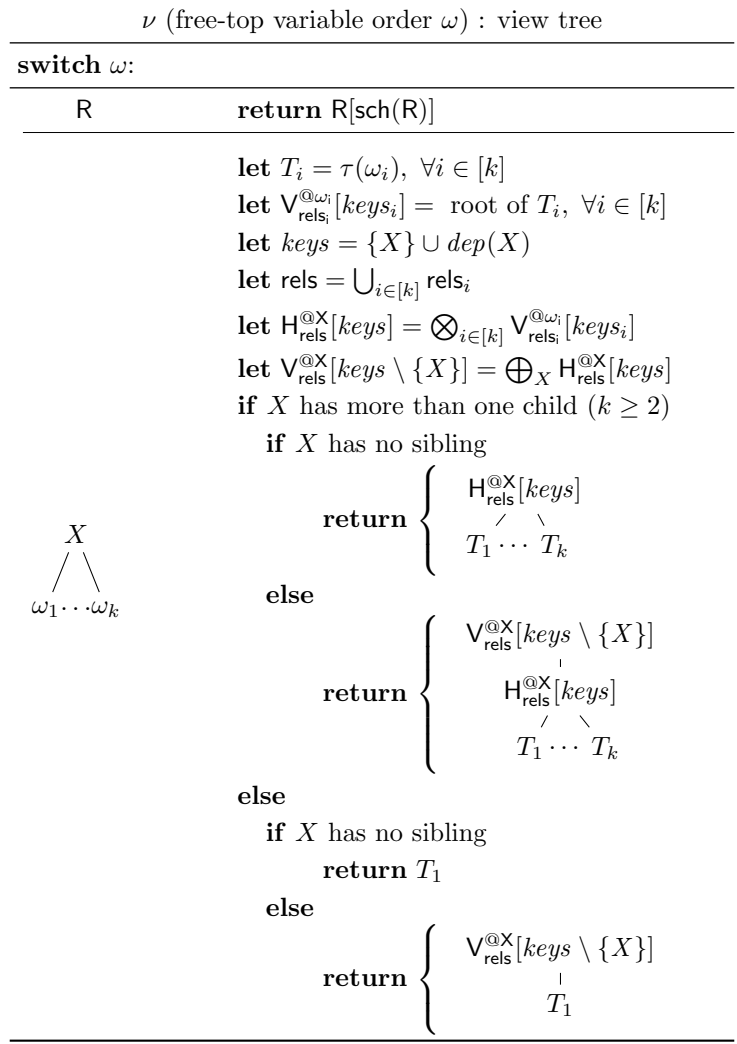

We next detail how \DF achieves the complexity from Theorem~\ref{th:special-cases} for a free-connex acyclic query $\VIEW[]{Q}$. 

\textbf{Preprocessing.} 
In the preprocessing phase, we create a view tree
that compactly represent the result of $\VIEW[]{Q}$.
Given a variable order, the function $\tau$ in 
Figure~\ref{fig:static_view_tree_algo} constructs a view tree where
the root view consists of all tuples over the free variables.
While this view allows for constant enumeration delay, it may 
require superlinear computation and maintenance time as the free variables may originate from different input relations. We would like to avoid this super-linearity.

To keep the preprocessing and update times linear, we proceed as follows.
We construct view trees such that the query result is kept and maintained factorized over several views at the top of the view tree.
This approach still allows for constant enumeration delay, using a known enumeration approach for factorized representations~\cite{Olteanu:FactBounds:2015:TODS}.
We construct the view tree following a free-top variable order of the query $\VIEW[]{Q}$
and materialize a view over the schema $\{X\} \cup \mathit{dep}(X)$ for each variable $X$ in the variable order. 
A key insight is that every free-connex acyclic query admits a free-top variable order 
where for each variable $X$, the set $\{X\} \cup \mathit{dep}(X)$ 
is covered by the variables of a single relation~\cite{BerkholzGS20}. 
This ensures linear preprocessing and maintenance time for all views in view trees
following such variable orders.

The function  $\nu$ in Figure~\ref{fig:static_view_tree_algo_free-connex}
constructs a view tree for a given free-top variable 
order of a free-connex query.
If a variable $X$ has at least two children, it proceeds as follows.
It creates at $X$ 
a view $\VIEW[]{H^{@X}_{rels}}$ 
with schema $\{X\} \cup \mathit{dep}(X)$
that joins the child views of $X$.
If $X$ has at least one sibling, it additionally 
creates a view $\VIEW[]{V^{@X}_{rels}}$ on top of $\VIEW[]{H^{@X}_{rels}}$
obtained from $\VIEW[]{H^{@X}_{rels}}$ by marginalizing  $X$.
 The first view 
enables efficient enumeration of $X$-values in the query result given a value tuple 
 over $\mathit{dep}(X)$; the second view enables efficient updates 
coming from the subtrees rooted at siblings of $X$.  
If $X$ has only one child, the creation of the view $\VIEW[]{H^{@X}_{rels}}$ is not needed for efficient enumeration.
In this case, the function creates  a view $\VIEW[]{V^{@X}_{rels}}$ marginalizing $X$ in the child view if 
$X$ has siblings.  
 
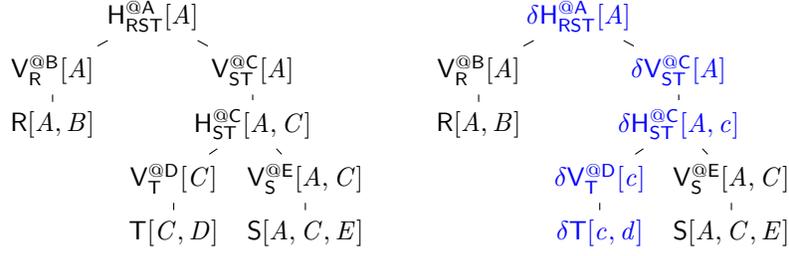
\begin{figure}[t]
\centering
  \begin{tikzpicture}[xscale=0.7, yscale=0.24]

    \node at (-0.4, 3) (A) {$\VIEW[A]{H^{@A}_{RST}}$};     
     \node at (1.5, 0) (C) {$\VIEW[A]{V^{@C}_{ST}}$} edge[-] (A);      
     \node at (1.5, -3) (C') {$\VIEW[A,C]{H^{@C}_{ST}}$} edge[-] (C);      
     \node at (2.5, -6) (E) {$\VIEW[A,C]{V^{@E}_{S}}$} edge[-] (C');
      \node at (2.5, -9) {$\VIEW[A,C,E]{S}$} edge[-] (E);
    
    \node at (0, -6) (D) {$\VIEW[C]{V^{@D}_{T}}$} edge[-] (C');
      \node at (0, -9) {$\VIEW[C,D]{T}$} edge[-] (D);  

      \node at (-2.3, -0) (B) {$\VIEW[A]{V^{@B}_{R}}$} edge[-] (A);
      \node at (-2.3, -3) {$\VIEW[A,B]{R}$} edge[-] (B);  
\end{tikzpicture}
\hspace{0.5cm}
  \begin{tikzpicture}[xscale=0.7, yscale=0.24]

    \node at (-0.4, 3) (A) {\color{blue} $\delta\VIEW[A]{H^{@A}_{RST}}$};     
     \node at (1.5, 0) (C) {\color{blue} $\delta\VIEW[A]{V^{@C}_{ST}}$} edge[-] (A);      
     \node at (1.5, -3) (C') {\color{blue} $\delta\VIEW[A,c]{H^{@C}_{ST}}$} edge[-] (C);      
     \node at (2.5, -6) (E) {$\VIEW[A, C]{V^{@E}_{S}}$} edge[-] (C');
      \node at (2.5, -9) {$\VIEW[A, C ,E]{S}$} edge[-] (E);
    
    \node at (0, -6) (D) {$\color{blue} \delta\VIEW[c]{V^{@D}_{T}}$} edge[-] (C');
      \node at (0, -9) {$\color{blue} \delta\VIEW[c,d]{T}$} edge[-] (D);  

      \node at (-2.3, -0) (B) {$\VIEW[A]{V^{@B}_{R}}$} edge[-] (A);
      \node at (-2.3, -3) {$\VIEW[A,B]{R}$} edge[-] (B);  
\end{tikzpicture}
\caption{(left) View tree constructed by the function $\nu$ in 
Figure~\ref{fig:static_view_tree_algo_free-connex} for the variable 
order $\omega$ in Figure~\ref{fig:example_payloads};
(right) Delta view tree for a single-tuple update to $\VIEW{T}$.}
\label{fig:free-connex_view_tree}
\end{figure}

\begin{example}\label{ex:free-connex-viewtree}
Consider the free-connex acyclic query 
$\VIEW[]{Q}$ from Example~\ref{ex:acyclic}.
Figure~\ref{fig:example_payloads} gives a free-top variable order $\omega$ for $\VIEW[]{Q}$. 
Figure~\ref{fig:free-connex_view_tree} (left) depicts 
the view tree $\nu(\omega)$. 
The view $\VIEW[]{H_{ST}^{@C}}$ can be computed by iterating 
over the $(A,C)$-tuples in $\VIEW[]{V_{S}^{@E}}$ and multiplying 
the payload of each such tuple with the payload of the matching 
$C$-value in $\VIEW[]{V_{T}^{@D}}$.   
Since each such $(A,C)$-tuple  must be in $\VIEW[]{S}$, we need to iterate 
over only linearly many such tuples.
Similarly, the view $\VIEW[]{H_{RST}^{@A}}$ can be computed by iterating 
over the $A$-values in one of the child views and doing lookups in the other child view to retrieve the payloads. For the computation of both views $\VIEW[]{H_{ST}^{@C}}$ and  $\VIEW[]{H_{RST}^{@A}}$,
we iterate over linearly  many 
tuples and do a constant-time lookup for each such tuple. 
All other views are obtained by marginalizing one variable from their child views. 
Hence, all views can be computed in linear time. 
\punto
\end{example}

\textbf{Updates.} 
 The construction of delta view trees under single-tuple updates
 is exactly as described 
by the function $\Delta$ in Figure~\ref{fig:dynamic_view_tree_algo} (Section~\ref{sec:factorized_IVM}).
Since the view trees can be constructed in linear time, 
the delta view trees can also be constructed in linear time.

\begin{example}
Continuing Example~\ref{ex:free-connex-viewtree}, we consider a single-tuple update $\delta\VIEW{T}[c,d]$ to relation $\VIEW{T}$.
Figure~\ref{fig:free-connex_view_tree} depicts the original view tree (left) and the delta view tree for updates to $\VIEW{T}$ (right). The difference is that along the path from $\VIEW{T}$ to the root, we now have delta views.
The delta view $\delta\VIEW[]{V^{@D}_{T}}$
results from  $\delta\VIEW[c,d]{T}$ by marginalizing $D$, which takes  
constant time since $D$ is fixed to the constant $d$.
To compute $\delta\VIEW[]{H^{@C}_{ST}}$, we iterate over all $A$-values 
paired with $c$ in $\VIEW[]{V^{@E}_{S}}$. This operation takes linear time with the support of an index on variable $C$ built for this view.
We obtain $\delta\VIEW[]{V^{@C}_{ST}}$ from $\delta\VIEW[]{H^{@C}_{ST}}$
by marginalizing the variable $C$. This requires constant time because $C$ is fixed to
the constant $c$. 
The top delta view $\delta\VIEW[]{H^{@A}_{RST}}$ is obtained by intersecting the two child views, e.g., by iterating over $\delta\VIEW[]{V^{@C}_{ST}}$ and doing lookups in $\VIEW[]{V^{@B}_R}$. This requires linear time. We conclude that the delta views can be computed in linear time. 
\punto
\end{example}

\textbf{Enumeration.} 
Consider a view tree $\tau$ constructed using the function $\nu$ from 
Figure~\ref{fig:static_view_tree_algo_free-connex} for a free-top variable order of a 
query $\VIEW[]{Q}$. 
We first describe how to enumerate with constant delay the distinct tuples in the result of $\VIEW{Q}$ using $\tau$. Then, we explain how to compute the payload of each result tuple in constant time.

Let $X_1, \ldots, X_n$ be an ordering of the free variables of the query
that is compatible with a top-down traversal of the free-top variable order. 
We use the views $\VIEW[]{V_1}, \ldots , \VIEW[]{V_n}$
to enumerate the distinct tuples
in the result of $\VIEW[]{Q}$, where 
$\VIEW[]{V_j}$ is $\VIEW[]{H^{@X_j}_{rels}}$ if $X_j$ has at least two children
and it is the child view of $X_j$ otherwise.  
We retrieve from $\VIEW[]{V_1}$ the first $X_1$-value in the 
result. 
When we arrive at a view  $\VIEW[]{V_j}$ with $j > 1$, we have already fixed 
the values of the variables above $X_j$ in the variable order. 
We retrieve from $\VIEW[]{V_j}$ the first $X_j$-value paired with these values. 
Once the values over all free variables 
are fixed, we have a complete result tuple that we output.
Then, we iterate over the remaining distinct $X_n$-values in $\VIEW[]{V_n}$
paired with the fixed values over the ancestor variables of $X_n$
and output a new tuple for each such value.
After all $X_n$-values are exhausted, we backtrack, i.e., we move to the next $X_{n-1}$-value
and restart the iteration of the matching $X_n$-values.

\begin{figure}[t]
\centering
\setlength{\tabcolsep}{3pt}
\begin{tabular}{@{}c@{}c@{~~~}l}
  \toprule
  \multicolumn{3}{c}{$payload$(\text{view tree} $\tau$, \text{tuple} \textvec{t}): \text{payload}} \\
  \midrule
  \multicolumn{3}{l}{\MATCH $\tau$:} \\
  \midrule 
  \phantom{a} & $\VIEW{R}$\hspace*{2.5em} & \RETURN $\VIEW{R}(\textvec{t})$ \\
  \cmidrule{2-3} \\[-6pt] 
  &
  \begin{minipage}[b]{1.5cm}
    \begin{tikzpicture}[xscale=0.4, yscale=1]
      \node at (0,-2)  (n4) {$\VIEW[\mathcal{X}]{V}$};
      \node at (-1,-3)  (n1) {$\tau_1$} edge[-] (n4);
      \node at (0,-3)  (n2) {$\ldots$};
      \node at (1,-3)  (n3) {$\tau_k$} edge[-] (n4);
      \node at (0,-4.5) {~};
    \end{tikzpicture}
    \vspace{-0.9cm} 
  \end{minipage}
  &
\begin{minipage}[b]{6.2cm} 
\IF $\mathcal{X} = \sch(\textvec{t})$\\[0.5ex]
\TAB \RETURN $\VIEW{V}(\textvec{t})$\\[0.5ex]
\ELSE \ // $\mathcal{X} \subset \sch(\textvec{t})$ \\[0.5ex]
   \TAB \LET $\mathcal{V}_i =$ variables in $\tau_i$, \ $\forall i \in [k]$ \\[0.5ex]
    \TAB \RETURN $\prod_{i \in [k]} payload(\tau_i, \pi_{\mathcal{V}_i}\textvec{t})$
  \end{minipage}
  \\
  \bottomrule
\end{tabular}
\caption{Computing the payload of a tuple from a view tree.}
\label{fig:payload_computation}
\end{figure}

Given a complete tuple $\textvec{t}$ constructed from the view tree $\tau$, we use the 
function $payload$ from Figure~\ref{fig:payload_computation} to compute its payload. 
The function first checks whether the schema of the root view is exactly the schema 
$\sch(\textvec{t})$ of $\textvec{t}$. If so, it returns the payload of $\textvec{t}$ in this view. Otherwise, the root view covers only a subset of the schema of the tuple. 
In this case, the function recursively computes the payload for each subtree $\tau_i$ of the root view
and the projection of $\textvec{t}$ onto the variables in $\tau_i$.
The final payload is the product of the payloads returned for the subtrees. 
The returned payloads are from the lowest views in the view tree whose schemas consist of free variables only. If all variables are free, then these lowest views are the input relations themselves.

\begin{remark}
The enumeration procedure needs the payloads of the lowest views whose schemas consist of free variables. The payloads from the views above these views thus need not be maintained, beyond keeping track of the multiplicities of each of their tuples. The maintenance of multiplicities is important for correctness, as it tells whether a tuple is to be removed from a view or still has at least one possible derivation from the input. 
For expensive payloads, such as those introduced in Section~\ref{sec:applications}, it is therefore more efficient to only maintain them for the views from the input relations up to the views used to compute the payloads. Their ancestor views only need maintenance of tuple multiplicities.\punto
\end{remark}

\begin{example}
We enumerate the distinct result tuples of the query
$\VIEW[A,B,C]{Q}$ 
from Example~\ref{ex:acyclic} using the view tree in Figure~\ref{fig:free-connex_view_tree} (left). We iterate with constant delay 
over the $A$-values in $\VIEW[A]{H^{@A}_{RST}}$. For each such $A$-value
$a$, we iterate with constant delay over the $B$-values in 
$\VIEW[a,B]{R}$ and over the $C$-values in 
$\VIEW[a,C]{H^{@C}_{ST}}$. Each triple $(a,b,c)$
obtained in this way is a result tuple of $\VIEW[]{Q}$.
Its payload is $\VIEW[a,b]{R} \cdot \VIEW[a,c]{H^{@C}_{ST}}$.  
\punto
\end{example}
\begin{remark}
  To efficiently support enumeration and updates, we may need several indices for the views in a view tree for a free-connex acyclic query. Each view  (and input relation) in the view tree in Figure~\ref{fig:free-connex_view_tree} (left) needs an index that can retrieve the payload for a given tuple of values over its variables. This is a primary index. For (top-down) enumeration, we may also need a secondary index per view to lookup for tuples that have as prefix a tuple of values over the variables shared with its parent view. Yet in case of some views, we may also need a tertiary index to support updates, which are propagated bottom-up. 
  For instance, the view $\VIEW[A,C]{V^{@E}_{S}}$ requires: a primary index to retrieve the payload for each $(A,C)$-tuple; a secondary index to enumerate the $C$-values paired with a given $A$-value fixed by the parent view; and a tertiary index to obtain all $A$-values paired with a given $C$-value $c$ fixed by the delta of its left sibling $\delta\VIEW[c]{V^{@D}_{T}}$. All other views only require primary and secondary indices and no tertiary index.\punto
\end{remark}

\subsection{$Q$-Hierarchical Queries}
\label{sec:q-hierarchical}

$Q$-hierarchical queries form a strict subclass of the free-co\-nnex acyc\-lic queries.
They admit linear preprocessing time, constant update time, and constant enumeration delay~\cite{Nicole:PODS:2017}. Under widely-held complexity theoretic assumptions, there is no algorithm that achieves constant update time and enumeration delay for queries that are not $q$-hierarchical and have no repeating relation symbols~\cite{Nicole:PODS:2017}.
\DF recovers the aforementioned complexities using exactly the same approach as for free-connex acyclic queries detailed in Section~\ref{sec:free-connex}. This directly implies linear preprocessing time and  constant enumeration delay. Constant update time follows from the following observation. Every $q$-hierarchical query admits a free-top variables order, where each root-to-leaf path consists of variables that represent precisely the schema of a relation in the query. A single-tuple update to that relation then sets all these variables to constants, effectively making each delta view along that path of constant size. Our view tree construction also ensures that the computation of each delta view only requires one constant-time lookup per child view.

We first define $q$-hierarchical queries and then show how \DF achieves constant-time update for them. 
For a variable $X$ in a query, we denote by $\textsf{rels}(X)$ the set of relations that contain $X$ in their schema.
\begin{definition}[\cite{Suciu:PDB:11,Nicole:PODS:2017}]
   A query is \emph{hierarchical} if for any two variables $X$ and $Y$, it holds $\textsf{rels}(X) \subseteq \textsf{rels}(Y)$, $\textsf{rels}(Y) \subseteq \textsf{rels}(X)$, or $\textsf{rels}(X) \cap \textsf{rels}(Y) = \emptyset$. 

   A query is \emph{$q$-hierarchical} if it is hierarchical and for any  two variables $X$ and $Y$,
   it holds: if $\textsf{rels}(X) \supset \textsf{rels}(Y)$ and $Y$ is free, then $X$ is free.
\end{definition}

 Every $q$-hierarchical query admits a {\em canonical free-top} variable order, where (i) each root-to-leaf path consists of variables that form the schema of a relation and (2) no bound variable is above a free variable~\cite{KNOZ20}.
We can construct such a variable order in polynomial time in the query size as follows.
We start with the empty variable order.
For each relation $R$, we add to the variable order a root-to-leaf path made up of $R$'s variables ordered  as follows:  
a variable $X$ is before a variable $Y$ if (1) $\textsf{rels}(X) \supset \textsf{rels}(Y)$
or (2) $\textsf{rels}(X) \not\supset \textsf{rels}(Y)$, $\textsf{rels}(X) \not\subset \textsf{rels}(Y)$, $X$ is free, and $Y$ is bound. 

 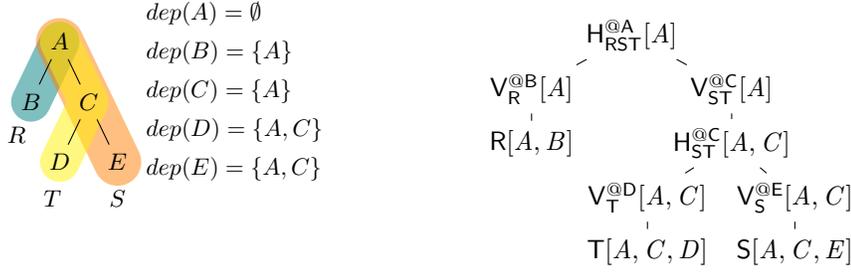
\begin{figure}[t]
    \centering
    \begin{minipage}[b]{0.3\linewidth}
      \begin{tikzpicture}[xscale=0.96, yscale=0.8]
        \node at (0, 0.0) (A) {\small  $A$};
        \node at (-0.4, -1.0) (B) {\small $B$}  edge[-] (A);
        \node at (0.4, -1.0) (C) {\small $C$} edge[-] (A);
        \node at (0.8, -2.0) (E) {\small $E$} edge[-] (C);
        \node at (0, -2.0) (D) {\small $D$} edge[-] (C);
        \node at (0.8, -2.6) (S) {\small $S$};
        \node at (-0.6, -1.55) (R) {\small  $R$};
        \node at (-0.1, -2.6) (T) {\small  $T$};
                \node at (2.3, -1.5) {\scalebox{0.9} {
        \begin{tabular}{@{~~~~}l}
          $dep(A) = \emptyset$\\[1ex]
          $dep(B)=\{A\}$\\[1ex]
          $dep(C)=\{A\}$\\[1ex]
          $dep(D)=\{A,C\}$\\[1ex]
          $dep(E)=\{A,C\}$\\[8ex]
        \end{tabular}
      }};
        \begin{pgfonlayer}{background}
          \draw[opacity=.5,fill opacity=.5,line cap=round, line join=round, line width=15pt,color=teal] (0,0.0) -- (-0.4,-1);
          \draw[opacity=.5,fill opacity=.5,line cap=round, line join=round, line width=18pt,color=orange] (0,0.0) -- (0.8,-2.0);
          \draw[opacity=.5,fill opacity=.5,line cap=round, line join=round, line width=15pt,color=yellow] (0,0.0) -- (0.4,-1) -- (0, -2.0);
        \end{pgfonlayer}
      \end{tikzpicture}
    \end{minipage}
  \hspace{1.2cm}
    \begin{minipage}[b]{0.4\linewidth}
  \begin{tikzpicture}[xscale=0.7, yscale=0.24]

    \node at (-0.4, 3) (A) {$\VIEW[A]{H^{@A}_{RST}}$};     
     \node at (1.5, 0) (C) {$\VIEW[A]{V^{@C}_{ST}}$} edge[-] (A);      
     \node at (1.5, -3) (C') {$\VIEW[A,C]{H^{@C}_{ST}}$} edge[-] (C);      
     \node at (2.7, -6) (E) {$\VIEW[A,C]{V^{@E}_{S}}$} edge[-] (C');
      \node at (2.7, -9) {$\VIEW[A,C,E]{S}$} edge[-] (E);
    
    \node at (-0.1, -6) (D) {$\VIEW[A,C]{V^{@D}_{T}}$} edge[-] (C');
      \node at (-0.1, -9) {$\VIEW[A,C,D]{T}$} edge[-] (D);  

      \node at (-2.3, -0) (B) {$\VIEW[A]{V^{@B}_{R}}$} edge[-] (A);
      \node at (-2.3, -3) {$\VIEW[A,B]{R}$} edge[-] (B);  
    \end{tikzpicture}
    \end{minipage}
    \caption{(left) Canonical free-top variable order of the query $\VIEW[]{Q_h}$ from Example 
    \ref{ex:hierarchical}; 
    (right) Corresponding view tree.}
    \label{fig:q-hierarchical_view_tree}
    \end{figure}
    
\begin{example}
\label{ex:hierarchical}
The free-connex acyclic query 
$\VIEW[A,B,C]{Q}$ $=$ 
$\VSUM_{D}\VSUM_{E}$ $\VIEW[A,B]{R} \VPROD \VIEW[A,C,E]{S} \VPROD \VIEW[C,D]{T}$
from Example~\ref{ex:acyclic} is not hierarchical: the sets $\textsf{rels}(A)= \{\VIEW[]{R} , \VIEW[]{S}\}$
$\textsf{rels}(C)= \{\VIEW[]{S} , \VIEW[]{T}\}$ are not disjoint, nor one is included in the other.
By extending the schema of $\VIEW[]{T}$ with $A$, we obtain the $q$-hierarchical query 
$\VIEW[A,B,C]{Q_h}$ $=$ 
$\VSUM_{D}\VSUM_{E}$ $\VIEW[A,B]{R} \VPROD \VIEW[A,C,E]{S} \VPROD \VIEW[A,C,D]{T}$
whose canonical free-top variable order is given in Figure~\ref{fig:q-hierarchical_view_tree} (left).
The variant of the query,  
where variable $A$ is bound is hierarchical but not $q$-hierarchical because  
the set $\mathsf{rels}(A)= \{\VIEW[]{R}, \VIEW[]{S}, \VIEW[]{T}\}$ 
for the  \emph{bound} variable  $A$ is a strict superset of the set 
$\mathsf{rels}(B)= \{\VIEW[]{R}\}$ for the \emph{free} variable $B$.
\punto
\end{example}
     
We next exemplify  how \DF achieves constant-time update for a $q$-hierarchical query.       
   
\begin{figure}[t]
    \hspace{-0.15cm}
    \centering
    \begin{minipage}[b]{0.3\linewidth}
  \begin{tikzpicture}[xscale=0.7, yscale=0.24]

    \node at (-0.4, 3) (A) {$\color{blue} \delta \VIEW[a]{H^{@A}_{RST}}$};     
     \node at (1.5, 0) (C) {$\VIEW[A]{V^{@C}_{ST}}$} edge[-] (A);      
     \node at (1.5, -3) (C') {$\VIEW[A,C]{H^{@C}_{ST}}$} edge[-] (C);      
     \node at (2.7, -6) (E) {$\VIEW[A,C]{V^{@E}_{S}}$} edge[-] (C');
      \node at (2.7, -9) {$\VIEW[A,C,E]{S}$} edge[-] (E);
    
    \node at (-0.1, -6) (D) {$\VIEW[A,C]{V^{@D}_{T}}$} edge[-] (C');
    \node at (-0.1, -9) {$\VIEW[A,C,D]{T}$} edge[-] (D);  

      \node at (-2.3, -0) (B) {$\color{blue}\delta \VIEW[a]{V^{@B}_{R}}$} edge[-] (A);
      \node at (-2.3, -3) {$\color{blue}\delta \VIEW[a,b]{R}$} edge[-] (B);  
\end{tikzpicture}
    \end{minipage}
  \hspace{1.5cm}
    \begin{minipage}[b]{0.4\linewidth}
  \begin{tikzpicture}[xscale=0.7, yscale=0.24]

    \node at (-0.4, 3) (A) {$\color{blue} \delta \VIEW[a]{H^{@A}_{RST}}$};     
     \node at (1.5, 0) (C) {$\color{blue}\delta \VIEW[a]{V^{@C}_{ST}}$} edge[-] (A);      
     \node at (1.5, -3) (C') {\color{blue}$\delta \VIEW[a,c]{H^{@C}_{ST}}$} edge[-] (C);      
     \node at (2.7, -6) (E) {$\VIEW[A,C]{V^{@E}_{S}}$} edge[-] (C');
      \node at (2.7, -9) {$\VIEW[A,C,E]{S}$} edge[-] (E);
    
    \node at (-0.1, -6) (D) {$\color{blue}\delta \VIEW[a,c]{V^{@D}_{T}}$} edge[-] (C');
    \node at (-0.1, -9) {$\color{blue}\delta \VIEW[a,c,d]{T}$} edge[-] (D);  

    \node at (-2.3, -0) (B) {$\VIEW[A]{V^{@B}_{R}}$} edge[-] (A);
    \node at (-2.3, -3) {$\VIEW[A,B]{R}$} edge[-] (B);  
\end{tikzpicture}
    \end{minipage}
    \caption{Delta view trees derived from the view tree in Figure~\ref{fig:q-hierarchical_view_tree} 
for single-tuple updates to relations $\VIEW[]{R}$ (left) and $\VIEW[]{T}$ (right).}
    \label{fig:q-hierarchical_delta_view_trees}
    \end{figure}
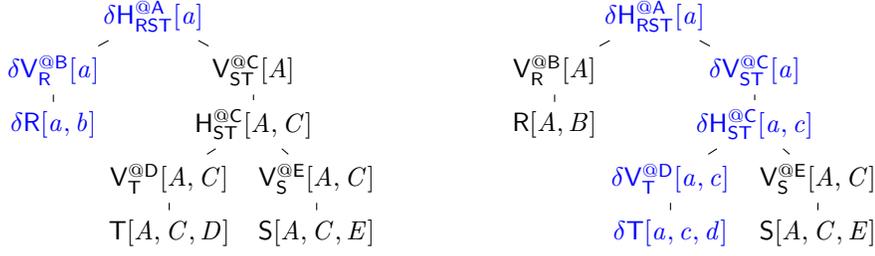

\begin{example}\label{ex:qhierarchical-update}
Figure~\ref{fig:q-hierarchical_view_tree} shows the view tree (right)
modeled on the canonical free-top variable order (left) of the 
$q$-hierarchical query $\VIEW[]{Q_h}$ in Example~\ref{ex:hierarchical}.
Figure~\ref{fig:q-hierarchical_delta_view_trees} shows the
delta view trees under single-tuple updates to $\VIEW[]{R}$  and $\VIEW[]{T}$.

In the delta view tree for $\VIEW[]{R}$, the delta view
$\delta \VIEW[]{H^{@A}_{RST}}$ can be computed by  a constant-time lookup in 
$\VIEW[]{V^{@C}_{ST}}$.   
In the delta view tree for $\VIEW[]{T}$, the delta views
$\delta \VIEW[]{H^{@C}_{ST}}$ and $\delta \VIEW[]{H^{@A}_{RST}}$ 
can be computed by constant-time lookups in 
$\VIEW[]{V^{@E}_{S}}$ and $\VIEW[]{V^{@B}_{R}}$, respectively. 
All other delta views are computed by marginalizing a variable with a single value. 
\punto
\end{example}

\begin{remark}
  $Q$-hierarchical queries admit view trees who\-se views only need primary indices to support payload lookup and updates and possibly secondary indices to support enumeration.
  Consider the view tree in Figure~\ref{fig:q-hierarchical_view_tree}. Enumeration proceeds top-down: We iterate over the $A$-values in the top view and for each such value $a$, 
  we look up in $\VIEW[a,B]{R}$ to enumerate over all the $B$-values paired with $a$, and also look up into $\VIEW[a,C]{H^{@C}_{ST}}$ to enumerate over all $C$-values paired with $a$. 
  All these look-ups require primary or secondary indices.

  Figure~\ref{fig:q-hierarchical_delta_view_trees} shows the delta view trees for single-tuple updates to $\VIEW[]{R}$ and $\VIEW[]{T}$. To compute a delta view along the path from the delta relation to the root of the delta view tree, we either perform a projection on a delta view or a lookup in the primary index of a sibling view (so with all keys of the index set to constants).\punto
\end{remark}

\subsection{Queries under Functional Dependencies}

Non-hierar\-chical queries may become hierarchical under functional dependencies (fds)~\cite{OlteanuHK09}.

Given a set $\Sigma$ of fds, we denote by $\textsf{CLOSURE}_\Sigma(\calS)$ the closure of the set $\calS$ of variables under $\Sigma$~\cite{AbiteboulHV95}. For instance, given the fds $\Sigma=\{A\rightarrow D;BD \rightarrow E\}$, we have $\textsf{CLOSURE}_\Sigma(\{A,B,C\})=\{A,B,C,D,E\}$.

\begin{definition}[adapted from~\cite{OlteanuHK09}]
  Given a set $\Sigma$ of fds and a query $\VIEW[\calS]{Q} = \VSUM_{\calB} \VIEW[\calS_1]{R_1}\VPROD\cdots\VPROD\VIEW[\calS_n]{R_n}$, the \emph{$\Sigma$-reduct} of $\VIEW[]{Q}$ under $\Sigma$ is:
\begin{align*}
  \;
  \VIEW[\textsf{CLOSURE}_\Sigma(\calS)]{Q} = \VSUM_{\calB} &\VIEW[\textsf{CLOSURE}_\Sigma(\calS_1)]{R_1}\VPROD\cdots\VPROD \VIEW[\textsf{CLOSURE}_\Sigma(\calS_n)]{R_n}
\end{align*}
\end{definition}
The $\Sigma$-reduct of a query is thus another query, where the schema of each relation is extended to include all variables in the closure of this schema under $\Sigma$. Since the added variables are functionally determined by the original schema, they do not add more information. So, we could extend these schemas and the underlying database without increasing the number of tuples in the relations. For any database $D$ with fds $\Sigma$ and a query $\VIEW[]{Q}$, the query result $\VIEW[]{Q}(D)$ is the same as the result of its $\Sigma$-reduct over the extended database.
The benefit of this rewriting is that queries may admit free-connex acyclic or even $q$-hierarchical $\Sigma$-reducts. We need not physically extend the database to reap this benefit. Instead, we use the $\Sigma$-reduct of $\VIEW[]{Q}$ to infer a free-top variable order or even a canonical free-top variable order \emph{for} $\VIEW[]{Q}$ in case the $\Sigma$-reduct is free-connex acyclic or $q$-hierarchical, respectively.
Using this variable order, we construct a view tree for $\VIEW[]{Q}$ that enjoys the preprocessing, update, and enumerate times as for its $\Sigma$-reduct.

Theorem~\ref{th:special-cases} can be generalized to account for fds.

\begin{theorem}\label{th:special-cases-fds}
  Let a query $\VIEW[]{Q}$ and a database of size $N$ and with a set $\Sigma$ of functional dependencies.

  \DF can maintain $\VIEW[]{Q}$ with $O(N)$ preprocessing, $O(1)$ enumeration delay, and $O(N)$ single-tuple updates in case the $\Sigma$-reduct of $\VIEW[]{Q}$ is free-connex acyclic.
    
  \DF can maintain $\VIEW[]{Q}$ with $O(N)$ preprocessing, $O(1)$ enumeration delay, and $O(1)$ single-tuple updates in case the $\Sigma$-reduct of $\VIEW[]{Q}$ is $q$-hierarchical.
\end{theorem}

  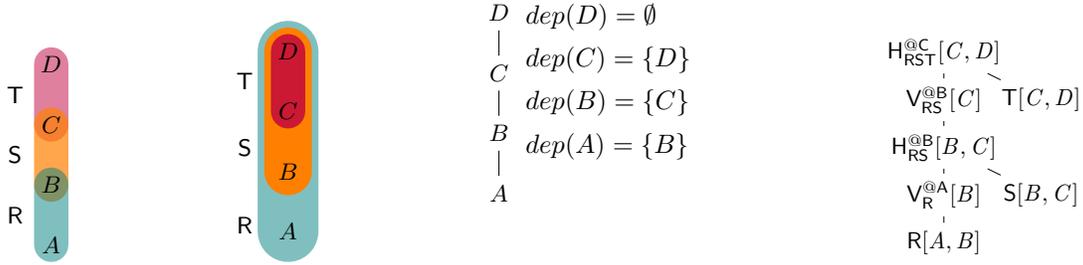
\begin{figure}[t]
   \hspace{-0.1cm}
   \centering
    \begin{minipage}[b]{0.2\linewidth}
      \begin{tikzpicture}[xscale=0.96, yscale=0.8]
        \node at (0, 0) (D) {\small  $D$};
        \node at (0, -1) (C) {\small $C$};
        \node at (0, -2) (B) {\small $B$};
        \node at (0, -3) (A) {\small $A$};
        
    \node at (-0.5, -0.5) (T) {\small  $\VIEW{T}$};    
        \node at (-0.5, -1.5) (S) {\small  $\VIEW{S}$};    
            \node at (-0.5, -2.5) (R) {\small  $\VIEW{R}$};    
\begin{pgfonlayer}{background}
\draw[opacity=.5,fill opacity=.5,line cap=round, line join=round, line width=13pt,color=purple] (0,0) -- (0,-1);\draw[opacity=0.7,fill opacity=0.7,line cap=round, line join=round, line width=13pt,color=orange] (0,-1) -- (0,-2.0);
\draw[opacity=.5,fill opacity=.5,line cap=round, line join=round, line width=13pt,color=teal] (0,-2) -- (0,-3);
        \end{pgfonlayer}
      \end{tikzpicture}
    \end{minipage}
  \hspace{-0.5cm}
\begin{minipage}[b]{0.2\linewidth}
      \begin{tikzpicture}[xscale=0.96, yscale=0.8]
        \node at (0, 0) (D) {\small  $D$};
        \node at (0, -1) (C) {\small $C$};
        \node at (0, -2) (B) {\small $B$};
        \node at (0, -3) (A) {\small $A$};
        
            \node at (-0.6, -0.5) (T) {\small  $\VIEW{T}$};    
        \node at (-0.6, -1.6) (S) {\small  $\VIEW{S}$};    
            \node at (-0.6, -2.9) (R) {\small  $\VIEW{R}$};    
\begin{pgfonlayer}{background}
\draw[opacity=.5,fill opacity=1,line cap=round, line join=round, line width=23pt,color=teal] (0,0) -- (0,-3);
\draw[opacity=1,fill opacity=1,line cap=round, line join=round, line width=18pt,color=orange] (0,0) -- (0,-2);
\draw[opacity=0.8,fill opacity=0.8,line cap=round, line join=round, line width=13pt,color=purple] (0,0) -- (0,-1);
        \end{pgfonlayer}
      \end{tikzpicture}
    \end{minipage}    
    \hspace{-0.2cm}
    \begin{minipage}[b]{0.3\linewidth}
      \begin{tikzpicture}[xscale=0.96, yscale=0.8]
        \node at (0, 0) (D) {\small  $D$};
        \node at (0, -1) (C) {\small $C$}edge[-] (D);
        \node at (0, -2) (B) {\small $B$}edge[-] (C);
        \node at (0, -3) (A) {\small $A$}edge[-] (B);
                   \node at (1.5, -1.9) {
        \begin{tabular}{@{~~}l}
          $dep(D)=\emptyset$\\[1ex]
          $dep(C)=\{D\}$\\[1ex]
          $dep(B)=\{C\}$\\[1ex]
          $dep(A)=\{B\}$\\[8ex]
        \end{tabular}
      }; 
      \end{tikzpicture}
    \end{minipage}
    \hspace{0cm}
    \begin{minipage}[b]{0.2\linewidth}
    \scalebox{0.88}{
  \begin{tikzpicture}[xscale=0.7, yscale=0.24]    
      \node at (-2.3, 6) (C) {$\VIEW[C,D]{H^{@C}_{RST}}$};
      \node at (-2.3, 3) (B) {$\VIEW[C]{V^{@B}_{RS}}$}edge[-] (C);
      \node at (-2.3, 0) (B') {$\VIEW[B,C]{H^{@B}_{RS}}$} edge[-] (B);
      \node at (-2.3, -3) (A) {$\VIEW[B]{V^{@A}_{R}}$}edge[-] (B');
      \node at (-2.3, -6) {$\VIEW[A,B]{R}$} edge[-] (A);  
      
            \node at (-0.2, -3) (S) {$\VIEW[B,C]{S}$}edge[-] (B');
      \node at (-0.2, 3) (T) {$\VIEW[C,D]{T}$}edge[-] (C);
\end{tikzpicture}
}
    \end{minipage}
    \caption{From left to right: Hypergraph of the query 
    $\VIEW{Q}$ and its $\Sigma$-reduct for $\Sigma = \{B \rightarrow C, C \rightarrow D\}$
    from Example~\ref{ex:q_hierarchical_rewriting}; 
    canonical variable order $\omega$ for  $\VIEW{Q}$; view tree modeled on $\omega$.}
    \label{fig:q-hierarchical_rewriting} 
    \end{figure}

    
  \begin{figure}[t]
 \hspace{-0.2cm}
 \centering
    \begin{minipage}[b]{0.2\linewidth}
  \begin{tikzpicture}[xscale=0.7, yscale=0.24]    
      \node at (-2.3, 6) (C) {\color{blue}$\delta\VIEW[c,d]{H^{@C}_{RST}}$};
      \node at (-2.3, 3) (B) {\color{blue}$\delta\VIEW[c]{V^{@B}_{RS}}$}edge[-] (C);
      \node at (-2.3, 0) (B') {$\color{blue}\delta\VIEW[b,c]{H^{@B}_{RS}}$} edge[-] (B);
      \node at (-2.3, -3) (A) {$\color{blue}\delta\VIEW[b]{V^{@A}_{R}}$}edge[-] (B');
      \node at (-2.3, -6) {$\color{blue}\delta\VIEW[a,b]{R}$} edge[-] (A);  
      
            \node at (-0.2, -3) (S) {$\VIEW[B,C]{S}$}edge[-] (B');
      \node at (-0.2, 3) (T) {$\VIEW[C,D]{T}$}edge[-] (C);
\end{tikzpicture}
    \end{minipage}
    \hspace{1cm}
    \begin{minipage}[b]{0.2\linewidth}
  \begin{tikzpicture}[xscale=0.7, yscale=0.24]    
      \node at (-2.3, 6) (C) {\color{blue} $\delta\VIEW[c,d]{H^{@C}_{RST}}$};
      \node at (-2.3, 3) (B) {\color{blue}$\delta\VIEW[c]{V^{@B}_{RS}}$}edge[-] (C);
      \node at (-2.3, 0) (B') {\color{blue}$\delta\VIEW[b,c]{H^{@B}_{RS}}$} edge[-] (B);
      \node at (-2.3, -3) (A) {$\VIEW[B]{V^{@A}_{R}}$}edge[-] (B');
      \node at (-2.3, -6) {$\VIEW[A,B]{R}$} edge[-] (A);  
      
            \node at (-0.2, -3) (S) {\color{blue}$\delta\VIEW[b,c]{S}$}edge[-] (B');
      \node at (-0.2, 3) (T) {$\VIEW[C,D]{T}$}edge[-] (C);
\end{tikzpicture}
    \end{minipage}
    \hspace{1cm}
    \begin{minipage}[b]{0.2\linewidth}
  \begin{tikzpicture}[xscale=0.7, yscale=0.24]    
      \node at (-2.3, 6) (C) {\color{blue}$\delta\VIEW[c,d]{H^{@C}_{RST}}$};
      \node at (-2.3, 3) (B) {$\VIEW[C]{V^{@B}_{RS}}$}edge[-] (C);
      \node at (-2.3, 0) (B') {$\VIEW[B,C]{H^{@B}_{RS}}$} edge[-] (B);
      \node at (-2.3, -3) (A) {$\VIEW[B]{V^{@A}_{R}}$}edge[-] (B');
      \node at (-2.3, -6) {$\VIEW[A,B]{R}$} edge[-] (A);  
      
            \node at (-0.2, -3) (S) {$\VIEW[B,C]{S}$}edge[-] (B');
      \node at (-0.2, 3) (T) {\color{blue}$\delta\VIEW[c,d]{T}$}edge[-] (C);
\end{tikzpicture}
    \end{minipage}
    \caption{Delta view trees derived from the view tree in Figure~\ref{fig:q-hierarchical_rewriting} for single-tuple updates to 
$\VIEW{R}$, $\VIEW{S}$, and $\VIEW{T}$ (left to right). 
The values $b$ and $c$ functionally determine $c$ and $d$, respectively.}
    \label{fig:q-hierarchical_rewriting_delta}
    \end{figure}
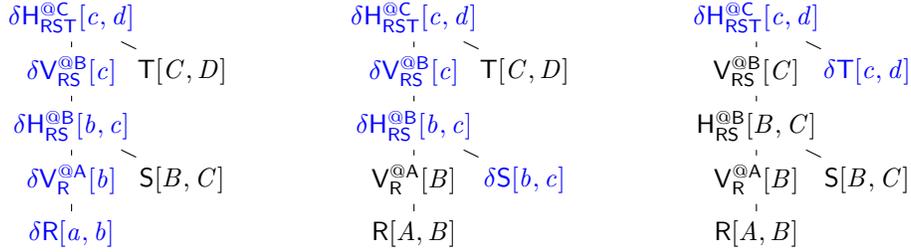
    
\begin{example} 
\label{ex:q_hierarchical_rewriting}
Consider $\Sigma = \{B \rightarrow C, C \rightarrow D\}$ and the free-connex acyclic but not hierarchical query
\begin{align*}
\quad  \VIEW[A,B,C,D]{Q} =\VIEW[A,B]{R} \VPROD \VIEW[B,C]{S} \VPROD \VIEW[C,D]{T}.
\end{align*}
The $\Sigma$-reduct of $\VIEW[]{Q}$ is 
\begin{align*}
\VIEW[A,B,C,D]{Q'} \hspace{-0.1em} =\VIEW[A,B,C,D]{R} \VPROD \VIEW[B,C,D]{S} \VPROD \VIEW[C,D]{T}.
\end{align*}
Figure~\ref{fig:q-hierarchical_rewriting} depicts the hypergraphs
of $\VIEW[]{Q}$ and $\VIEW[]{Q'}$ (left), a free-top variable order for 
$\VIEW[]{Q}$ that is also canonical for $\VIEW[]{Q'}$ (middle), and 
the view tree for $\VIEW[]{Q}$ modeled on this variable order (right).
Since $\VIEW{Q}$ is free-connex acylic, we can compute the view tree in linear time and enumerate the result tuples of $\VIEW{Q}$ with constant delay, as explained in Section~\ref{sec:free-connex}. 
We next describe how to achieve constant-time update by exploiting 
  the fds. Figure~\ref{fig:q-hierarchical_rewriting_delta} shows the delta view trees obtained from the view tree for $\VIEW{Q}$ for single-tuple updates to $\VIEW{R}$, $\VIEW{S}$, and $\VIEW{T}$.

Consider first the update $\delta\VIEW[a,b]{R}$ to relation $\VIEW[]{R}$. The delta view $\delta\VIEW[b]{V^{@A}_{R}}$ is just a projection of the update tuple. The delta view $\delta\VIEW[b,c]{H^{@B}_{RS}}$ requires a lookup in $\VIEW[B,C]{S}$ for $B=b$. In general, there may be many $C$-values paired with $b$. However, under the fd $B\rightarrow C$, there is at most one $C$-value $c$ paired with $b$. Hence, the construction of this delta view takes constant time. Similarly, the delta view $\delta\VIEW[c,d]{H^{@C}_{RST}}$ requires a lookup in $\VIEW[C,D]{T}$ for $C=c$. Again, there may be many $D$-values paired with $c$, yet under the fd $C\rightarrow D$, there is at most one $D$-value $d$ paired with $c$. Hence, the construction of this delta view takes constant time, too.

Similar reasoning applies to the update $\delta\VIEW[b,c]{S}$. To compute the delta view $\delta\VIEW[c,b]{H^{@B}_{RS}}$, we need a constant-time lookup in the view $\VIEW[B]{V^{@A}_R}$ with $B=b$. Computing $\delta\VIEW[c,d]{H^{@C}_{RST}}$ takes constant time due to the fd $C\rightarrow D$, as with updates to  $\VIEW{R}$. 
Processing the update $\delta\VIEW[c,d]{T}$ takes constant time without exploiting the fds: it only requires a lookup in the view $\VIEW[C]{V^{@B}_{RS}}$ with $C=c$. \punto
\end{example} 

%% file: cyclic.tex
\subsection{Cyclic Queries}
\label{sec:cyclic_queries}
Our framework supports arbitrary conjunctive queries. Whereas for an acyclic join query the size of each view is asymptotically upper-bounded by the size of the query result, for a cyclic query views may be larger in size than the 
query result. 
In prior work~\cite{FIVM:SIGMOD:2018}, we show how to reduce the size of intermediate views
for cyclic queries  by extending view trees with indicator projections~\cite{FAQ:PODS:2016}.  Such projections have no effect on the query result but can constrain view definitions (e.g., create cycles) and bring asymptotic savings in space and time. 

\begin{figure}\centering
  \begin{tabular}[c]{l@{~~~}l}  
    \begin{minipage}[b]{4cm}
      \begin{tikzpicture}[scale=0.5]
        \node at (-2.4, -2) (r) {$R$};
        \node at (2.4, -2) (r) {$T$};
        \node at (0, -5.75) (r) {$S$};
        \draw[rotate=60,line width=0.1mm,fill opacity=0.8,fill=magenta!40] (-2.75,-0.4) ellipse (2.45cm and 0.8cm);
       \draw[rotate=-60,line width=0.1mm,fill opacity=0.8,fill=green!60] (2.75,-0.4) ellipse (2.45cm and 0.8cm);
        \draw[rotate=0,line width=0.1mm,fill opacity=0.8,fill=blue!40] (0,-4.4) ellipse (2.45cm and 0.8cm);
        \node at (0, -1) (A) {A};
        \node at (-2, -4.35) (B) {B};
        \node at (2, -4.35) (C) {C};
      \end{tikzpicture}
      \vspace{2mm}
    \end{minipage}
    &
    \begin{minipage}[b]{5cm}
      \begin{tikzpicture}[xscale=0.9,yscale=0.9]
        \node at (0, 0) (A) {$\VIEW[~]{V^{@A}_{RST}}$};
        \node at (0, -1) (B) {$\VIEW[A]{V^{@B}_{RST}}$} edge[-] (A);
        \node at (0, -2) (C) {$\VIEW[A,B]{V^{@C}_{ST}}$} edge[-] (B);

        \node at (2, -2) (R) {$\VIEW[A,B]{R} \makebox[0pt][l]{$\phantom{\VIEW[A,B]{V^{@C}}}$} $} edge[-] (B);
        \node at (-1.5, -3) (S) {$\VIEW[B,C]{S}$} edge[-] (C);
        \node[color=red] at (-0, -3.15) (S) {$\VIEW[A,B]{\displaystyle\VEXISTS{A,B}{R}}$} edge[red,dashed] (C);
        \node at (1.5, -3) (T) {$\VIEW[C,A]{T}$} edge[-] (C);
      \end{tikzpicture}
    \end{minipage}
    \end{tabular}
\caption{(left) Hypergraph of the triangle query $Q_{\vartriangle}$; (right) View tree for the variable order $A - B - C$ with an indicator projection $\exists_{A,B}\VIEW{R}$.}
\label{fig:triangle_hypergraph_viewtree}
\end{figure}
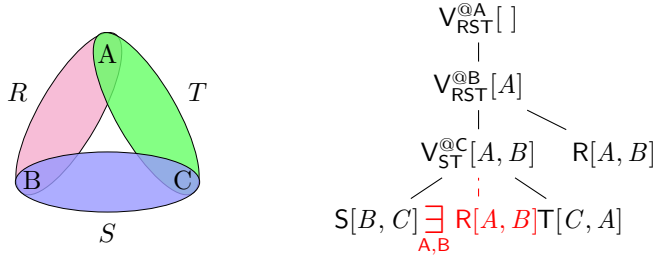

\begin{example}\label{ex:triangle_query_ivm}
We consider the triangle query: 
%
$$
\VIEW[~]{Q_{\vartriangle}} = \VSUM_{A}\VSUM_{B}\VSUM_{C} \VIEW[A,B]{R} \VPROD \VIEW[B,C]{S} \VPROD \VIEW[C,A]{T} 
$$

Figure~\ref{fig:triangle_hypergraph_viewtree} shows the hypergraph of $Q_{\vartriangle}$ and the view tree constructed for the variable order $A-B-C$ by placing each relation directly under its lowest variable. We assume all {relations} are of size $\bigO{N}$. Computing the triangle query from scratch using a worst-case optimal join algorithm takes $\bigO{N^{3/2}}$ time~\cite{Ngo:SIGREC:2013}.

In the given view tree (without the view in red), we first join $\VIEW{S}$ and $\VIEW{T}$ and then marginalize out $C$.
This view at node $C$ may contain $\bigO{N^2}$ pairs of $(A,B)$ values, which is larger than the worst-case size $\bigO{N^{3/2}}$. 
However, by materializing the view at $C$, we enable single-tuple updates to $R$ in constant time; single-tuple updates to other relations take $\bigO{N}$ time.

To avoid the large intermediate result at variable $C$, we can change the view tree by placing the relation $R$ under variable $C$. Then, joining all three relations at node $C$ takes $\bigO{N^{3/2}}$ time. Updates to any relation now cause recomputation of a $3$-way join, like in first-order IVM. For single-tuple updates, recomputing deltas takes $\bigO{N}$ as only two of the three variables are bound to constants. In contrast, the first approach trades off space for time: We need $\bigO{N^2}$ space but then support $\bigO{1}$ updates to one of the three relations.
\punto
\end{example}

The above example demonstrates how placing a relation under a different node in a view tree can create a cycle of relations and constrain the size of a view. This strategy, however, might not be always feasible or efficient: One relation might form multiple cycles of relations in different parts of a view tree -- for example, in the cyclic $4$-loop
 query $\VIEW[~]{Q_{\boxslash}}$ $=$ 
 $\VSUM_{A}\VSUM_{B}\VSUM_{C} \VSUM_{D}$ $\VIEW[A,B]{R}\VPROD \VIEW[B,C]{S} \VPROD \VIEW[C,D]{T}
 \VPROD \VIEW[D,A]{U} \VPROD \VIEW[A,C]{W}$ the chord relation $\VIEW{W}$ is part of two triangle subqueries. Since this relation cannot be duplicated in multiple subtrees (for correctness reasons so as to avoid multiplying the same payload several times instead of using it once), one would have to evaluate these subqueries in sequence, which yields a view tree that is higher and more expensive to maintain. 

\paragraph{\textbf{Indicator Projections.}}
Instead of moving relations in a view tree, we extend the tree with indicator projections that identify the active domains of these relations~\cite{FAQ:PODS:2016}. Such projections have no effect on the query result but can constrain view definitions (e.g., create cycles) and bring asymptotic savings in space and time. 

We define a new unary operation $\VEXISTS{\mathcal{A}}{\VIEW{R}}$ that, given a relation $\VIEW{R}$ over schema 
$\mathcal{S}$ with payloads from a ring $(\RING, \RINGPLUS, \RINGPROD, \RINGZERO, \RINGONE)$, and a set of attributes $\mathcal{A} \subseteq \mathcal{S}$, projects tuples from $\VIEW{R}$ with non-$\RINGZERO$ payload on $\mathcal{A}$ and assigns to these tuples the payload $\RINGONE$.

\begin{definition}[Indicator Projection]
   For a relation R over schema $\calS$ and $\calA \subseteq \calS$, 
   the indicator projection $\VEXISTS{\mathcal{A}}{\VIEW{R}}$ is a relation over $\calA$ such that
   $\forall \vecnormal{t} \in \Dom(\calA)$:
\begin{align*}
  \left( \textstyle\VEXISTS{\mathcal{A}}{\VIEW{R}} \right)[\vecnormal{t}] = 
  \begin{cases} 
    \RINGONE & \exists\vecnormal{s} \in\Dom(\mathcal{S}), \vecnormal{s}\in \VIEW{R}, \vecnormal{t} = \pi_{\mathcal{A}}(\vecnormal{s})\\
    \RINGZERO & \text{otherwise}
  \end{cases}
\end{align*}
\end{definition}

\nop{
The delta rule for $\VEXISTS{\mathcal{A}}$ is $\delta(\VEXISTS{\mathcal{A}}{\VIEW{R}}) = \VEXISTS{\mathcal{A}}(\VIEW{R} + \delta{\VIEW{R}}) - \VEXISTS{\mathcal{A}}{\VIEW{R}}$. This rule recomputes the query twice, once to insert new contents and once to delete the old contents, which clearly defeats the purpose of incremental computation. Note that some tuples might be unaffected by a given update $\delta{\VIEW{R}}$, so inserting those tuples and deleting them again is wasted work.

We observe that $\delta(\VEXISTS{\mathcal{A}}{\VIEW{R}})$ might change in the output only those tuples from $\VEXISTS{\mathcal{A}}{\delta{\VIEW{R}}}$. We exploit this restriction to refine the delta rule as: $\delta(\VEXISTS{\mathcal{A}}{\VIEW{R}}) = \VEXISTS{\mathcal{A}}{\delta{\VIEW{R}}} \VPROD (\VEXISTS{\mathcal{A}}(\VIEW{R} + \delta{\VIEW{R}}) - \VEXISTS{\mathcal{A}}{\VIEW{R}})$.
}

Indicator projections may change with updates to input relations. For instance, adding a tuple with a 
new $\mathcal{A}$-value to $\VIEW{R}$ enlarges the result of $\VEXISTS{\mathcal{A}}\VIEW{R}$; similarly, deleting the last tuple with the given $\mathcal{A}$-value reduces the result. One change in the input may cause at most one change in the output: $|\delta{(\VEXISTS{\mathcal{A}}\VIEW{R})}| \le |\delta{\VIEW{R}}|$.

To facilitate the computation of $\delta{(\VEXISTS{\mathcal{A}}\VIEW{R})}$, we keep track of how many tuples with non-$\RINGZERO$ payloads project on each $\mathcal{A}$-value. For updating the payload of a tuple in $\VIEW{R}$ from $\RINGZERO$ to non-$\RINGZERO$ (or vice versa), we increase (decrease) the count corresponding to the given $\mathcal{A}$-value. If this count changes from $0$ to $1$ (meaning the $\mathcal{A}$-value is unique) or from $1$ to $0$ (meaning there are no more tuples with the $\mathcal{A}$-value), then $\delta{(\VEXISTS{\mathcal{A}}\VIEW{R})}$ contains a tuple of $\mathcal{A}$-values with the payload of $\RINGONE$ or $-\RINGONE$, respectively; otherwise, the delta is empty.

\begin{example}
Consider a relation $\VIEW{R}$ over schema $\{A,B\}$ and 
with payloads from a ring $(\RING, \RINGPLUS, \RINGPROD, \RINGZERO, \RINGONE)$. We want to maintain the result of the query $\VIEW[A]{Q} = \VEXISTS{A}\VIEW[A,B]{R}$. To compute $\VIEW[A]{\delta{Q}}$ for updates to $\VIEW{R}$ efficiently, we count the tuples from $\VIEW{R}$ with non-$\RINGZERO$ payloads for each $A$-value, denoted by $\VIEW[A]{CNT_{Q}}$. For example:
\begin{align*}
  \begin{tabular}[t]{@{}l|@{~}c@{~}c@{~}c@{~}c@{}}
    $\VIEW{R}$ & A & B \\
    \midrule
    & $a_1$ & $b_1$ & $\to$ & $r_1$ \\
    & $a_1$ & $b_2$ & $\to$ & $r_2$ \\  
    & $a_2$ & $b_3$ & $\to$ & $r_3$ \\
  \end{tabular}
  \qquad
  \begin{tabular}[t]{@{}l|@{~}c@{~}c@{~}c@{}}
    $\VIEW{CNT_{Q}}$ & A \\
    \midrule
    & $a_1$ & $\to$ & $2$ \\
    & $a_2$ & $\to$ & $1$ \\
  \end{tabular}
  \qquad
  \begin{tabular}[t]{@{}l|@{~}c@{~}c@{~}c@{}}
    $\VIEW{Q}$ & A \\
    \midrule
    & $a_1$ & $\to$ & $\RINGONE$ \\
    & $a_2$ & $\to$ & $\RINGONE$ \\
  \end{tabular}
\end{align*}
where $r_1$, $r_2$, and $r_3$ are non-$\RINGZERO$ payloads from $\RING$.
An update $\VIEW{\delta{R}} = \{ \tuple{a_1, b_2} \to -r_2 \}$ removes the tuple $\tuple{a_1,b_2}$ from $\VIEW{R}$, which in turn decreases $\VIEW[\text{$a_1$}]{CNT_{Q}}$ by $1$. Since there is still a tuple in $\VIEW{R}$ that projects on $a_1$, the result of $\VIEW{Q}$ remains unchanged. 
A subsequent update $\{ \tuple{a_1, b_1} \to -r_1 \}$ to $\VIEW{R}$ drops the count for $a_1$ to $0$, which triggers a change in the output, $\VIEW{\delta{Q}} = \{ \tuple{a_1} \to -\RINGONE \}$.
\punto
\end{example}

\begin{figure}[t]
	\centering
	\setlength{\tabcolsep}{3pt}
	\begin{tabular}[t]{@{}c@{}c@{}l@{}}
		\toprule
		\multicolumn{3}{c}{$indicators(\text{view tree } \tau)$ : view tree}   \\
		\midrule
		\multicolumn{3}{l}{\MATCH $\tau$:}                       \\
		\midrule
		\phantom{ab} & $\VIEW{R}(\calS))$ \hspace*{2.5em} & 

		  \RETURN $\VIEW{R}(\calS)$ \\
		\cmidrule{2-3} \\[-6pt]
		             &
		\begin{minipage}[t]{1.5cm}
			\vspace{-1.5em}
			\hspace*{-0.55cm}
			\begin{tikzpicture}[xscale=0.5, yscale=1]
				\node at (0,-2)  (n4) {$\VIEW{V_{rels}^{@X}}[keys]$};
				\node at (-1,-3)  (n1) {$\tau_1$} edge[-] (n4);
				\node at (0,-3)  (n2) {$\ldots$};
				\node at (1,-3)  (n3) {$\tau_k$} edge[-] (n4);
			\end{tikzpicture}
		\end{minipage}
		             &
		\begin{minipage}[t]{5.8cm}
			\vspace{-0.4cm}
			 \LET $\hat{\tau}_i = indicators(\tau_i)$ \ $\forall i\in[k]$ \\[0.5ex]
		          \LET $\calR $ be the set of all relation symbols  \\[0.5ex]
		           \LET $\calI = \{\VEXISTS{pk}\VIEW{R}\mid \VIEW{R} \in \calR \setminus \textsf{rels} \text{ and } \\[0.5ex]
		           \TAB\TAB\TAB\STAB pk = \sch(\VIEW{R}) \cap keys \neq \emptyset \}$ \\[0.5ex]
			 \LET $\{I_1, \ldots , I_{\ell}\} = \textsf{GYO}^*(\calI,\textsf{rels})$ \\[0.5ex] 
		  \RETURN $\left\{
				\begin{array}{@{~~}c@{~~}}
					\tikz {
						\node at (3.6,-1)  (n4) {$X$};
					        \node at (2.2,-1.75)  (n1) {$\hat{\tau}_1$} edge[-] (n4);
						\node at (2.65,-1.75)  (n2) {$\ldots$};
						\node at (3.2,-1.75)  (n3) {$\hat{\tau}_k$} edge[-] (n4);
						\node at (4.0,-1.75)  (n3) {$I_1$} edge[-] (n4);
						\node at (4.4,-1.75)  (n2) {$\ldots$};
						\node at (4.9,-1.75)  (n3) {$I_\ell$} edge[-] (n4);
					}
				\end{array}  \right.$
		\end{minipage}                                              \\[2.75ex]
		\bottomrule
	\end{tabular}
	\caption{Adding indicator projections to a view tree $\tau$. 
	Each view in $\tau$ gets as new children the indicator projections of relations that do not occur in the subtree rooted at the view  but form a cycle with those that occur. 
	$\textsf{GYO}^*$ is based on the GYO reduction~\cite{BeeriFMY83}.}
	\label{fig:indicator_projections_algo}
\end{figure}
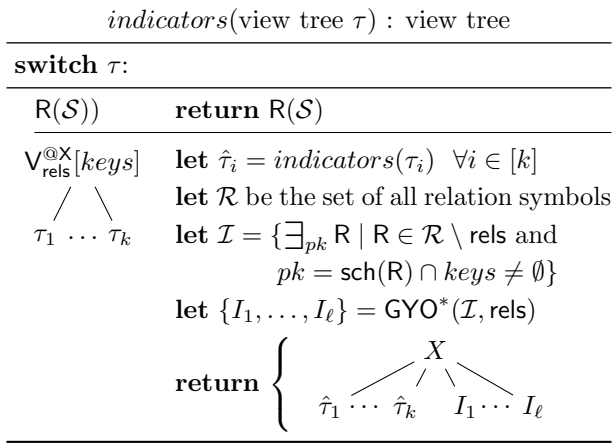

\paragraph{\textbf{View Trees with Indicator Projections.}}
Figure~\ref{fig:indicator_projections_algo} gives an algorithm that traverses a given view tree 
recursively and extends it with indicator projections. At each view $\VIEW{V^{@X}_{rels}}$, the algorithm first computes a set 
$\calI$ of indicator projections for those relations that share common variables with $\VIEW{V^{@X}_{rels}}$ and 
do not appear in \textsf{rels}, hence do not take part 
in the view definition.
Then, it chooses from this set those indicator projections 
that form a cycle with the relations 
in the subtree rooted at $\VIEW{V^{@X}_{rels}}$. To achieve this,
it uses a variant of the \textsf{GYO} reduction~\cite{BeeriFMY83}.  
  Given the hypergraph formed by 
the hyperedges representing the indicator projections $\calI$ and the relations \textsf{rels}, 
\textsf{GYO} repeatedly applies two rules until it
reaches a fixpoint: (1) Remove a node that only appears in one hyperedge; (2) Remove a
hyperedge that is included in another hyperedge. If the result of \textsf{GYO} is a hypergraph with
no nodes and one empty hyperedge, then the input hypergraph is acyclic. Otherwise,
the input hypergraph is cyclic and the output of \textsf{GYO} is a hypergraph with cycles. The \textsf{GYO}
variant, dubbed $\textsf{GYO}^*$ in the procedure in Figure~\ref{fig:indicator_projections_algo}, returns the 
hyperedges that originated from the indicator
projections in $\calI$ and contribute to this non-empty output hypergraph. The chosen indicator
projections become children of $\VIEW{V^{@X}_{rels}}$.

In a view tree with indicator projections, changes in one relation may propagate along multiple leaf-to-root paths. We propagate them in sequence, that is, updates to one relation are followed by a sequence of updates to its indicator projections. 

\begin{example}
The algorithm from Figure~\ref{fig:indicator_projections_algo} extends the view tree of the triangle query with an indicator projection $\VEXISTS{A,B}\VIEW[A,B]{R}$ placed below the view $\VIEW{V^{@C}_{ST}}$. This view at $C$ is now a cyclic join of the three relations, which can be computed in $\bigO{N^{3/2}}$ time. The indicator projection also reduces the size of this view to $\bigO{N}$.

Single-tuple updates to $S$ and $T$ still take linear time; however, bulk updates of size $\bigO{N}$ can now be processed in $\bigO{N^{3/2}}$ time, same as reevaluation. Updates to $R$ might affect the indicator projection: If a single-tuple update $\VIEW{\delta{R}}$ causes no change in the projection, then incremental maintenance takes constant time; otherwise, joining a tuple $\delta({\VEXISTS{A,B}\VIEW{R}})$ with $\VIEW{S}$ and $\VIEW{T}$ at node $C$ takes linear time. Bulk updates $\VIEW{\delta{R}}$ of size $\bigO{N}$ can also be processed in $\bigO{N^{3/2}}$ time. We conclude that using indicator projections in this query takes the best of both approaches from Example~\ref{ex:triangle_query_ivm}, namely faster incremental maintenance and more succinct view representation.
\punto
\end{example}

%% file: rings.tex
\section{Applications}
\label{sec:applications}

This section highlights four applications of \DF, including learning regression models, building Chow-Liu trees, 
computing listing or factorized representations of the results of conjunctive queries, 
and multiplying a sequence of matrices.
They behave the same in the key space, yet differ in the rings used to define the payloads.

\subsection{Covariance Matrix and Linear Regression}
\label{sec:application-lr}

We next introduce the covariance matrix ring used for training linear regression models.

\paragraph{\textbf{Linear Regression.}}
Consider a training dataset that consists of $k$ samples with $(X_i)_{i\in[m-1]}$ features and a label $X_m$ arranged into a design matrix ${\bf M}$ of size $k \times m$; in our setting, this design matrix is the result of a join query. The goal of linear regression is to learn the parameters $\Th = \TR{[\theta_1 \ldots \theta_m]}$ of a linear function\footnote{We consider wlog: $\theta_1$ is the bias parameter and then $X_1=1$ for all tuples in the input data; $\theta_m$ remains fixed to $-1$ and corresponds to the label/response $X_m$ in the data.} $f(X_1,...,X_{m-1}) = \sum_{i\in[m-1]}\theta_iX_i$ best satisfying ${\bf M} \Th \approx {\bf 0}_{k\times 1}$, where ${\bf 0}_{k\times 1}$ is the zero matrix of size $k \times 1$.

We can solve this optimization problem using batch gradient descent. This method iteratively updates the model parameters in the direction of the gradient to decrease the squared error loss and eventually converge to the optimal value. Each convergence step iterates over the entire training dataset to update the parameters, $\Th := \Th - \alpha\TR{\bf M}{\bf M}\Th$, where $\alpha$ is an adjustable step size. The complexity of each step is $\bigO{mk}$. The {\em covariance matrix}  $\TR{\bf M}{\bf M}$ quantifies the degree of correlation for each pair of features (or feature and label) in the data. Its computation can be done once for all convergence steps~\cite{SOC:SIGMOD:2016}. This is crucial for performance in case $m \ll k$ as each iteration step now avoids processing the entire training dataset and takes time $\bigO{m^2}$. 

We next show how to compute the covariance matrix assuming all features have continuous domains; we consider the case with categorical features later on. 

The covariance matrix $\TR{\bf M}{\bf M}$ accounts for the interactions {\tt SUM(X*Y)} of variables $X$ and $Y$ with continuous domains.
We can factorize their computation over training datasets defined by arbitrary join queries~\cite{SOC:SIGMOD:2016}. We can further share their computation by casting the covariance matrix computation as the computation of one compound aggregate. 
This compound aggregate is a triple 
$(\LRringC,\LRringS,\LRringQ)$, where $\LRringC$ is the number of tuples in the training dataset (size $k$ of the design matrix), $\LRringS$ is an $m\times 1$ matrix (or vector) with one sum of values per variable, and $\LRringQ$ is an $m\times m$ matrix of sums of products of values for any two variables. The covariance matrix computation can be captured by a ring.
\begin{definition}
  \label{def:lr_ring_parameterized}
  Fix a ring $(\RING, +, *, \RINGZERO, \RINGONE)$ and $m \in \mathbb{N}$.
  Let $\mathsf{C}$ denote the set of triples $({\bf D}, {\bf D}^{m}, {\bf D}^{m \times m})$,
  $\RINGZERO^{\mathsf{C}} = (\RINGZERO, \RINGZERO_{m \times 1}, \RINGZERO_{m \times m})$, and 
  $\RINGONE^{\mathsf{C}} = (\RINGONE, \RINGZERO_{m \times 1}, \RINGZERO_{m \times m})$, 
  where $\RINGZERO_{m \times n}$ is an $m \times n$ matrix with all zeros from $\RING$. 
  For $a = (\LRringC_a, \LRringS_a, \LRringQ_a) \in {\mathsf{C}}$ and $b = (\LRringC_b, \LRringS_b, \LRringQ_b) \in {\mathsf{C}}$, define the operations $+^{\mathsf{C}}$ and $*^{\mathsf{C}}$ over ${\mathsf{C}}$ as:
  \begin{align*}
  & a +^{\mathsf{C}} b = (\LRringC_a {\,\scriptstyle+\,} \LRringC_b,\; \LRringS_a {\,\scriptstyle+\,} \LRringS_b,\; \LRringQ_a {\,\scriptstyle+\,} \LRringQ_b) \\
  &a *^{\mathsf{C}} b \hspace{-0.05em}=\hspace{-0.05em} (\LRringC_a \LRringC_b,\; \LRringC_b \LRringS_a {\,\scriptstyle+\,} \LRringC_a \LRringS_b,\; \LRringC_b \LRringQ_a {\,\scriptstyle+\,} \LRringC_a \LRringQ_b {\,\scriptstyle+\,} \LRringS_a \TR{\LRringS_b} {\,\scriptstyle+\,} \LRringS_b \TR{\LRringS_a}) 
  \end{align*}
  using matrix addition, scalar multiplication, and matrix multiplication over $\RING$. 
  We refer to $(\mathsf{C}, +^{\mathsf{C}}, *^{\mathsf{C}}, \RINGZERO^\mathsf{C}, \RINGONE^\mathsf{C})$ as the {\em covariance structure of degree $m$ over $\RING$}.
\end{definition}

\begin{theorem}
  For $m\in\mathbb{N}$ and a ring $\RING$, the covariance structure of degree $m$ over $\RING$ forms a commutative ring. 
\end{theorem}

\begin{definition}\label{def:lr_ring}
  The {\em continuous covariance ring of degree $m$} is the covariance structure of degree $m$ over $\mathbb{R}$.
\end{definition}

We next show how to use this ring to compute the covariance matrix over a training dataset defined by a join with relations $(\VIEW{R_i})_{i\in[n]}$ over variables $(X_j)_{j\in[m]}$. The payload of each tuple in a relation is the identity $\RINGONE^{\mathsf{C}}$ from the continuous covariance ring of degree $m$. The query computing the covariance matrix is:
\begin{align*}
\quad \VIEW{Q} = \textstyle\VSUM_{X_1}{} \cdots \VSUM_{X_m}{\VPRODBIG_{i \in [n]} \VIEW[\mathit{\sch(R_i)}]{R_i}}
\end{align*}
For each $X_j$-value $x$, the lifting function is $g_{X_j}(x) = (1, \LRringS, \LRringQ)$, where $\LRringS$ is an $m \times 1$ vector with all zeros except the value of $x$ at position $j$, i.e., $\LRringS_j=x$, and $\LRringQ$ is an $m \times m$ matrix with all zeros except the value $x^2$ at position $(j,j)$: $\LRringQ_{(j,j)}=x^2$.

\begin{example}
\label{ex:gradient-computation}
We show how to compute the covariance matrix using the join and view tree from 
Figure~\ref{fig:example_payloads} and the database from Figure~\ref{fig:count}.
We assume alphabetical order of the five variables in the covariance matrix. The leaf relations $\VIEW{R}$, $\VIEW{S}$, and $\VIEW{T}$ map tuples to $\RINGONE^{\mathsf{C}}$ from the continuous covariance ring of degree 5. 

In the view $\VIEW{V^{@D}_{T}}$, each $D$-value $d$ is lifted to a triple $(1, \LRringS, \LRringQ)$, where $\LRringS$ is a $5\times 1$ vector with one non-zero element $\LRringS_4=d$, and $\LRringQ$ is a $(5 \times 5)$ matrix with one non-zero element $\LRringQ_{(4,4)} = d^2$. Those covariance triples with the same key $c$ are summed up, yielding:

\vspace{-8pt}
{\small
\begin{align*}
\quad \VIEW{V^{@D}_{T}}[c_1] &= (1,\LRringS_4=d_1, \LRringQ_{(4,4)}=d_1^2) \\
\quad \VIEW{V^{@D}_{T}}[c_2] &= (2,\LRringS_4=d_2+d_3, \LRringQ_{(4,4)}=d_2^2+d_3^2) \\
\quad \VIEW{V^{@D}_{T}}[c_3] &= (1,\LRringS_4=d_4, \LRringQ_{(4,4)}=d_4^2)
\end{align*}
}

The views $\VIEW{V^{@B}_{R}}$ and $\VIEW{V^{@E}_{S}}$ are computed similarly.
The view $\VIEW{V^{@C}_{ST}}$ joins $\VIEW{V^{@D}_{T}}$ and $\VIEW{V^{@E}_{S}}$ and marginalizes $C$. For instance, the payload for the key $a_2$ is:

{\setlength{\arraycolsep}{1.35pt}
\begin{align*}
\VIEW{V^{@C}_{ST}}[a_2] &= \VIEW[\mathit{c_2}]{V^{@D}_{T}} *^{\mathsf{C}} \VIEW[\mathit{a_2, c_2}]{V^{@E}_{S}} *^{\mathsf{C}} g_{C}(c_2) \\[0.5ex]
&=
\VIEW[\mathit{c_2}]{V^{@D}_{T}}
*^{\mathsf{C}}
\left(\!
    1,
    \begin{vmatrix}
    0 \\ 0 \\ 0 \\ 0 \\ e_4
    \end{vmatrix},
    \begin{vmatrix}
    0 & 0 & 0 & 0 & 0 \\
    0 & 0 & 0 & 0 & 0 \\
    0 & 0 & 0 & 0 & 0 \\
    0 & 0 & 0 & 0 & 0 \\
    0 & 0 & 0 & 0 & e_4^2
    \end{vmatrix}
\right) 
\hspace{-0.2em} *^{\mathsf{C}} \hspace{-0.2em}
\left(\!
    1,
   \begin{vmatrix}
     0 \\ 0 \\ c_2 \\ 0 \\ 0
    \end{vmatrix},
    \begin{vmatrix}
    0 & 0 & 0 & 0 & 0 \\
    0 & 0 & 0 & 0 & 0 \\
    c_2^2 & 0 & 0 & 0 & 0\\
    0 & 0 & 0 & 0 & 0 \\
    0 & 0 & 0 & 0 & 0
    \end{vmatrix}
\right)\\[0.5ex]
&= 
\left(
  2,
  \begin{vmatrix}
    0 \\ 0 \\ 2 c_2 \\ d_2 + d_3 \\ 2 e_4
  \end{vmatrix},
  \begin{vmatrix}
    0 & 0 & 0 & 0 & 0 \\
    0 & 0 & 0 & 0 & 0 \\
    0 & 0 & 2c_2^2 & c_2(d_2 + d_3) & 2 c_2 e_4 \\
    0 & 0 & c_2(d_2 + d_3) & d_2^2 + d_3^2 & (d_2 + d_3)e_4 \\
    0 & 0 & 2 c_2 e_4 & (d_2 + d_3)e_4 & 2 e_4^2
  \end{vmatrix}
\right)
\end{align*}
}

The root view $\VIEW{V^{@A}_{RST}}$ maps the empty tuple to the ring element $\sum_{i\in[2]}\VIEW[a_i]{V^{@B}_{R}} *^{\mathsf{C}} \VIEW[a_i]{V^{@C}_{ST}} *^{\mathsf{C}} g_{A}(a_i)$.
This payload has aggregates for the entire join result: the count of tuples in the result, the vector with one sum of values per variable, and the covariance matrix.
\punto
\end{example}


\paragraph{\textbf{Linear Regression with Categorical Variables.}}
Real-world datasets consists of both continuous and categorical variables. The latter take on values from predefined sets of possible values (categories). It is common practice to one-hot encode categorical variables as indicator vectors. This encoding can blow up the size of the covariance matrix and increase its sparsity.

Instead of blowing up the covariance matrix with one-hot encoding, we can capture the interactions between continuous and categorical variables as group-by queries: 
{\tt SUM(X)} group by $Y$, when $X$ is continuous and $Y$ is categorical, and
{\tt SUM(1)} group by $X$ and $Y$, when $X$ and $Y$ are categorical.
Using the group-by queries ensures a compact representation of such interactions by considering only those categories and interactions that exist in the join result.
We can encode those interactions as values from the relational data ring, introduced next. 

\begin{definition}\label{def:relational_ring}
  Let $\mathbb{F}[\mathbb{R}]$ denote the set of relations over the $\mathbb{R}$ ring, 
  the zero $\RINGZERO$ in $\mathbb{F}[\mathbb{R}]$ is the empty relation $\{\}$, which maps every tuple to  $0\in\mathbb{R}$, and the identity ${\bf 1}$ is the relation $\{ () \rightarrow 1 \}$, which maps the empty tuple to $1 \in\mathbb{R}$ and all other tuples to $0 \in\mathbb{R}$. The structure $(\mathbb{F}[\mathbb{R}], \VPLUS, \VPROD, \RINGZERO, \RINGONE)$ forms the {\em relational data} ring.\footnote{
  To form a proper ring, we need a generalization~\cite{Koch:Ring:2010:PODS} of relations and join and union operators, where: 
  tuples have their own schemas; union applies to tuples with possibly different schemas; join accounts for multiple derivations of output tuples. For our needs this generalization is not necessary.}
\end{definition}

We generalize the continuous covariance ring from Definition~\ref{def:lr_ring} to uniformly treat continuous and categorical variables as follows: 
we use relations from the relational data ring as values in $c$, $\LRringS$, and $\LRringQ$ instead of scalars;
we use union and join instead of scalar addition and multiplication;
we use the empty relation $\RINGZERO$ instead of the zero scalar.
The operations $+^\mathsf{C}$ and $*^\mathsf{C}$ over triples $(c, \LRringS, \LRringQ)$ remain unchanged.

\begin{definition}
\label{def:generalized_lr_ring}
  The {\em generalized covariance ring of degree $m$} is the covariance structure of degree $m$ over $\mathbb{F}[\mathbb{R}]$.
\end{definition}

For clarity, we show the operations $+^\mathsf{C}$ and $*^\mathsf{C}$ of the generalized covariance ring $\mathsf{C}$ of degree $m$.
$$(\LRringC', \LRringS', \LRringQ') +^\mathsf{C} (\LRringC'', \LRringS'', \LRringQ'') = (\LRringC, \LRringS, \LRringQ)$$ 
where
$\LRringC = c' \VPLUS c''$,  
$\LRringS_j = \LRringS'_{j} \uplus \LRringS''_j$, 
$\LRringQ_{(i,j)} = \LRringQ''_{(i,j)} \uplus \LRringQ''_{(i,j)}$;
$$(\LRringC', \LRringS', \LRringQ') *^\mathsf{C} (\LRringC'', \LRringS'', \LRringQ'') = (\LRringC, \LRringS, \LRringQ)$$
where
$\LRringC = c' \VPROD c''$,  
$\LRringS_j = (\LRringC'' \VPROD \LRringS'_{j}) \uplus (\LRringC' \VPROD \LRringS''_j)$, and
$\LRringQ_{(i,j)} = (\LRringC'' \VPROD \LRringQ'_{(i,j)}) \uplus (\LRringC' \VPROD \LRringQ''_{(i,j)}) \uplus (\LRringS'_{i} \VPROD \LRringS''_{j}) \uplus (\LRringS''_{i} \VPROD \LRringS'_{j})$.

The lifting function $g_{X_j}$ now depends on whether $X_j$ is continuous or categorical.
For each $X_j$-value $x$, 
$g_{X_j}(x) = (\bm{1}, \LRringS, \LRringQ)$, 
where $\bm{1} = \{() \to 1\}$,
$\bm{s}$ is an $m\times1$ vector with all $\RINGZERO$s 
except $\LRringS_j = \{ () \to x \}$ if $X_j$ is continuous and $\LRringS_j =\{ x \to 1 \}$ otherwise, 
and $\LRringQ$ is an $m \times m$ matrix with all $\RINGZERO$s 
except $\LRringQ_{(j,j)} = \{ () \to x^2 \}$ if $X_j$ is continuous and $\bm{Q}_{(j,j)} = \{ x \to 1 \}$ otherwise.

\begin{example}\label{ex:covariance-matrix-mixed}
  We compute the covariance matrix using the view tree and database from 
  Example~\ref{ex:gradient-computation} assuming that $C$ is categorical. 
  Since $B$, $D$, and $E$ are continuous, 
  the contents of $\VIEW{V^{@B}_{R}}$, $\VIEW{V^{@D}_{T}}$, and $\VIEW{V^{@E}_{S}}$ are similar to those of Example~\ref{ex:gradient-computation}
  except that every scalar value $x$ in their payloads is replaced by the relation $\{ () \to x \}$.
  The view $\VIEW{V^{@C}_{ST}}$ marginalizes $C$, lifting every $C$-value $c$ to $(\RINGONE, \LRringS_3 = \{c \to 1\},  \LRringQ_{(3,3)} = \{ c \to 1\})$, and the other entries in $\LRringS$ and $\LRringQ$ are $\RINGZERO$s.
  The payload $\VIEW{V^{@C}_{ST}}[a_2]$ encodes 
  the result of {\tt SUM(1)} group by $C$ as $\LRringS_3 = \LRringQ_{(3,3)} = \{ c_2 \to 2 \}$,
  the result of {\tt SUM(D)} group by $C$ as $\LRringQ_{(3,4)} = \{ c_2 \to d_2 + d_3 \}$, and 
  the result of {\tt SUM(E)} group by $C$ as $\LRringQ_{(3,5)} = \{ c_2 \to 2e_4 \}$. The remaining entries in the payload $\VIEW{V^{@C}_{ST}}[a_2]$ are relations mapping the empty tuple to the same scalar value from $\VIEW{V^{@C}_{ST}}[a_2]$ in  
  Example~\ref{ex:gradient-computation}. 
  The root view $\VIEW{V^{@A}_{RST}}$ computes the payload associated with the empty tuple in the same manner as in the continuous-only case but under the generalized covariance ring.
  \punto
\end{example}

\begin{remark}
For performance reasons, we only store as payloads blocks of matrices with non-zero values and assemble larger matrices as the computation progresses towards the root of the view tree. We further exploit the symmetry of the covariance matrix to compute only the entries above and including the diagonal.
For the generalized covariance ring, we store relations, which have the empty tuple as key, as scalar values.
\end{remark}

\subsection{Mutual Information and Chow-Liu Tree}
\label{sec:mutual-information}
The mutual information (MI) of two random variables $X$ and $Y$ quantifies their degree of correlation~\cite{murphy2013}: 
\[
  I(X,Y) = \hspace{-0.3cm} \sum_{x \in \Dom{(X)}} \sum_{y \in \Dom{(Y)}} p_{XY}(x,y) \log \frac{p_{XY}(x,y)}{p_X(x)p_Y(y)}
\]
where $p_{XY}(x,y)$ is the joint probability of $X=x$ and $Y=y$, and $p_X(x)$ and $p_Y(y)$ are the marginal probabilities of $X = x$ and $Y = y$, respectively.
A value close to $0$ means the variables are almost independent, while a large value means they are highly correlated.
It can be used to identify variables that predict a given label variable and can thus be used for model selection~\cite{murphy2013}. 

In our case, we are given the joint probability of several categorical variables as a relation, or the join of several relations. The probabilities defining the MI of any pair of variables can be computed as group-by aggregates over this relation. Let 
$C_{\emptyset} = {\tt SUM(1)}$, $C_{X} = {\tt SUM(1)}$ group by $X$, $C_{Y} = {\tt SUM(1)}$ group by $Y$, and $C_{XY} = {\tt SUM(1)}$ group by $X,Y$. Then, 
$p_{XY}(x,y) = \frac{C_{XY}(x,y)}{C_{\emptyset}}$,
$p_{X}(x) = \frac{C_{X}(x)}{C_{\emptyset}}$,  
$p_{Y}(y) = \frac{C_{Y}(y)}{C_{\emptyset}}$, and 
\[
  I(X,Y) = \hspace{-0.4cm} \sum_{x \in \Dom{(X)}} \sum_{y \in \Dom{(Y)}} \frac{C_{XY}(x,y)}{C_{\emptyset}} \log \frac{C_\emptyset C_{XY}(x,y)}{C_{X}(x)C_{Y}(y)}
\]
The aggregates $C_\emptyset$, $C_X$, and $C_{XY}$ define the covariance matrix over categorical variables, so we can use the generalized covariance ring to compute and maintain them 
(Section~\ref{sec:application-lr}). To compute the MI for continuous variables, we first discretize their domains into finitely many bins, so we turn them into categorical variables.

Mutual information is used for learning the structure of Bayesian networks. 
 Let a graph with one node per variable and one edge per pair of variables weighted by their MI,
 a Chow-Liu tree is a maximum weight spanning tree.
The Chow-Liu algorithm~\cite{Chow-Liu-trees:1968} constructs such a tree  in several rounds: 
it starts with a single node in the tree and in each round it connects a new node to a node already in the tree such that their pairwise MI is maximal among all pairs of variables not chosen yet.

\subsection{Factorized Representation of Query Results}
\label{sec:relational-ring}

Our framework can also support scenarios where the view payloads are themselves relations representing results of conjunctive queries, or even their factorized representations. Factorized representations can be much smaller than the listing representation of a query result~\cite{Olteanu:FactBounds:2015:TODS}, with orders of magnitude size gaps reported in practice~\cite{SOC:SIGMOD:2016}. They nevertheless remain lossless and support constant-delay enumeration of the tuples in the query result as well as subsequent aggregate processing in one pass. Besides the factorized view computation and the factorizable updates, this is the third instance where our framework exploits factorization.

We store entire relations as payloads using a variant of the relational data ring (c.f. Definition~\ref{def:relational_ring}) where values are relations over the $\mathbb{Z}$ ring. We denote this ring as $\mathbb{F}[\mathbb{Z}]$.
When marginalizing a variable, we move its values from the key space to the payload space. The tuple payloads in a view are now relations over the same schema. These relations have themselves payloads in the $\mathbb{Z}$ ring used to maintain the multiplicities of their tuples. 

We model conjunctive queries as count queries that marginalize {\em every} variable but use different lifting functions for the free and bound variables. 
For a free variable $X$ and any of its values $x$, we define $g_{X}(x) = \{ x \to 1 \}$, i.e., the lifting function 
maps $x$ to the unary relation that consists of the single value $x$ whose payload is $1$.
In case $X$ is bound, we define $g_{X}(x) = \RINGONE = \{ () \to 1 \}$, i.e.,
the lifting function maps $x$ to the identity element $\RINGONE$ of the relational data ring. 
This element is the unique relation that consist of the empty tuple whose payload is $1$.
 We have relational operations occurring at two levels: for keys, we join views and marginalize variables as before; for payloads, we interpret multiplication and addition of payloads as join and union of relations.

\begin{example}
\label{ex:relational_ring}
Consider the conjunctive query
\begin{align*}
Q(A,B,C,D) = R(A,B), S(A,C,E), T(C,D)
\end{align*}
over the three relations from Figure~\ref{fig:count}, where each tuple gets the identity payload $\{ \tuple{} \to 1 \} \in \mathbb{F}[\mathbb{Z}]$. The corresponding view marginalizes all the variables:
\begin{align*}
 \VIEW[~]Q = \textstyle\VSUM_{A}\ldots\VSUM_{E} \VIEW[A,B]{R} \VPROD \VIEW[A,C,E]{S} \VPROD \VIEW[C,D]{T}
\end{align*}
The lifting function for $E$ maps each value to $\{() \to 1 \}$, while the lifting functions for all other variables map value $x$ to $\{x \to 1 \}$.

Figure~\ref{fig:factorized_listing_ring} shows the contents of the views with relational data payloads (in black and red) for the view tree from Figure~\ref{fig:example_payloads} and the database from Figure~\ref{fig:count}. The view keys gradually move to payloads as the computation progresses towards the root. The view definitions are identical to those of the {\tt COUNT} query (but under a different ring!). The view $\VIEW{V^{@D}_{T}}$ lifts each $D$-value $d$ from $\VIEW{T}$ to the relation $\{ d \to 1 \}$ over schema $\{D\}$, multiplies (joins) it with the payload $\RINGONE$ of each tuple, and sums up (union) all payloads with the same $c$-value. The views at $\VIEW{V_R^{@B}}$ and $\VIEW{V_S^{@E}}$ are computed similarly, except the latter lifts $e$-values to $\RINGONE$ since $E$ is a bound variable. 
The view {\color{red}$\VIEW{V^{@C}_{ST}}$} assigns to each $A$-value a payload that is a union of Cartesian products of the payloads of its children and the lifted $C$-value. The root view {\color{red}$\VIEW{V^{@A}_{RST}}$} similarly computes the payload of the empty tuple, which represents the query result (both views are at the right).
\punto
\end{example}

\begin{figure}[t]
    \begin{minipage}{\linewidth}
      \centering
      \small
      \begin{tikzpicture}[xscale=1.8, yscale=1.01]
  
        \node [text=blue, anchor=north west] at (3, 1) {
          \begin{tabular}{@{}l@{\,} @{\,}c@{\,}c@{\,}l@{}}
            & $()$ & $\to$ & $\VIEW[\;]{V^{@A}_{RST}}$ \\[1ex]\toprule 
             & $\tuple{}$ & $\rightarrow$ &
              \begin{tabular}{@{}l@{\,}!{\vrule width 0.03em}@{\,}c@{}c@{}c@{}}
                & $\mathsf{A}$ & & \\
                \specialrule{.03em}{0em}{0em} 
                & $a_1$ & $\rightarrow$ & $8$ \\
                & $a_2$ & $\rightarrow$ & $2$ \\
              \end{tabular}\\\bottomrule 
          \end{tabular}
        };
  
        \node [text=red, anchor=north east] at (6.7, 1) {
          \begin{tabular}{@{}l@{\,}  @{\,}c@{\,}c@{\,}l@{}}
            & $()$ & $\rightarrow$ & \ $\VIEW[\;]{V^{@A}_{RST}}$ \\[1ex]\toprule
              & $\tuple{}$ & $\rightarrow$ & 
              \begin{tabular}{@{}l@{\,}!{\vrule width 0.03em}@{\,}c@{\,}c@{\,}c@{\,}c@{}c@{}c@{}}
                & $\mathsf{A}$ & $\mathsf{B}$ & $\mathsf{C}$ & $\mathsf{D}$ & & \\
                \specialrule{.03em}{0em}{0em} 
                & $a_1$ & $b_1$ & $c_1$ & $d_1$ & $\rightarrow$ & $2$ \\
                & $a_1$ & $b_1$ & $c_2$ & $d_2$ & $\rightarrow$ & $1$ \\
                & $a_1$ & $b_1$ & $c_2$ & $d_3$ & $\rightarrow$ & $1$ \\
                & $a_1$ & $b_2$ & $c_1$ & $d_1$ & $\rightarrow$ & $2$ \\
                & $a_1$ & $b_2$ & $c_2$ & $d_2$ & $\rightarrow$ & $1$ \\
                & $a_1$ & $b_2$ & $c_2$ & $d_3$ & $\rightarrow$ & $1$ \\
                & $a_2$ & $b_3$ & $c_2$ & $d_2$ & $\rightarrow$ & $1$ \\
                & $a_2$ & $b_3$ & $c_2$ & $d_3$ & $\rightarrow$ & $1$ \\
              \end{tabular}\\\bottomrule
          \end{tabular}
        };
  
        \node [text=blue, anchor=north west] at (3, -1.4) {
          \begin{tabular}{@{}l@{\,} @{\,}c@{\,}c@{\,}l@{}}
            & $\mathsf{A}$ & $\rightarrow$ & $\VIEW[A]{V^{@C}_{ST}}$ \\[1ex]\toprule
             & $a_1$ & $\rightarrow$ &
              \begin{tabular}{@{}l@{\,}!{\vrule width 0.03em}@{\,}c@{\,}c@{\,}c@{}}
                & $\mathsf{C}$ & & \\
                \specialrule{.03em}{0em}{0em} 
                & $c_1$ & $\rightarrow$ & $2$ \\
                & $c_2$ & $\rightarrow$ & $2$ \\
              \end{tabular} \\
            \rule{0mm}{4mm} & $a_2$ & $\rightarrow$ &
              \begin{tabular}{@{}l@{\,}!{\vrule width 0.03em}@{\,}c@{\,}c@{\,}c@{}}
                  & $\mathsf{C}$ & & \\
                  \specialrule{.03em}{0em}{0em} 
                  & $c_2$ & $\rightarrow$ & $2$ \\
              \end{tabular}\\\bottomrule 
          \end{tabular}
        };
  
        \node [text=red, anchor=south east] at (6.7, -7.2) {
          \begin{tabular}{@{}l@{\,} @{\,}c@{\,}c@{\,}l@{}}
            & $\mathsf{A}$ & $\to$ & \ $\VIEW[A]{V^{@C}_{ST}}$ \\[1ex]\toprule
             & $a_1$ & $\rightarrow$ & 
              \begin{tabular}{@{}l@{\,}!{\vrule width 0.03em}@{\,}c@{\,}c@{\,}c@{\,}c@{}}
                & $\mathsf{C}$ & $\mathsf{D}$ & & \\
                \specialrule{.03em}{0em}{0em} 
                & $c_1$ & $d_1$ & $\rightarrow$ & $2$ \\
                & $c_2$ & $d_2$ & $\rightarrow$ & $1$ \\
                & $c_2$ & $d_3$ & $\rightarrow$ & $1$ \\
              \end{tabular}\\
            \rule{0mm}{6mm} & $a_2$ & $\rightarrow$ & 
              \begin{tabular}{@{}l@{\,}!{\vrule width 0.03em}@{\,}c@{\,}c@{\,}c@{\,}c@{}}
                  & $\mathsf{C}$ & $\mathsf{D}$ & & \\
                  \specialrule{.03em}{0em}{0em} 
                  & $c_2$ & $d_2$ & $\rightarrow$ & $1$ \\
                  & $c_2$ & $d_3$ & $\rightarrow$ & $1$ \\
              \end{tabular}\\\bottomrule
          \end{tabular}
        };
  
        \node [anchor=south west] at (3, -7.2) {
          \begin{tabular}{@{}l@{\,} @{\,}c@{\,}c@{\,}c@{\,}c@{}}
            & $\mathsf{A}$ & $\mathsf{C}$ & $\to$ & $\VIEW[A,C]{V^{@E}_{S}}$ \\[1ex]\toprule
             & $a_1$ & $c_1$ & $\rightarrow$ & 
              \begin{tabular}{@{}l@{\,}!{\vrule width 0.03em}@{\,}c@{\,}c@{\,}c@{}}
                & & & \\[-2ex]
                \specialrule{.03em}{0em}{0em} 
                & $\tuple{}$ & $\rightarrow$ & $2$ \\
              \end{tabular} \\
            \rule{0mm}{3mm} & $a_1$ & $c_2$ & $\rightarrow$ & 
              \begin{tabular}{@{}l@{\,}!{\vrule width 0.03em}@{\,}c@{\,}c@{\,}c@{}}
                  & & & \\[-2ex]
                  \specialrule{.03em}{0em}{0em} 
                  & $\tuple{}$ & $\rightarrow$ & $1$ \\
              \end{tabular}\\
            \rule{0mm}{3mm} & $a_2$ & $c_2$ & $\rightarrow$ &
              \begin{tabular}{@{}l@{\,}!{\vrule width 0.03em}@{\,}c@{\,}c@{\,}c@{}}
                  & & & \\[-2ex]
                  \specialrule{.03em}{0em}{0em} 
                  & $\tuple{}$ & $\rightarrow$ & $1$ \\
              \end{tabular}\\\bottomrule
          \end{tabular}
        };
  
        \node [anchor=north west] at (1, 1) {
          \begin{tabular}{@{}l@{\,} @{\,}c@{\,}c@{\,}c@{}}
            & $\mathsf{A}$ & $\to$ & $\VIEW[A]{V^{@B}_{R}}$ \\[1ex]\toprule
             & $a_1$ & $\rightarrow$ &
              \begin{tabular}{@{}l@{\,}!{\vrule width 0.03em}@{\,}c@{\,}c@{\,}c@{}}
                & $\mathsf{B}$ & & \\
                \specialrule{.03em}{0em}{0em} 
                & $b_1$ & $\rightarrow$ & $1$ \\
                & $b_2$ & $\rightarrow$ & $1$ \\
              \end{tabular} \\
            \rule{0mm}{4mm} & $a_2$ & $\rightarrow$ &
              \begin{tabular}{@{}l@{\,}!{\vrule width 0.03em}@{\,}c@{\,}c@{\,}c@{}}
                  & $\mathsf{B}$ & & \\
                  \specialrule{.03em}{0em}{0em} 
                  & $b_3$ & $\rightarrow$ & $1$ \\
              \end{tabular} \\
            \rule{0mm}{4mm} & $a_3$ & $\rightarrow$ &
              \begin{tabular}{@{}l@{\,}!{\vrule width 0.03em}@{\,}c@{\,}c@{\,}c@{}}
                  & $\mathsf{B}$ & & \\
                  \specialrule{.03em}{0em}{0em} 
                  & $b_4$ & $\rightarrow$ & $1$ \\
              \end{tabular}\\\bottomrule 
          \end{tabular}
        };
  
        \node [anchor=south west] at (1, -7.2) {
          \begin{tabular}{@{}l@{\,} @{\,}c@{\,}c@{\,}c@{}}
            & $\mathsf{C}$ & $\to$ & $\VIEW[C]{V^{@D}_{T}}$ \\[1ex]\toprule
             & $c_1$ & $\rightarrow$ &
              \begin{tabular}{@{}l@{\,}!{\vrule width 0.03em}@{\,}c@{\,}c@{\,}c@{}}
                  & $\mathsf{D}$ & & \\
                  \specialrule{.03em}{0em}{0em} 
                  & $d_1$ & $\rightarrow$ & $1$ \\
              \end{tabular} \\
            \rule{0mm}{6mm} & $c_2$ & $\rightarrow$ & 
              \begin{tabular}{@{}l@{\,}!{\vrule width 0.03em}@{\,}c@{\,}c@{\,}c@{}}
                & $\mathsf{D}$ & & \\
                \specialrule{.03em}{0em}{0em} 
                & $d_2$ & $\rightarrow$ & $1$ \\
                & $d_3$ & $\rightarrow$ & $1$ \\
              \end{tabular} \\
            \rule{0mm}{4.5mm} & $c_3$ & $\rightarrow$ &
              \begin{tabular}{@{}l@{\,}!{\vrule width 0.03em}@{\,}c@{\,}c@{\,}c@{}}
                  & $\mathsf{D}$ & & \\
                  \specialrule{.03em}{0em}{0em} 
                  & $d_4$ & $\rightarrow$ & $1$ \\
              \end{tabular}\\\bottomrule 
          \end{tabular}
        };
      \end{tikzpicture}
    \end{minipage}
  \caption{
  Computing the query from Example~\ref{ex:relational_ring} 
  over the database
   in Figure~\ref{fig:count} 
   and the relational ring, where $\forall i\in[12]: p_i=\{ () \to 1 \}$.
   The computation uses 
   the view tree $\tau$
  in Figure~\ref{fig:example_payloads}.
   The red views (rightmost column) have payloads storing the listing representation of the intermediate and final query results. The blue views (top two views in the middle column) encode a factorized representation of these results distributed over their payloads. The remaining (black) views remain the same for both representations.
  }
  \label{fig:factorized_listing_ring}
\end{figure}

We next show how to construct a factorized representation of the query result. In contrast to the scenarios discussed above, this representation is {\em not} available as one payload at the root view, but {\em distributed} over the payloads of all views. This hierarchy of payloads, linked via the keys of the views, becomes the factorized representation. A further difference lies with the multiplication operation. For the listing representation, the multiplication is the Cartesian product. For a given view, it is used to concatenate payloads from its child views. For the factorized representation, we further project away values for all but the marginalized variable. More precisely, for each view $\VIEW[\mathcal{S}]{V^{@X}_{rels}}$ and each of its keys $a_{\mathcal{S}}$, let $\VIEW[\mathcal{T}]{P} = \VIEW[a_\mathcal{S}]{V^{@X}_{rels}}$ be the corresponding payload relation. Then, instead of computing this payload, we compute $\VSUM_{Y\in \mathcal{T}-\{X\}}\VIEW[\mathcal{T}]{P}$ by marginalizing the variables in $\mathcal{T}-\{X\}$ and summing up the multiplicities of the tuples in $\VIEW[\mathcal{T}]{P}$ with the same $X$-value.

\begin{example}
\label{ex:factorized_ring}
We continue Example~\ref{ex:relational_ring}. 
Figure~\ref{fig:factorized_listing_ring} shows the contents of the views with factorized payloads (first two columns in black and blue). 
Each view stores relational payloads that have the schema of the marginalized variable. 
Together, these payloads form a factorized representation over the variable order $\omega$ used to define the view tree in Figure~\ref{fig:example_payloads}. At the top of the factorization, we have a union of two $A$-values: $a_1$ and $a_2$. This is stored in the payloads of (middle) {\color{blue}$\VIEW[\;]{V^{A}_{RST}}$}. The payloads of (middle) {\color{blue}$\VIEW[A]{V^{@C}_{ST}}$} store a union of $C$-values $c_1$ and $c_2$ under $a_1$, and a singleton union of $c_2$ under $a_2$. The payloads of $\VIEW[A]{V^{@B}_R}$ store a union of $B$-values $b_1$ and $b_2$ under $a_1$ and a singleton union of $b_3$ under $a_2$. Note the (conditional) independence of the variables $B$ and $C$ given a value for $A$. This is key to succinctness of factorization. In contrast, the listing representation explicitly materializes all pairings of $B$ and $C$-values for each $A$-value, as shown in the payload of (right) {\color{red}$\VIEW[\;]{V^{A}_{RST}}$}. Furthermore, the variable $D$ is independent of the other variables {\em given} $C$. This is a further source of succinctness in the factorization: Even though $c_2$ occurs under both $a_1$ and $a_2$, the relations under $c_2$, in this case the union of $d_2$ and $d_3$, is only stored once in $\VIEW[C]{V^{@D}_T}$. Each value in the factorization keeps a multiplicity, that is, the number of its derivations from the input data. This is necessary for maintenance. 

This factorization is over a variable order that can be used for all queries with same body and different free variables: As long as their free variables sit on top of the bound variables, the variable order is valid and so is the factorization over it. For instance, if the variable $D$ were not free, then the factorization for the new query would be the same except that we would discard the $D$-values from the payload of the view $\VIEW{V^{@D}_{T}}$.\punto
\end{example}


\subsection{Matrix Chain Multiplication}
\label{sec:mcm}

Consider the problem of computing a product of a sequence of matrices $\bm{A}_1, \ldots, \bm{A}_n$ over some ring $\RING$, where matrix $\bm{A}_i[x_i, x_{i+1}]$ has the size $p_{i} \times p_{i+1}$, $i \in [n]$. The product $\bm{A} = \bm{A}_1 \cdots \bm{A}_n$ is a matrix of size $p_1 \times p_{n+1}$ and can be formulated as follows:

\begin{align*}
\bm{A}[x_1, x_{n+1}] = \sum_{x_2 \in[p_2]} \cdots \sum_{x_n\in [p_n]} \prod_{i\in[n]} \bm{A}_i[x_i, x_{i+1}]
\end{align*}

We model a matrix $\bm{A}_i$ as a relation $\VIEW[X_i, X_{i+1}]{A_i}$ with the payload carrying matrix values. The query that computes the matrix $\bm{A}$ is:
\begin{align*}
\VIEW[X_1, X_{n+1}]{A} = \VSUM_{X_2} \cdots \VSUM_{X_n} \VPRODBIG_{i \in [n]} \VIEW[X_i, X_{i+1}]{A_i}
\end{align*}
where each of the lifting functions $\{g_{X_j}\}_{j \in [2,n]}$ maps any key value to payload $\RINGONE\in\RING$.
Different variable orders lead to different evaluation plans for matrix chain multiplication. The optimal variable order corresponds to the optimal sequence of matrix multiplications that minimizes the overall multiplication cost, which is the textbook Matrix Chain Multiplication problem~\cite{Cormen:2009:Algorithms}.

\begin{example}
\label{ex:MCM-factorized-update}
Consider a multiplication chain of $4$ matrices of equal size $p \times p$ encoded as relations $\VIEW[X_i, X_{i+1}]{A_i}$. Let $\mathcal{F} = \{ X_1, X_5 \}$ be the set of free variables and $\omega$ be the variable order $X_1 - X_5 - X_3 - \{ X_2, X_4 \}$, i.e., $X_2$ and $X_4$ are children of $X_3$, with the matrix relations placed below the leaf variables in $\omega$. The view tree $\tau(\omega, \mathcal{F})$ has the following views (from bottom to top; the views at $X_5$ and $X_1$ are equivalent to the view at $X_3$):
\begin{align*}
\VIEW[X_1,X_3]{V^{@X_2}_{A_1A_2}} &= \textstyle\VSUM_{X_2} \VIEW[X_1,X_2]{A_1} \VPROD \VIEW[X_2,X_3]{A_2} \\
\VIEW[X_3,X_5]{V^{@X_4}_{A_3A_4}} &= \textstyle\VSUM_{X_4} \VIEW[X_3,X_4]{A_3} \VPROD \VIEW[X_4,X_5]{A_4} \\
\VIEW[X_1,X_5]{V^{@X_3}_{A_1A_2A_3A_4}} &=\hspace{-0.05em} \textstyle\VSUM_{X_3}\hspace{-0.05em} \VIEW[X_1,X_3]{V^{@X_2}_{A_1A_2}} \hspace{-0.05em}\VPROD\hspace{-0.05em} \VIEW[X_3,X_5]{V^{@X_4}_{A_3A_4}}
\end{align*}
Recomputing these views from scratch for each update to an input matrix takes $\bigO{p^3}$ time. A single-value change in any input matrix causes changes in one row or column of the parent view, and propagating them to compute the final delta view takes $\bigO{p^2}$ time. 
Updates to $\VIEW{A_2}$ and $\VIEW{A_3}$ change every value in $\VIEW{A}$. 
In case of a longer matrix chain, propagating $\VIEW{\delta{A}}$ further requires $\bigO{p^3}$ matrix multiplications, same as recomputation.

We exploit factorization to contain the effect of such changes. For instance, if $\VIEW{\delta{A_2}}$ is a factorizable update  expressible as 
$\VIEW[X_2,X_3]{\delta{A_2}} = \VIEW[X_2]{u} \VPROD \VIEW[X_3]{v}$ (see Section~\ref{sec:factorizable_updates}), then we can propagate deltas more efficiently, as products of 
subexpressions:
\begin{align*}
\quad&\VIEW[X_1,X_3]{\delta{V}^{@X_2}_{A_1A_2}} = \underbrace{\left( \textstyle\VSUM_{X_2} \VIEW[X_1,X_2]{A_1} \VPROD \VIEW[X_2]{u} \right)}_{\VIEW[X_1]{u_2}} \VPROD \VIEW[X_3]{v} \\
&\VIEW[X_1,X_5]{\delta{V}^{@X_3}_{A_1A_2A_3A_4}} = \VIEW[X_1]{u_2} \VPROD \left( \textstyle\VSUM_{X_3} \VIEW[X_3]{v} \VPROD \VIEW[X_3,X_5]{V^{@X_4}_{A_3A_4}} \right)
\end{align*}
Using such factorizable updates enables the incremental computation in $\bigO{p^2}$ time. The final delta is also in factorized form, suitable for further propagation. 

In general, for a chain of $k$ matrices of size $p \times p$, using a binary view tree of the lowest depth, incremental maintenance with factorizable updates takes $\bigO{p^2\log{k}}$ time, while reevaluation takes $\bigO{p^3 k}$ time. The space needed in both cases is $\bigO{p^2 k}$.
\punto
\end{example}

The above example recovers the main idea of LINVIEW~\cite{NEK:SIGMOD:2014}: use factorization in the incremental computation of linear algebra programs where matrix changes are encoded as vector outer products, $\delta{A} = u \TR{v}$. Such rank-$1$ updates can capture many practical update patterns such as perturbations of one complete row or column, or even changes of the whole matrix when the same vector is added to every row or column. \DF generalizes this idea to arbitrary join-aggregate queries.

%% file: experiments.tex

\section{Experiments}
\label{sec:experiments}

This section reports our experimental findings with our system \DF and three competitors: first-order IVM (\IVM), DBToaster's higher-order IVM (\DBT), and Apache Flink. We first summarize our findings.

\begin{enumerate}
  \item For maintaining covariance matrices over continuous variables, \DF outperforms \DBT and \IVM by up to three orders of magnitude. This is primarily due to the use of the covariance ring in \DF, which can capture the maintenance for an entire covariance matrix of 100-800 entries with under ten views. In contrast, \DBT requires 600-3,000 views, while \IVM needs as many delta queries as matrix entries (136 - 820).
  A similar conclusion holds for maintaining covariance matrices over continuous and categorical variables and also only over categorical variables, albeit the performance gap becomes smaller.  Thanks to the covariance ring, \DF also has a low memory footprint, on par with \IVM and 4-16x less than \DBT.
  \item Maintaining linear regression models over the covariance matrices takes insignificant time if the batch gradient descent resumes with the values for the model parameters computed after the previous update batch.
  \item Maintaining mutual information and Chow-Liu trees over the covariance matrices requires recomputation after every update batch and this can decrease the throughput of \DF by up to one order of magnitude.
  \item For $q$-hierarchical queries,  \DF is the fastest approach in case the updates are followed occasionally by a request to enumerate the query result. \DF pushes the updates from the leaves to the root view in the view tree, yet keeps the result factorized. This ensures update time and enumeration delay per tuple proportional to the payload size. We confirmed experimentally that \DBT and \IVM cannot achieve constant time for both update and enumeration.
  \item For path queries of up to 20 joins over the Twitter and TikTok graph datasets, \DF's throughput remains at least an order of magnitude larger than of competitors. \IVM and Apache Flink do not manage to process one 1K-batch within four hours for paths of more than 10 joins.
\end{enumerate}

Our conference paper~\cite{NO17} reports further experiments with \DF showing that:
(1) \DF outperforms competitors in maintaining one sum aggregate over joins;
(2) Using batches with $1,000-10,000$ tuples performs best in maintaining the covariance matrix;  
(3) Factorized updates lead to two orders of magnitude speedup for \DF over competitors for matrix chain multiplication; and
(4) For conjunctive query evaluation, factorized payloads can speed up view maintenance and reduce memory by up to two orders of magnitude compared to the listing representation of payloads.
For convenience, we include these experiments in Sections~\ref{sec:covariance-regression} (last two paragraphs), \ref{sec:sum_aggregate_maintenance}, \ref{sec:matrix_chain_multiplication}, and~\ref{sec:factorized_conjunctive}.

\subsection{Experimental Settings}
\label{sec:experiments-settings}

{\bf Competitors.}
The three maintenance strategies use DBToaster v2.3~\cite{DBT:VLDBJ:2014}, a system that compiles SQL queri\-es into code that maintains the query result under updates to input relations. The generated code represents an in-memory stream processor that is standalone and independent of any database system. DBToaster's performance on decision support and financial workloads can be several orders of magnitude better than state-of-the-art commercial databases and stream processing systems~\cite{DBT:VLDBJ:2014}. 
DBToaster natively supports \DBT and \IVM.
We use the intermediate language of DBToaster to encode \DF that maintains a set of materialized views for a given variable order and a set of updatable relations. We feed this encoding into the code generator of DBToaster. 
Unless stated otherwise, all approaches use the same runtime and store views as multi-indexed maps with memory-pooled records. The algorithms and record types used in these approaches can differ greatly. We also report on the performance of Apache Flink v1.17.1~\cite{Flink:2015} (via Table API), configured to utilize all  cores and main memory of our machine.


\begin{figure}[t]
  \centering
  \setlength{\tabcolsep}{2pt}
  \renewcommand{\arraystretch}{1.2}
  \begin{tabular}{ccccc}
  Dataset & \#Tuples & \#Relations & \#JoinVars & \# Non-JoinVars \\
  \hline
  Housing & 1.4M & 6 & 1 & 26 \\
  Retailer & 85M & 5 & 4 & 39 \\
  Favorita & 125M & 6 & 3 & 15 \\
  Twitter & 1.7M & 1 & 2 & 0 \\
  TikTok & 0.6M & 1 & 2 & 0 \\
  \hline
  \end{tabular}
  \caption{{Characteristics of the input datasets.}}
  \label{fig:datasets}
\end{figure}

{\bf Datasets.}
Figure~\ref{fig:datasets} summarizes our datasets:
\begin{itemize}[leftmargin=0.5em,itemindent=1.5em]
    \item {\em Housing} is a synthetic dataset modeling a house price market~\cite{SOC:SIGMOD:2016}.
    It consists of six relations: {\tt House}, {\tt Shop}, {\tt Institution}, {\tt Restaurant}, {\tt Demographics}, and {\tt Transport}, arranged into a star schema. The natural join of all relations is on the common variable (postcode) and has $26$ non-join variables, 14 continuous and 12 categorical. We consider a variable order where each root-to-leaf path consists of variables of one relation.

    \item {\em Retailer} is a real-world dataset used by a retailer to inform decision-making and forecast user demands \cite{SOC:SIGMOD:2016}. 
    It has a snowflake schema with one large fact relation {\tt Inventory} storing information about the inventory units for products in a location, at a given date.
    This relation joins along three dimension hierarchies: {\tt Item} (on product id), {\tt Weather} (on location and date), and {\tt Location} (on location) with its lookup relation {\tt Census} (on zip). 
    The natural join of these relations is acyclic and has 33 continuous and 6 categorical non-join variables. We use a variable order, where the variables of each relation form a distinct root-to-leaf path, and the partial order on join variables is: location - $\{$ date - $\{$ product id $\}$, zip $\}$.
     
    \item {\em Favorita} is a real-world dataset comprising sales data of items sold in grocery stores in Ecuador~\cite{Favorita:Dataset}.    
    It has a star schema with one large fact relation {\tt Sales}  storing information on sales transactions, including the date, store, item, and item quantity. This relation joins with five dimension tables: {\tt Stores} (on store id), {\tt Item} (on item id), {\tt Transaction} (on date and store id), {\tt Holiday} (on date), and {\tt Oil} (on date).
    The natural join has 3 continuous and 12 categorical non-join variables.
    We consider a variable order where the order on join variables is: date - store id - item id.

    \item 
      {\em Twitter~\cite{SNAP} and TikTok~\cite{TikTok}} are publicly available graph datasets.
    
\end{itemize}

We evaluate the maintenance strategies over data streams synthesized from the above datasets by interleaving insertions to the input relations in a round-robin fashion. 
These insertions arrive sorted following a top-down order of \DF's variable orders. This leads to improved runtimes of all systems relative to out-of-order insertions.
We group insertions into batches of 1000 tuples and place no restriction on the order of records in input relations. 
In all experiments, we use payloads defined over rings with additive inverse, thus processing deletions is similar to processing insertions.

\begin{figure}[t]
  \centering
  \renewcommand{\arraystretch}{1.2}
  \begin{tabular}{llc@{~}cc@{~}cc@{~}c}
    & & \multicolumn{2}{c}{\DF} & \multicolumn{2}{c}{\DBT} & \multicolumn{2}{c}{\IVM} \\
    \midrule
    \multirow{3}{*}{\scalebox{0.85}{\rotatebox{90}{\!\!\centering CONT}}} 
    & Housing & 11,570.2 & (7) & 953.7 &(626) & 0.7 & (384) \\
    & Retailer & 3,818.8 & (9) & 9.1$^{*}$ & (3,186) & 28.8 & (825) \\
    & Favorita & 1,411.3 & (9) & 33.2$^{*}$ & (615) & 182.0 & (142) \\
    \midrule
    \multirow{3}{*}{\scalebox{0.85}{\rotatebox{90}{\!\!\centering MIXED}}} 
    & Housing & 996.4 & (7) & 682.6 & (599) & 1.3 & (375) \\
    & Retailer & 1,255.8 & (9) & 7.2$^{*}$ & (3,144) & 21.7$^{*}$ & (819) \\
    & Favorita & 354.0 & (9) & 18.3$^{*}$ & (535) & 87.2 & (130) \\
    \bottomrule
  \end{tabular}
  \caption{The average throughput (in thousands of tuples/sec) and in parentheses the number of materialized views for the maintenance of the covariance matrix over datasets where features are treated as all continuous (CONT) and as a mix of continuous and categorical (MIXED). The symbol $^{*}$ denotes the one-hour timeout.}
  \label{fig:throughput_views_per_dataset}
\end{figure}

{\bf Queries.} 
We consider the following queries:
\begin{itemize}[leftmargin=0.5em,itemindent=1.5em]
\item {\em Covariance Matrix:}
For \DF, we use one query per dataset to compute one covariance aggregate over the natural join of the input relations. For instance, the query over the {\em Retailer} schema is:
\begin{lstlisting}[language=SQL,columns=flexible, mathescape, basicstyle=\linespread{1.1}\ttfamily\small]
  SELECT SUM(g$_1$(X$_1$) * ... * g$_{39}$(X$_{39}$))
  FROM Inv NATURAL JOIN It NATURAL JOIN W
            $\hspace{0.2mm}$NATURAL JOIN L NATURAL JOIN C;
\end{lstlisting}
where $\{ X_i \}_{i \in [39]}$ are all the non-join variables from the {\em Retailer} schema.
We consider three scenarios: (1) we treat all variables as continuous; (2) with a mix of continuous and categorical variables; and (3) with all categorical variables.
For the first, we use the continuous covariance ring of degree 39 
and the lifting function $g_i(x) = (\LRringC_i=1, \LRringS_i = x, \LRringQ_{(i,i)} = x^2)$ for each variable $X_i$, as in Example~\ref{ex:gradient-computation}.
For the other two, we use the generalized covariance ring with relational values, as in Example~\ref{ex:covariance-matrix-mixed}.
Similarly, the queries over {\em Housing} ({\em Favorita}) use the covariance rings of degree $26$ ($15$). 

For \DBT and \IVM, we use queries that compute scalar sum aggregates in the covariance matrix. When considering all variables as continuous, we use one query per dataset to compute 
$1 + n + \frac{n(n+1)}{2}$ sums, where $n$ is the number of variables; 
for {\em Housing}, {\em Retailer}, and {\em Favorita}, we compute $378$, $820$, and $136$ sums, respectively. 
When considering continuous and categorical variables, we use a batch of group-by aggregate queries as input to DBToaster. 
For {\em Housing}, {\em Retailer}, and {\em Favorita}, the number of queries with distinct group-by variables is $46$, $22$, and $79$, respectively.
  
\item
{\em $Q$-Hierarchical Queries:} We use the natural joins of all relations in each dataset.
For {\em Housing}, this is a star join query. For {\em Favorita}, the relation {\tt Stores}  violates the $q$-hierarchical property: Its variables form a strict subset of a root-to-leaf path in the canonical variable order. To ensure constant time for single-tuple updates, we require {\tt Stores} to be non-updatable.
For {\em Retailer}, the query is $q$-hierarchical due to (1) the  functional dependency ${\tt zip}\rightarrow {\tt location}$ in {\tt Census} and (2) requiring the relation {\tt Item} be non-updatable.

\item
{\em $k$-Path Queries:} These queries join $k$ copies \texttt{R}$_1$ to \texttt{R}$_k$ of the edge relation of the input graph: 
\begin{lstlisting}[language=SQL,columns=flexible, mathescape, basicstyle=\linespread{1.1}\ttfamily\small]
  SELECT A$_1$, A$_{k+1}$, SUM(W$_1$ * ... * W$_k$)
  FROM R$_1$ NATURAL JOIN ... NATURAL JOIN R$_k$
  GROUP BY A$_1$, A$_{k+1}$
\end{lstlisting}
Each relation \texttt{R}$_i$ has schema (\texttt{A}$_i$, \texttt{A}$_{i+1}$, \texttt{W}$_i$) and can be seen as the adjacency matrix of the input graph, with rows indexed by \texttt{A}$_i$, columns indexed by \texttt{A}$_{i+1}$, and the value \texttt{W}$_i=1$ in cell (\texttt{A}$_i$, \texttt{A}$_{i+1}$). The path query is then the $k$ times multiplication of the adjacency matrix.

\item 
\add{
{\em Matrix Chain Multiplication:}
The query in standard SQL is defined over tables $A_1(I,J,P_1)$, $A_2(J,K,P_2)$, $A_3(K,L,P_3)$:
}
\begin{lstlisting}[language=SQL,columns=flexible, basicstyle=\linespread{1.1}\ttfamily\small]
  SELECT A1.I, A3.L, SUM(A1.P1 * A2.P2 * A3.P3)
  FROM A1 NATURAL JOIN A2 NATURAL JOIN A3
  GROUP BY A1.I, A3.L;
\end{lstlisting}
\add{
In our formalism, each relation maps pairs of indices to matrix values, all lifting functions map values to $1$, and the query is: 
$\VIEW[I,L]{Q}=\VSUM_{J}\VSUM_{K} \VIEW[I,\textsf{$J$}]{A_1} \VPROD \VIEW[\textsf{$J$},K]{A_2} \VPROD \VIEW[K,L]{A_3}$.
}

\item
\add{
{\em Factorized Computation of Conjunctive Queries:}
We consider two full conjunctive queries joining all the relations in the {\em Retailer} and respectively {\em Housing} datasets. 
}

\end{itemize}

{\bf Experimental Setup. }
We run the first four experiments, Sections~\ref{sec:covariance-regression} (except the last two paragraphs), \ref{sec:MI}, \ref{sec:q-hier-experiment} and~\ref{sec:path-queries}, on a machine with Intel(R) Xeon(R) Silver 4214 CPU @ 2.20GHz, 188GB RAM, and Debian 10. 
We use DBToaster v2.3 for running \DBT and \IVM and generating code in \DF. 
The generated C++ code is single-threaded and compiled using g++ 8.3.0 with the -O3 flag. 

The remaining experiments, which are detailed in Sections~\ref{sec:covariance-regression} (last two paragraphs), \ref{sec:sum_aggregate_maintenance}, \ref{sec:matrix_chain_multiplication} and~\ref{sec:factorized_conjunctive}, are from the conference paper~\cite{FIVM:SIGMOD:2018}. They were run on a Microsoft Azure instance with Intel(R) Xeon(R) CPU E5-2620 v3 @ 2.40GHz, 32 GB RAM, and Ubuntu Server 14.04.
We used DBToaster v2.2 and the compiler g++ 6.3.0 with the -O3 flag.

All experiments are run single-threaded.
Unless stated otherwise, we set an one-hour timeout on query execution and report wall-clock times by averaging three best results out of four runs. 
We profile memory utilization using gperftools, not counting the memory used for storing input streams.

\begin{figure*}[t]
  \centering   
  \includegraphics[width=0.32\textwidth]{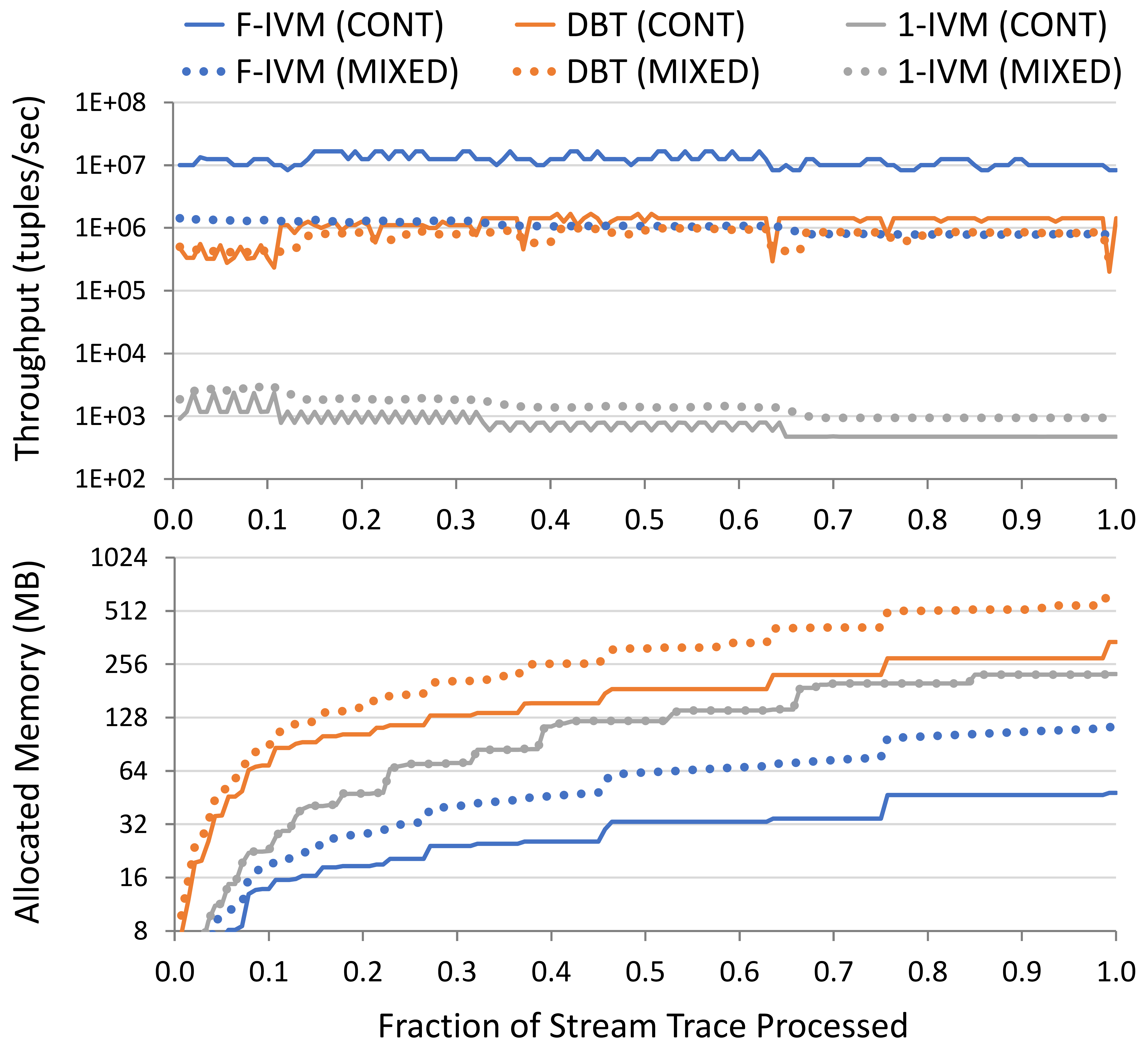}
  \;
  \includegraphics[width=0.32\textwidth]{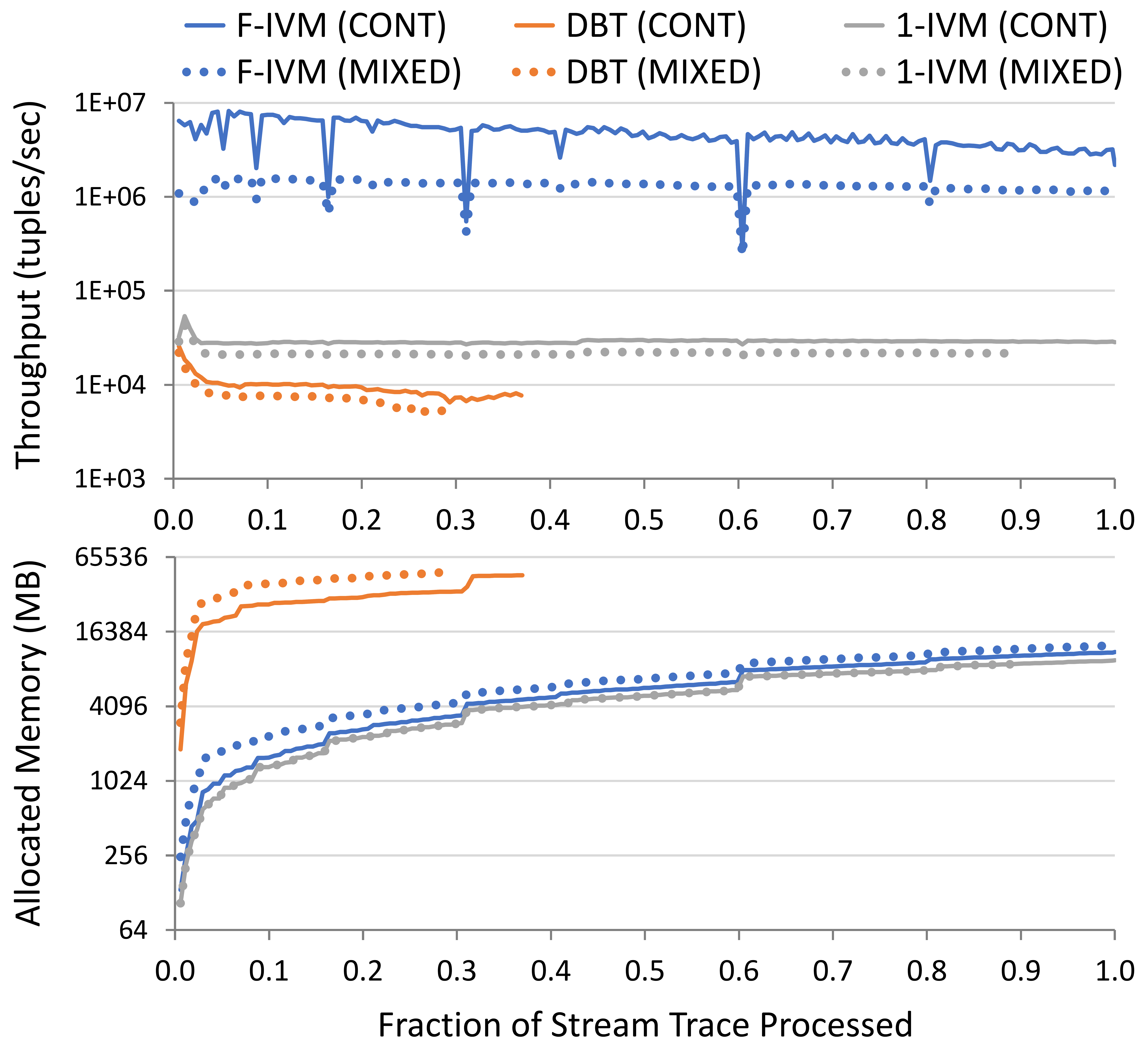}
  \;
  \includegraphics[width=0.32\textwidth]{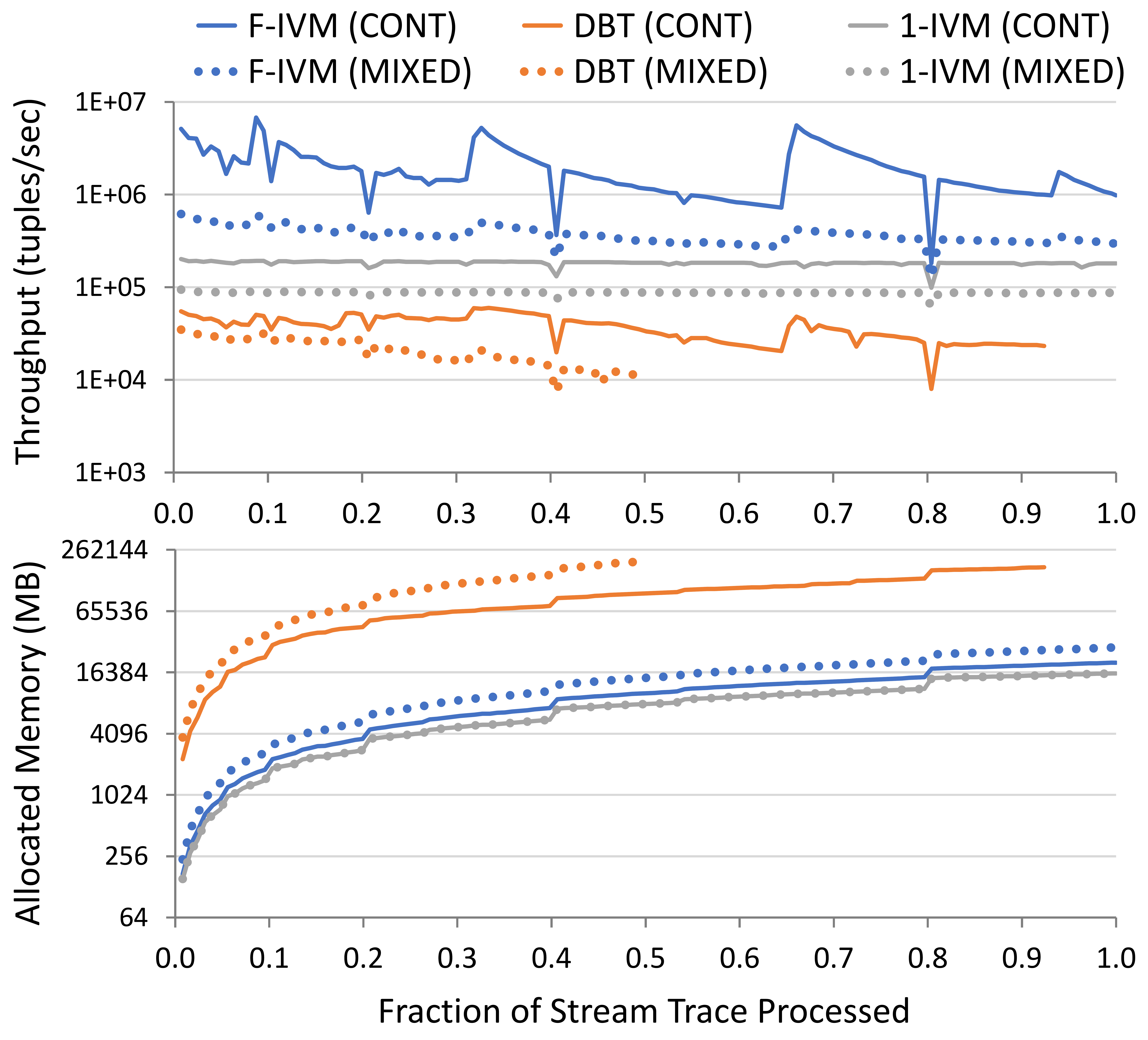}
  \caption{Incremental maintenance of the covariance matrix over the {\em Housing} dataset (left),  
  {\em Retailer} dataset (middle), and {\em Favorita} dataset (right) under updates of size $1,000$ to all relations with a one-hour timeout. The CONT plots consider all features as continuous, while the MIXED plots consider a mix of continuous and categorical features. }
  \label{fig:cofactor_IVM_trace_ALL}
\end{figure*}
\begin{figure*}[t]
  \centering   
  \includegraphics[width=0.32\textwidth]{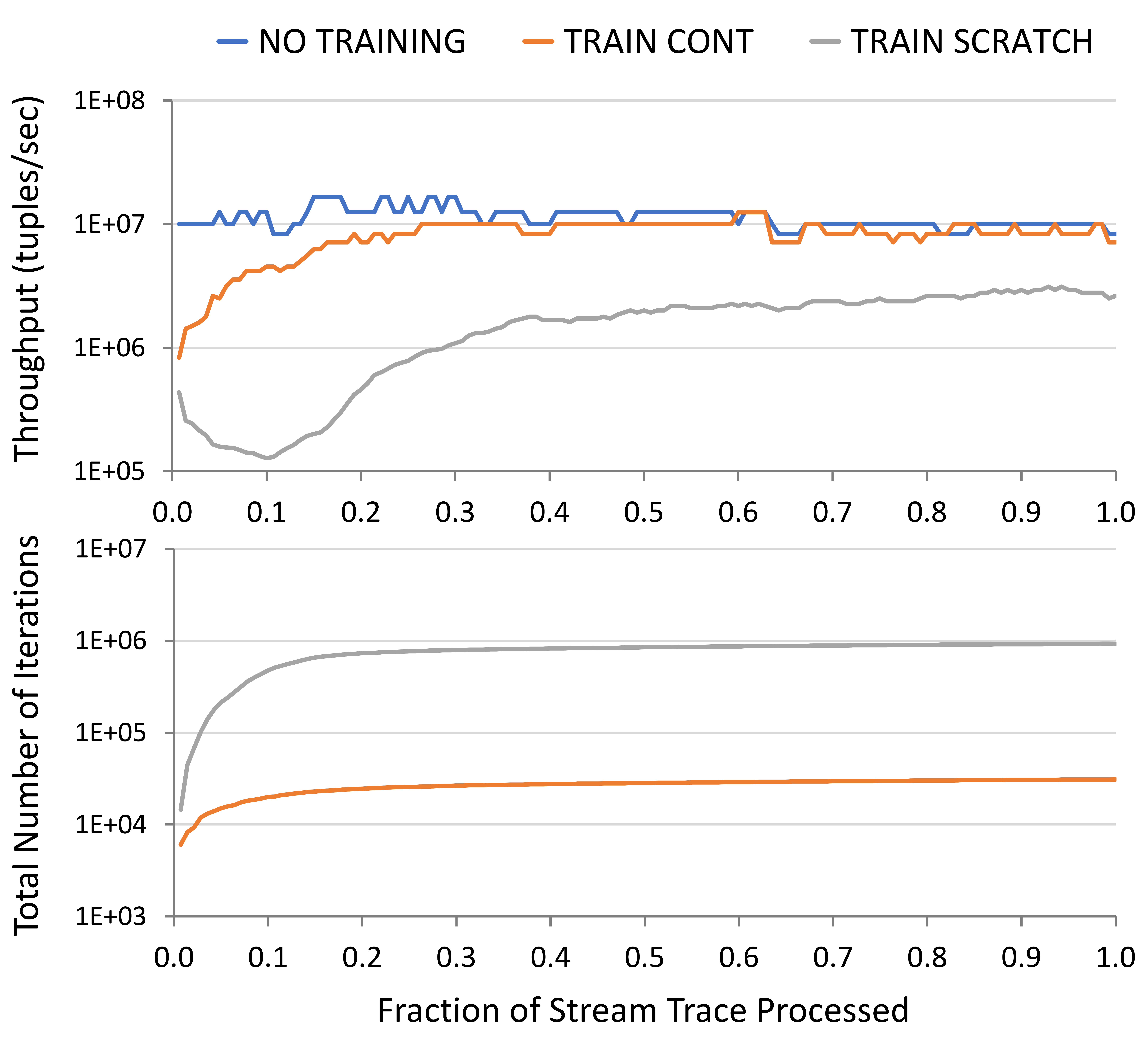}
  \;
  \includegraphics[width=0.32\textwidth]{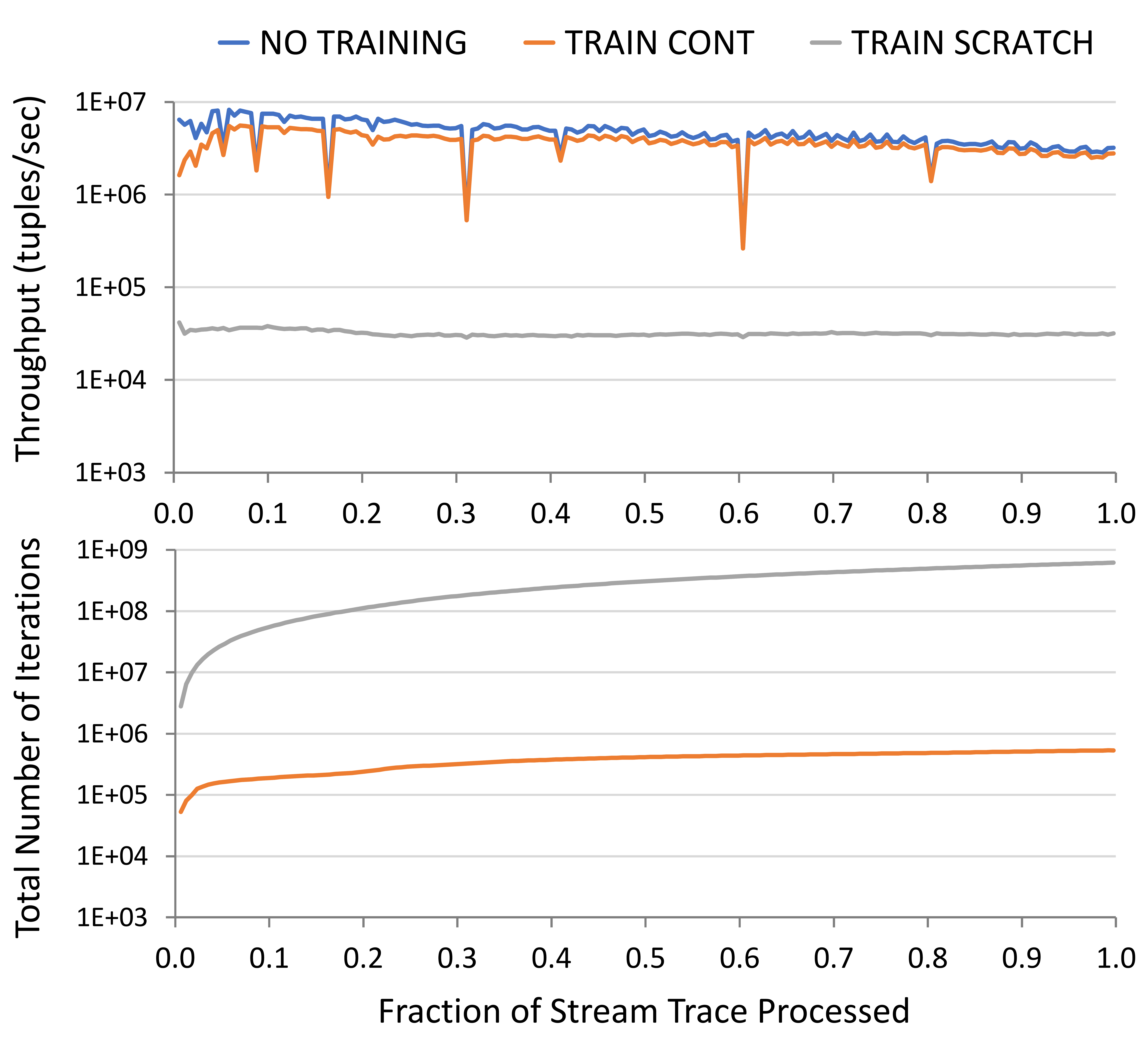}
  \;
  \includegraphics[width=0.32\textwidth]{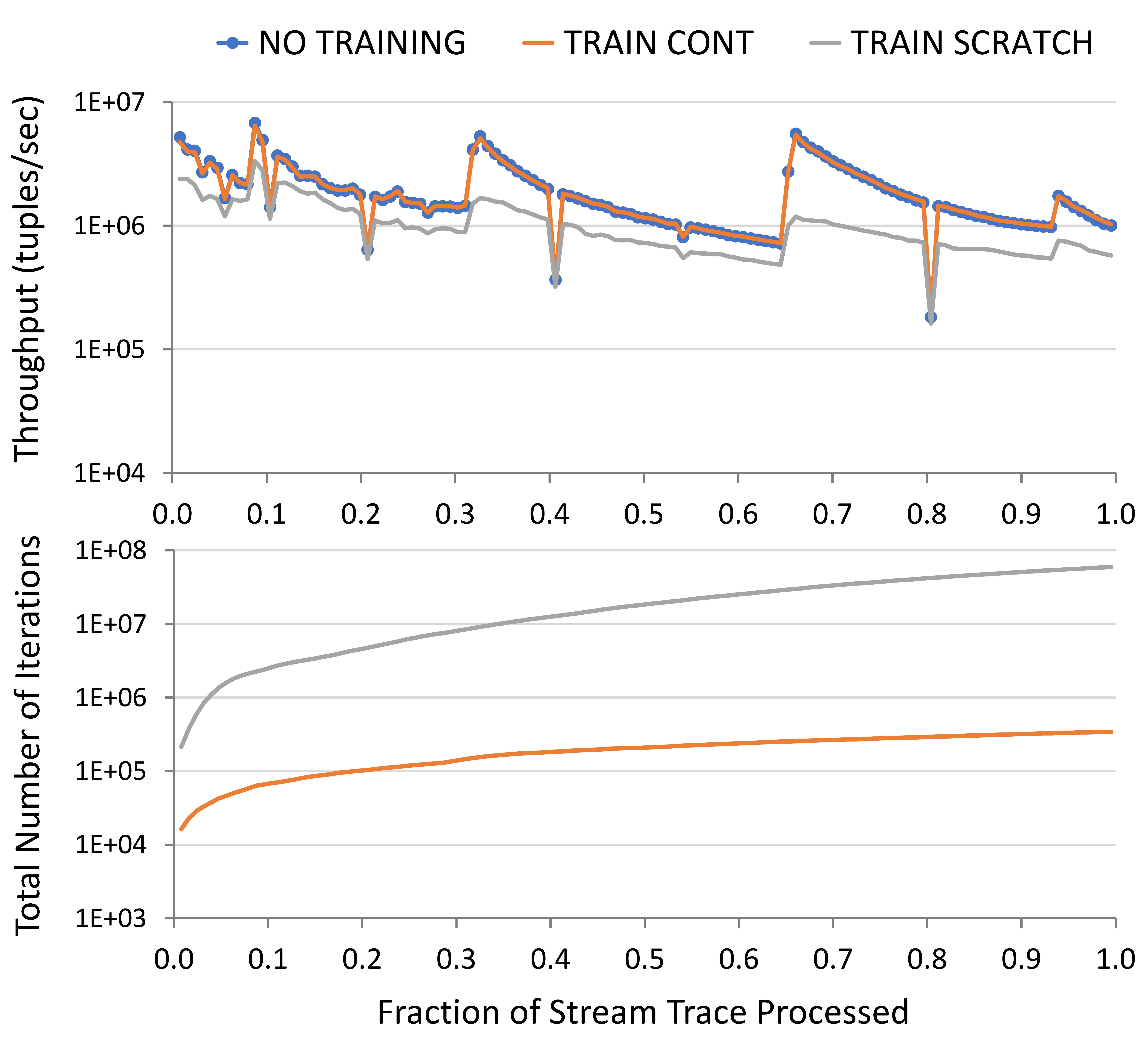}
  \caption{Maintaining linear regression models over the {\em Housing} dataset (left), {\em Retailer} dataset (middle), and {\em Favorita} dataset (right) under updates of size $1,000$ to all relations using \DF. Batch gradient descent is invoked after every update using previously learned parameters (TRAIN CONT) and parameters set to 0 (TRAIN SCRATCH). The NO TRAINING plots show the time to compute the covariance matrices only. The bottom charts show the cumulative numbers of iterations used by the batch gradient descent during the training phase. }
  \label{fig:cofactor_IVM_training}
\end{figure*}

\subsection{Covariance Matrix and Linear Regression} 
\label{sec:covariance-regression}

We benchmark the performance of maintaining a covariance matrix for learning regression models over natural joins. We consider updates to all input relations. We compute the covariance matrix over all non-join variables of the join query (i.e., over all non-join attributes in the input database), which suffices to learn linear regression models over {\em any label and set of features} that is a subset of the set of variables~\cite{OS:PVLDB:16}. This is achieved by specializing the convergence step in batch gradient descent to the relevant restriction of the covariance matrix. 
In our approach for learning linear regression models over database joins, the convergence step takes orders of magnitude less time compared to the data-dependent covariance matrix computation.

Figure~\ref{fig:throughput_views_per_dataset} shows the number of views materialized by \DF, \DBT, and \IVM for computing the covariance matrix.
\DF computes one aggregate query with payloads from a covariance ring. 
For {\em Housing}, where all relations join on one variable, \DF materializes seven views: one view per relation to marginalize out all non-join variables, and the root view to join these views. 
For {\em Retailer}, \DF materializes five views over the input relations, three intermediate views, and the root view; similarly, for {\em Favorita}.
These views have payloads from the continuous (generalized) covariance ring if all features are continuous (continuous and categorical). 

\DBT and \IVM maintain a batch of sum aggregate queries with scalar payloads. 
These materialization strategies fail to effectively share the computation of covariance aggregates, materializing linearly many views in the size of the covariance  matrix: for instance, when considering all variables as continuous, 
\DBT and \IVM materialize $626$ and respectively $384$ views to maintain $378$ scalar aggregates for {\em Housing}; similar reasoning holds for the other datasets and the scenarios with both continuous and categorical variables.

{\bf Throughput.}
Figure~\ref{fig:cofactor_IVM_trace_ALL} shows the throughput of \DF, \DBT, and \IVM as they process an increasing fraction of the stream of tuple inserts. 
Figure~\ref{fig:throughput_views_per_dataset} shows their average throughput after processing the entire stream. 
The throughput is higher when all features are continuous than for a  mix of continuous and categorical features.
This is expected as the latter computes additional group-by aggregates for the categorical features; in this case, the number of computed aggregates is data-dependent.
The occasional  hiccups in the throughput of \DF are due to doubling the memory allocated to the underlying data structures used for the views.


The query for {\em Housing} joins all relations on the common variable, which is the root in our variable order; thus, the query is hierarchical. \DF computes the covariance matrix using the query with no free variables in both scenarios (CONT and MIXED) and can process a single-tuple update to any input relation in time linear in the size of the payload. In the continuous-only scenario, the update time is $\bigO{m^2}$, where $m$ is the number of continuous features; in the mixed scenario, the update time depends on the size of the domain of the categorical features.
\DBT exploits the conditional independence in the derived deltas to materialize each input relation separately such that all non-join variables are aggregated away. In the case of all continuous features, each materialized view has $\bigO{1}$ maintenance cost per update tuple, but the large number of views in \DBT is the main reason for its poor performance. 
\IVM stores entire tuples of the input relations including non-join variables.
On each update, \IVM recomputes a batch of aggregates on top of the join of these input relations and the update tuple.
Since the update tuple binds the value of the common join variable, the hypergraph of the delta query consists of disconnected components. 
DBToaster first aggregates over each relation and then joins together the partial aggregates on the common variable.
Even with this optimization, \IVM takes time linear in the size of the dataset, which explains its poor performance.

For {\em Retailer}, the inserts are mostly into {\tt Inventory}.  Since the variables of this relation form a root-to-leaf path in the variable order, \DF can process single-tuple updates to this relation in $\bigO{1}$ time in data complexity in the continuous-only scenario. 
\DBT maintains up tp four views per scalar aggregate and fails to process the entire stream within a one-hour limit 
in both scenarios. 
\IVM maintains one view per scalar group-by aggregate but recomputes the delta query on each update, resulting in 132x (58x) lower throughput than \DF in the continuous-only (mixed) scenario.

For {\em Favorita}, \IVM achieves better performance than on {\em Retailer} but still 7.8x (4.1x) slower than \DF in the continuous (mixed) scenario. \DBT  fails to finish the entire stream within a one-hour timeout.

{\bf Memory Consumption.}
Figure~\ref{fig:cofactor_IVM_trace_ALL} shows that \DF achieves lower or comparable memory utilization on the three datasets, while providing orders of magnitude better performance than its competitors.  
The reason behind this memory efficiency is that \DF uses compound aggregates and factorization structures to express the covariance matrix computation over fewer views compared to \DBT and \IVM.

{\bf End-to-End Training.}
We next analyze the cost of learning linear regression models from the computed covariance matrices. 
We consider the scenario where all variables are continuous and the target label is house price ({\em Housing}), inventory units ({\em Retailer}), and sold units ({\em Favorita}).
Using batch gradient descent and the covariance matrix, the time needed to converge on the model parameters represents $0.24\%$, $0.2\%$, and $0.001\%$ of the time needed to process all updates in the stream for {\em Housing}, {\em Retailer}, and {\em Favorita}, respectively.

Figure~\ref{fig:cofactor_IVM_training} illustrates the performance of \DF for maintaining the covariance matrix in three scenarios:
1) without training the linear regression model (baseline);
2) with training after each batch update, starting from previously learned parameters (CONT); and 3) with training after every batch update, starting with value 0 for the  parameters (SCRATCH). 
Continuously refreshing the model after every update reduces the throughput of baseline by 41\%, 18\%, and 2\% for {\em Housing}, {\em Retailer}, and {\em Favorita}. 
In contrast, retraining from scratch after every update has  significantly higher overheads and reduces the baseline throughput by 95\%, 99\%, and 42\% for the three datasets. 
Decreasing the training frequency brings the throughput closer to the baseline. 

Figure~\ref{fig:cofactor_IVM_training} (bottom plots) shows for each dataset the cumulative number of iterations of batch gradient descent in the two training scenarios. 
Continuously improving learned parameters yields 30x, 1160x, and 175x fewer iterations compared to retraining from scratch for {\em Housing}, {\em Retailer}, and {\em Favorita}, respectively. 
This reflects in the throughput of the two training scenarios.

\add{
\begin{figure*}[t]
\begin{minipage}[t]{.475\textwidth}
  \centering   
  \includegraphics[width=\columnwidth]{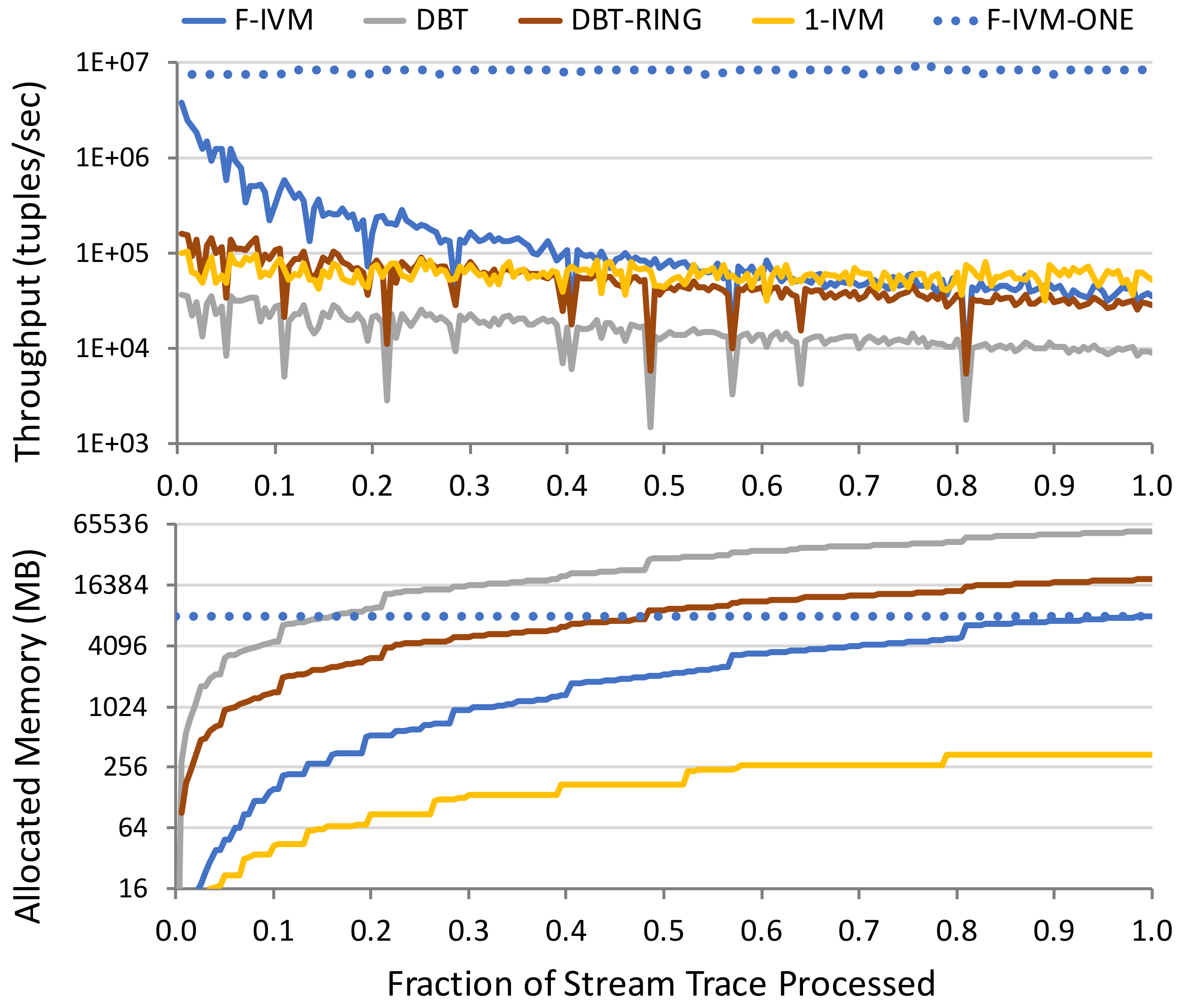}
  \caption{Incremental maintenance of the covariance matrix on top of the triangle query on {\em Twitter} for updates of size $1,000$ to all input relations.}
  \label{fig:cofactor_Triangle_IVM_trace_ALL}
\end{minipage}
\quad
\begin{minipage}[t]{.475\textwidth}
  \centering   
  \includegraphics[width=\columnwidth]{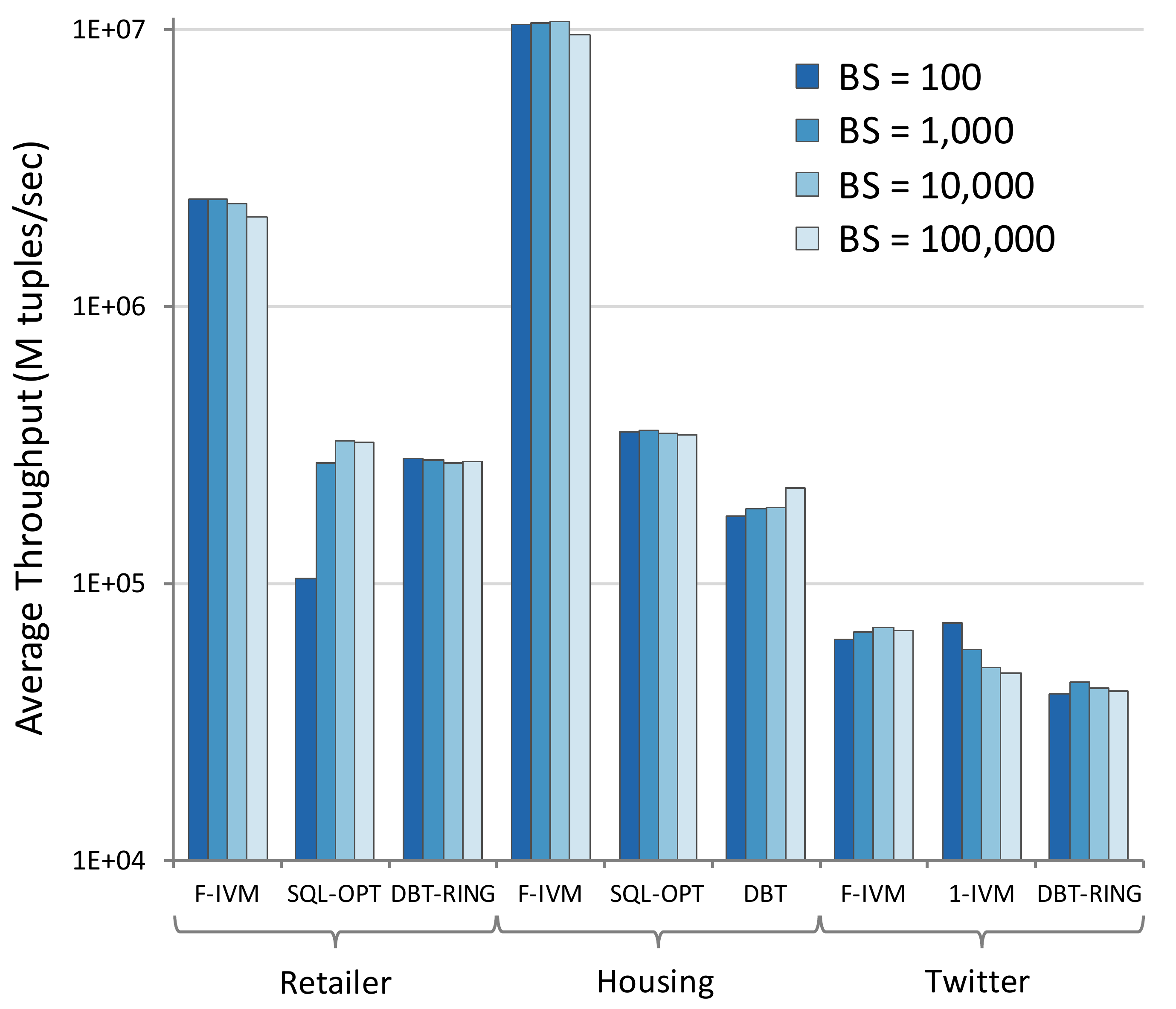}
  \caption{Incremental maintenance of the covariance matrix under batch updates of different sizes to all input relations. }
  \label{fig:cofactor_IVM_batch_sizes_ALL}
\end{minipage}%
\end{figure*}

{\bf Covariance Matrix Computation over the Triangle Query.}
We analyze the covariance matrix computation over the triangle query on the {\em Twitter} dataset and updates of size $1,000$ to all the relations.
In addition to the three incremental strategies from before, we now also benchmark 
\DFONE, which is \DF but under updates to relation $R$ only ($S$ and $T$ are non-updatable), 
\DBTRING, which is DBToaster's recursive IVM strategy with payloads from continuous covariance ring of degree $m$ (cf.~Section~\ref{sec:application-lr}) instead of scalars, 
and \SQLOPT, an optimized SQL encoding of covariance matrix computation. 
The \SQLOPT strategy arranges regression aggregates -- recall there are quadratically many such aggregates in the number of query variables -- into a {\em single} aggregate column indexed by the degree of each query variable. 
This encoding takes as input a variable order and constructs one SQL query that intertwines join and aggregate computation by pushing (partial) regression aggregates (counts, sums, and covariance matrices) past joins~\cite{Olteanu:FactorizedDB:2016:SIGREC}.

\DF uses the view tree from Figure~\ref{fig:triangle_hypergraph_viewtree}~(right) without the indicator projection and materializes the join of $S$ and $T$ of size $\bigO{N^2}$. 
Its time complexity for a single-tuple update to $R$ is $\bigO{1}$, but updating the join of $S$ and $T$ takes $\bigO{N}$. 
\DFONE uses the same view tree. For updates to $R$ only, \DFONE requires one lookup in the materialized join of the two non-updatable relations $S$ and $T$ per update, which takes $\bigO{1}$ time.
\DBTRING uses payloads from the continuous covariance ring of degree $3$ and materializes all three such pairwise joins, each requiring linear time maintenance.
\DBT uses scalar payloads and materializes $21$ views (to maintain $6$ aggregates), out of which $12$ views are over two relations. Its time complexity for processing single-tuple updates to either of the three relations is also $\bigO{N}$.
\IVM maintains just the input relations and recomputes the delta upon each update in linear time.

Figure~\ref{fig:cofactor_Triangle_IVM_trace_ALL} shows the throughputs of the three strategies on the {\em Twitter} dataset.
This experiment result is from the conference version of this paper~\cite{FIVM:SIGMOD:2018}. 
The throughput rate of the strategies that materialize views of quadratic size declines sharply as the input stream progresses. \DBT exhibits the highest processing and memory overheads caused by storing $12$ auxiliary views of quadratic size. \DBTRING underperforms \DF due to maintaining two extra views of quadratic size, which contribute to a $2.3$x higher peak memory utilization. \IVM exhibits a $42$\% decline in performance after processing the entire trace due to its linear time maintenance. The extent of this decrease is much lower compared to the other approaches with the quadratic space complexity. 
\DFONE has two orders of magnitude higher throughput than \IVM at the cost of using $23$x more memory.

Clique queries like triangles provide no factorization opportunities. Materializing auxiliary views to speed up incremental view maintenance increases memory and processing overheads. However, \DF can exploit indicator projections to bound the size of such materialized views, as described in Section~\ref{sec:cyclic_queries}.


{\bf The Effect of Batch Size on IVM.}
This experiment evaluates the performance of maintaining a covariance matrix for batch updates of different sizes. Figure~\ref{fig:cofactor_IVM_batch_sizes_ALL} shows the throughput of batched incremental processing for batch sizes varying from $100$ to $100,000$ on the {\em Retailer}, {\em Housing}, and {\em Twitter} datasets for updates to all relations.
We show only the best three approaches for each dataset.
This experiment result is from the conference version of this paper~\cite{FIVM:SIGMOD:2018}. 

We observe that using very large or small batch sizes can have negative performance effects: Iterating over large batches invalidates previously cached data resulting in future cache misses, whereas using small batches cannot offset the overhead associated with processing each batch. 
Using batches with $1,000-10,000$ tuples delivers best performance in most cases, except when needed to incrementally maintain a large number of views. This conclusion about covariance matrix computation is in line with similar findings on batched delta processing in decision support workloads~\cite{Nikolic:Batching:2016:SIGMOD}.

Batched incremental processing is also beneficial for one-off computation of the entire covariance matrix. Using medium-sized updates can bring better performance, cf.\@ Figure~\ref{fig:cofactor_IVM_batch_sizes_ALL}, but can also lower memory requirements and improve cache locality during query processing. 
For instance, incrementally processing the {\em Retailer} dataset in chunks of $1,000$ tuples can bring up to $2.45$x better performance compared to processing the entire dataset at once.

}

\subsection{Mutual Information and Chow–Liu Trees}
\label{sec:MI}

\begin{figure*}[t]
  \centering   
  \includegraphics[width=0.32\textwidth]{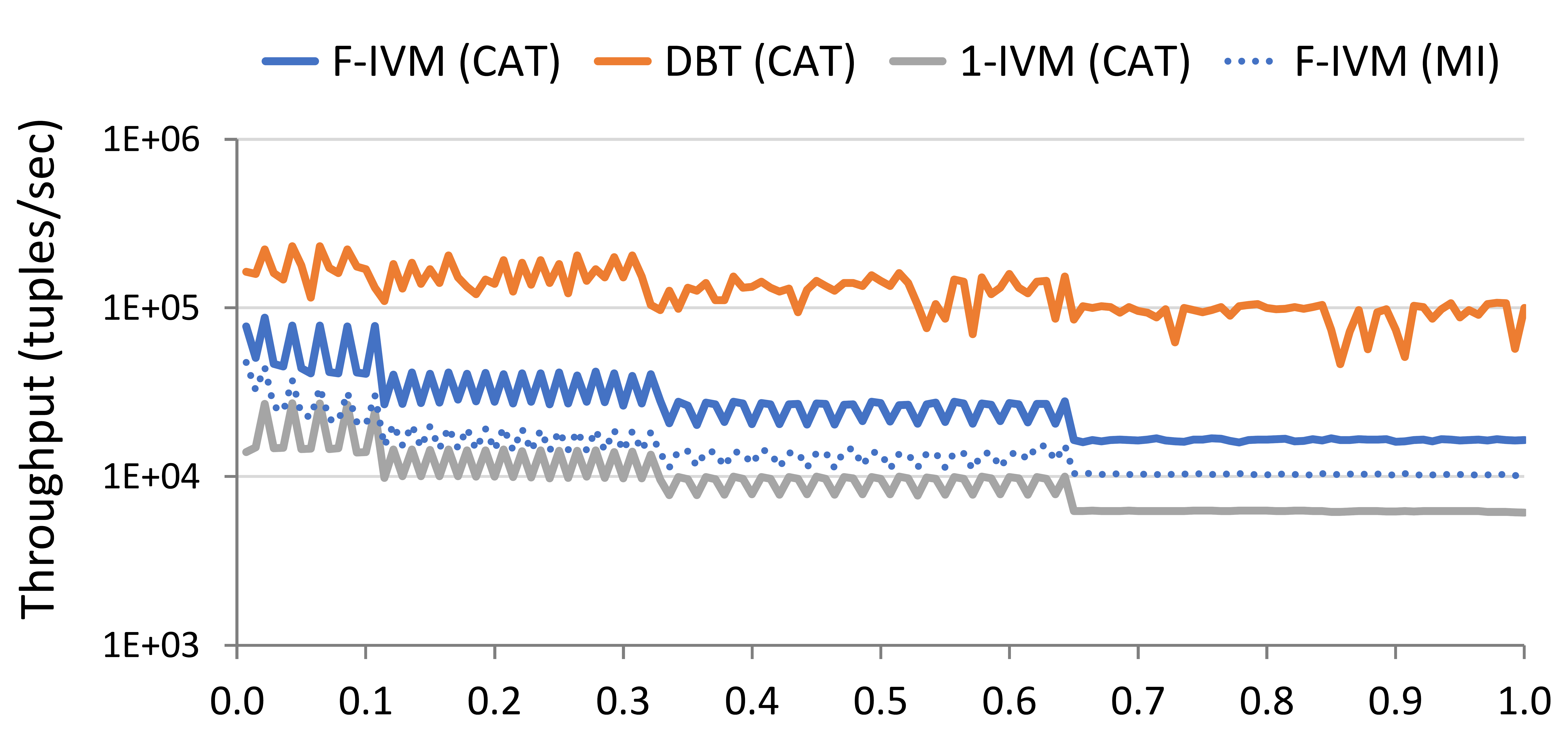}
  \;
  \includegraphics[width=0.32\textwidth]{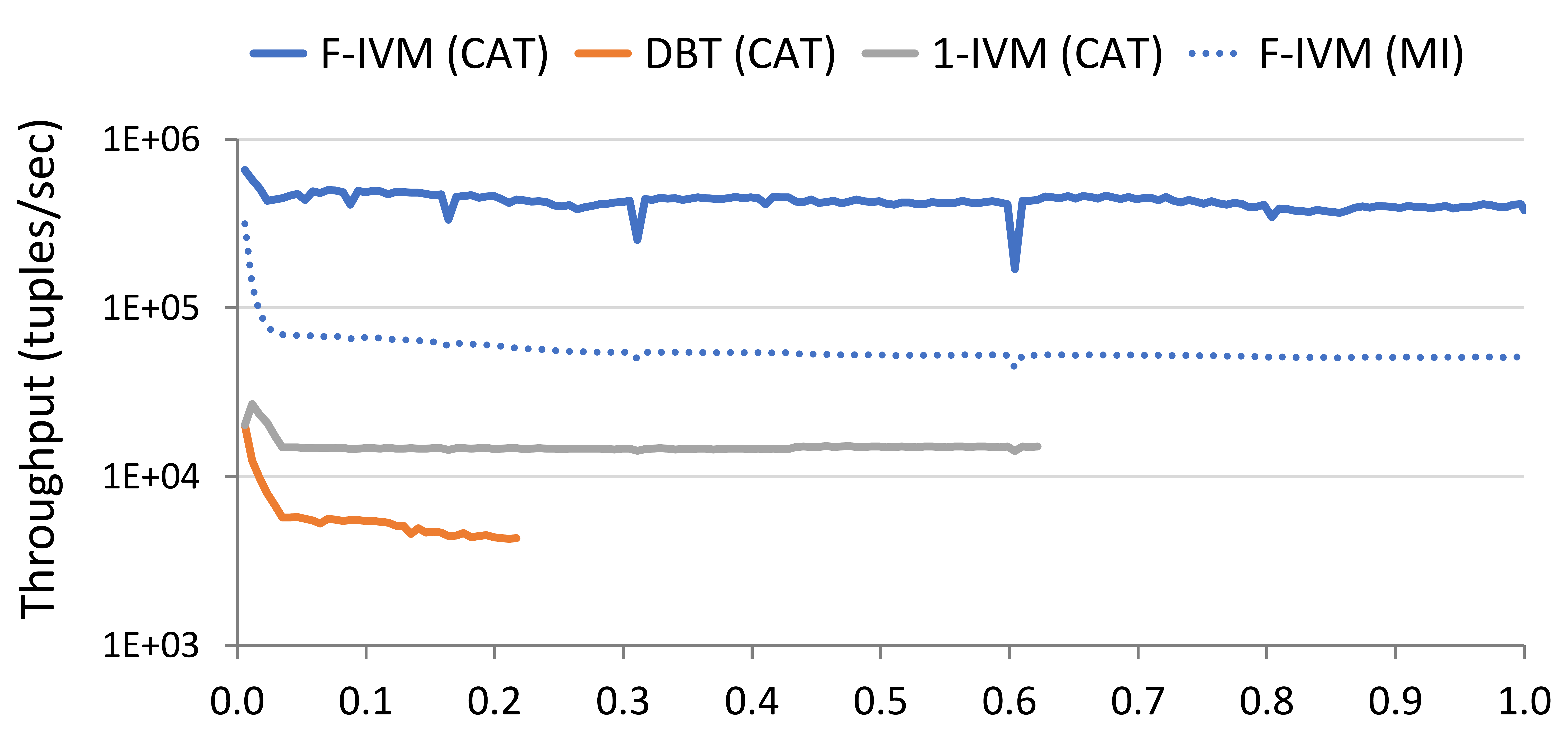}
  \;
  \includegraphics[width=0.32\textwidth]{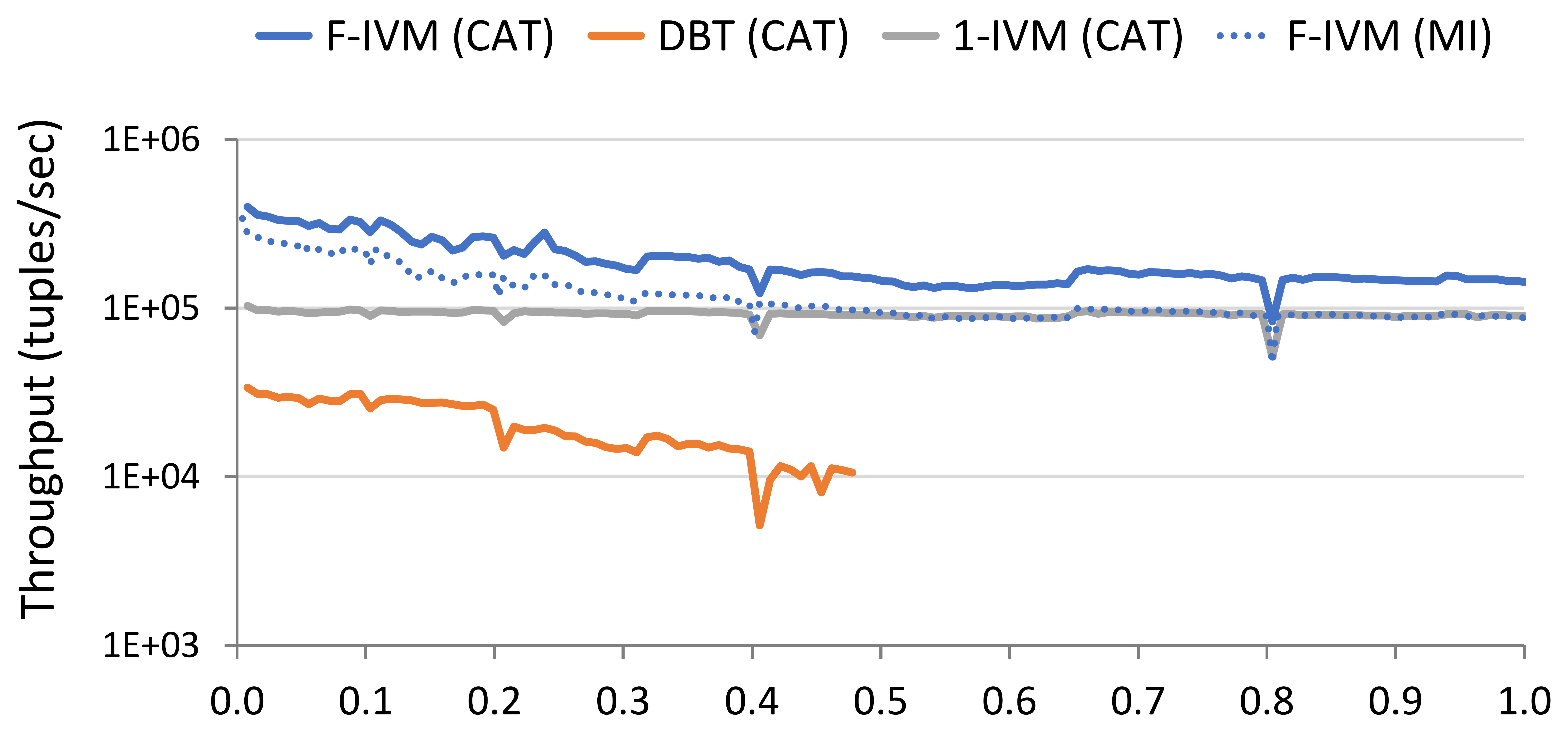}
  \caption{Solid lines: incremental maintenance of the covariance matrix over the {\em Housing} (left), {\em Retailer} (middle) and {\em Favorita} (right) datasets under batches of 1,000 updates to all relations with a one-hour timeout. All features are either categorical in the original dataset or made categorical by discretizing their domains into 100 buckets. 
  Dotted line: computation of the mutual information matrix and the Chow–Liu tree on top of the covariance matrix after each batch of 1,000 updates.}
  \label{fig:MI-all}
\end{figure*}

We benchmark the performance of maintaining the matrix of pairwise mutual information (MI) for the features representing  the non-join variables in our datasets and Chow-Liu trees on top of the MI matrices. 

As explained in Section~\ref{sec:mutual-information}, the MI matrix can be derived from the covariance matrix over categorical variables. We discretize the active domain of each continuous variable into 100 bins of equal size.
The {\em Housing}, {\em Retailer}, and {\em Favorita} datasets have $26$, $39$, and $15$ categorical variables, respectively. Insertions and deletions of values for a continuous variable are distributed into the appropriate bins, without changing the number of bins. Whereas the covariance matrix can be maintained incrementally under updates, the MI matrix needs to be recomputed from scratch after each update batch.

The view construction and maintenance are as in Section~\ref{sec:covariance-regression}, except that all variables are now categorical.
Figure~\ref{fig:MI-all} (solid lines) shows the throughput of \DF, \DBT, and \IVM for maintaining the covariance matrix as they process an increasing fraction of the stream of tuple updates. 
\DF is $74$x faster than \DBT and $28$x faster than \IVM for the {\em Retailer} dataset and $9.3$x and $2.1$x faster, respectively, for {\em Favorita}.
For {\em Housing}, \DF is $2.8$x faster than \IVM but $4.6$x slower than \DBT. This is because: (i) {\em Housing} is a relatively small dataset and the domain of the categorical variables is also small; (ii) \DBT has specific optimizations for group-by count over star joins such as in this case.

Computing the MI matrix from the covariance matrix takes time linear in the number of categories of the variables. Computing the Chow-Liu tree takes time $\bigO{m \log m}$, where $m$ is the number of variables.
Figure~\ref{fig:MI-all} (dotted line) shows the throughput of \DF when the MI matrix and Chow–Liu tree are computed after each update batch. This throughput is $46\%$, $86\%$, and $35\%$ smaller than the time to maintain the covariance matrix for {\em Housing}, {\em Retailer}, and {\em Favorita}, respectively.

\subsection{$Q$-Hierarchical Queries}
\label{sec:q-hier-experiment}

\begin{figure*}[t]
  \centering   
  \includegraphics[width=0.32\textwidth]{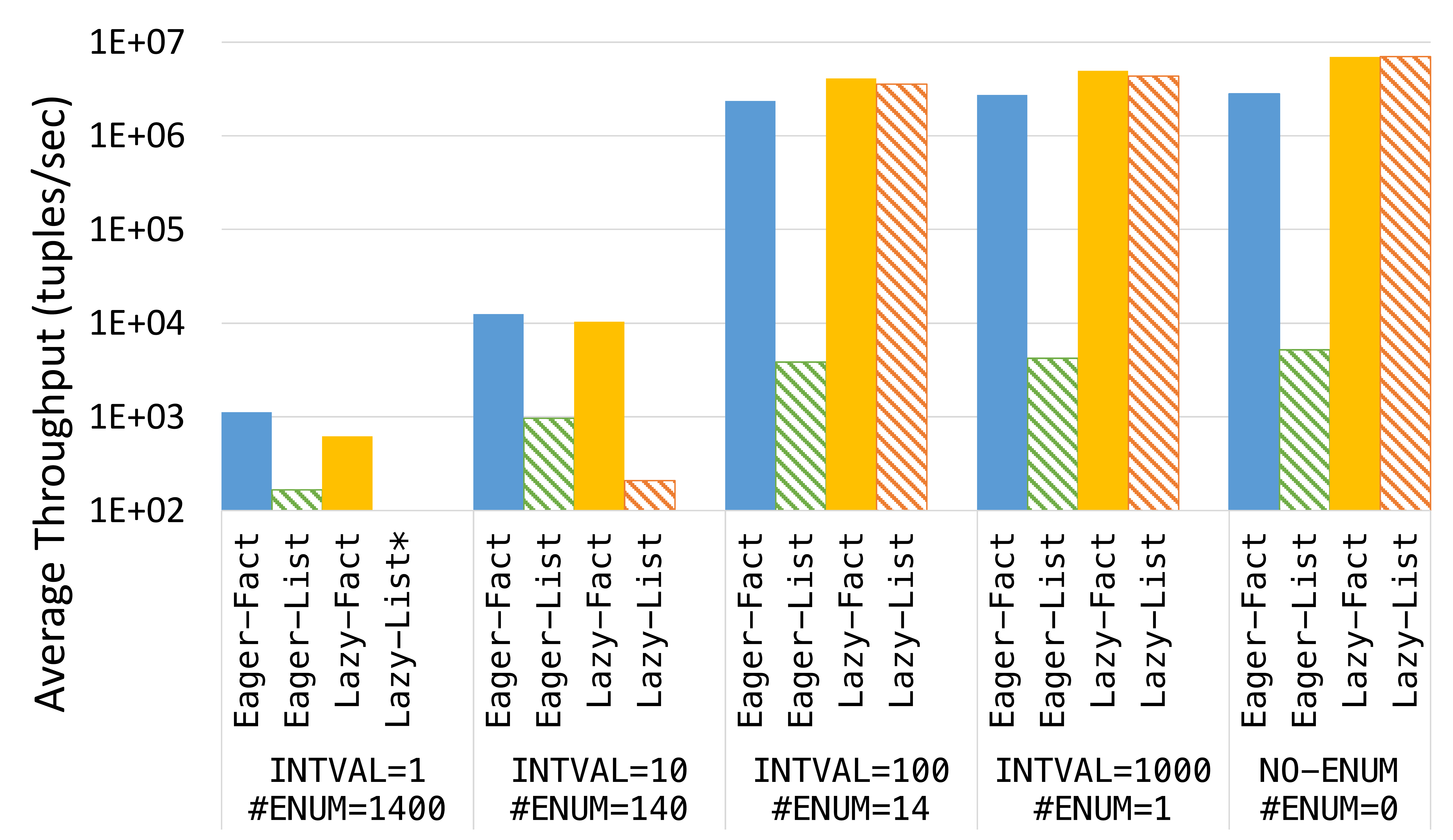}
  \;
  \includegraphics[width=0.32\textwidth]{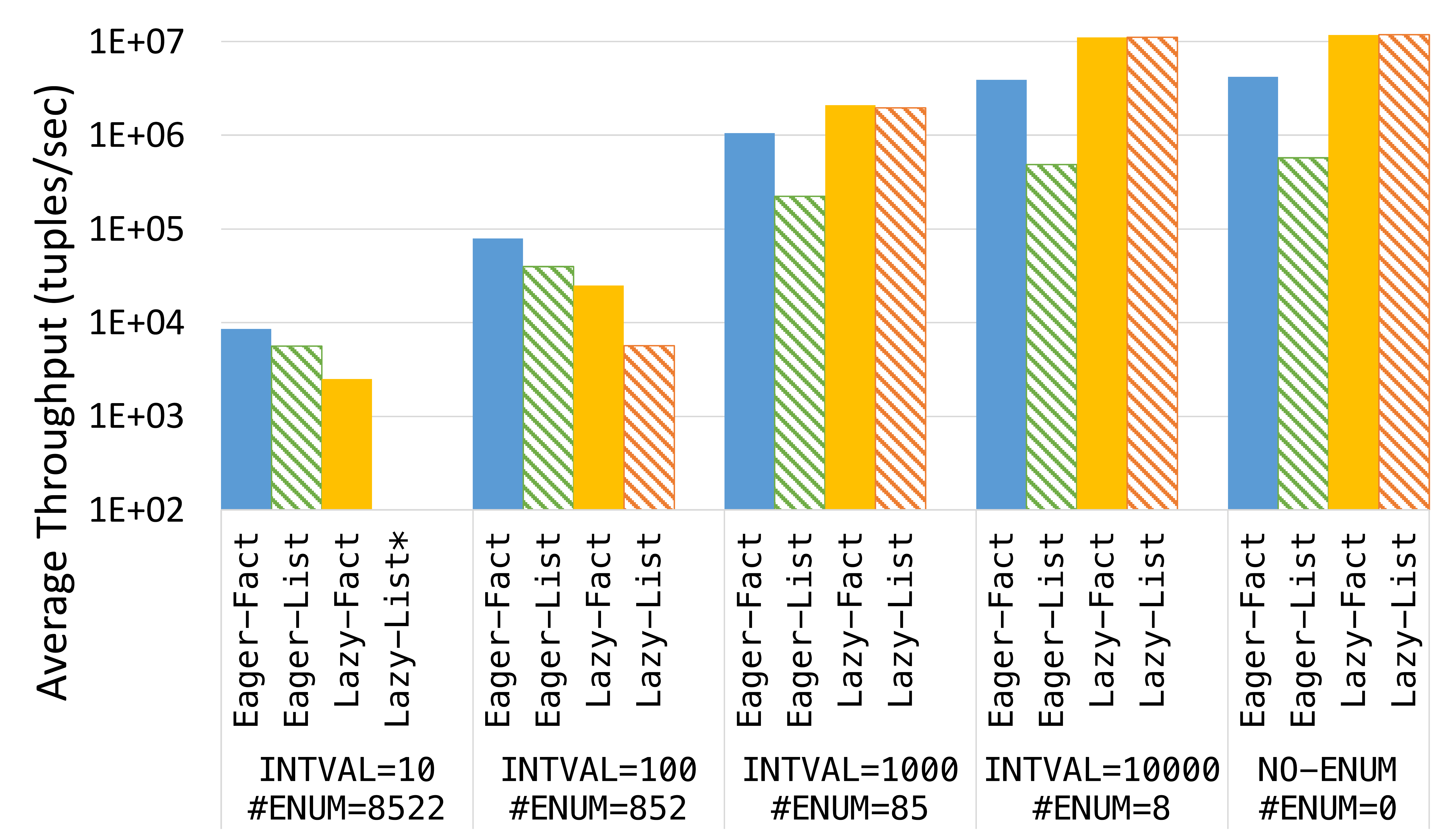}
  \;
  \includegraphics[width=0.32\textwidth]{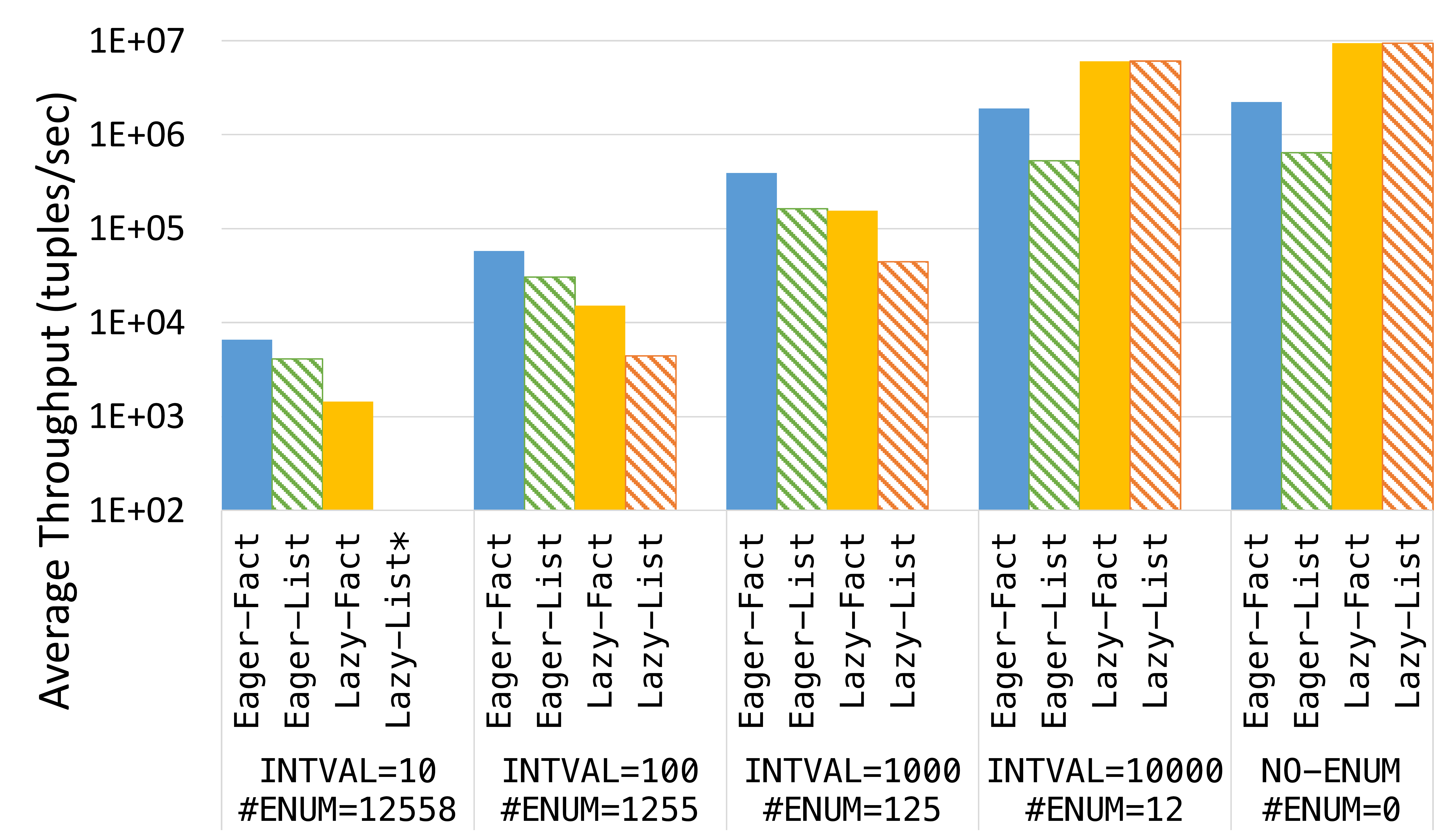}
  \caption{Incremental maintenance of the result of the $q$-hierarchical queries over the {\em Housing} (left),  
  {\em Retailer} (middle), and {\em Favorita} (right) datasets under update batches and requests to enumerate all tuples in the query results after every \texttt{INTVL} update batches;
  \texttt{\#ENUM} denotes the overall number of the enumeration requests. 
  The symbol $*$ denotes the case where an IVM variant did not finish within the time limit (50 hours) for this experiment. The throughput is not shown in this case.
  }
  \label{fig:q-hier-all}
\end{figure*}

The $q$-hierarchical queries are those queries that admit the lowest (i.e., constant) enumeration delay and single-tuple update time (Section~\ref{sec:query_classes}). We would like to understand how different IVM variants perform for such queries in practice. We consider one such query per dataset, as described in Section~\ref{sec:experiments-settings}. 

We construct a view tree modelled on the canonical free-top variable order for each of the three $q$-hierarchi\-cal queries. The query result is constructed and maintained in the payload space. We consider two dimensions. 
One dimension is whether we push the updates all the way to the result (eager) or we only update the input relations and only construct the query result on an enumeration request (lazy).
The other dimension is whether the query result has a listing representation (one tuple after the other) or a factorized representation. 
This defines four variants: eager-list (which is DBT), eager-fact (\DF's default strategy), lazy-list (1-IVM), and lazy-fact (a hybrid of \DF and 1-IVM).

Figure~\ref{fig:q-hier-all} shows the average throughput of the four variants on the three $q$-hierarchical queries. 
We report the overall runtime in case of update batches as in the previous experiments but where in addition we have requests for the enumeration of all tuples in the query result after every $\texttt{INTVAL}$ batches of updates.
We tried $\texttt{INTVAL}$ values $1, 10, 100, 1000,$ and $10000$. Each such value corresponds to different numbers of enumeration requests (\texttt{\#ENUM}) as the datasets have different sizes.
The lazy-list variant did not finish within the time limit of 50 hours (denoted by $*$ in Figure~\ref{fig:q-hier-all}). The lazy-list variant has the lowest throughput among the four variants in our experiment.

The two lazy variants are clear winners in case of none or very few enumeration requests. In this case, there is almost no difference between their throughputs since they spend most of their time updating the input relations. In case of more enumeration requests, however, the eager variants are the winners, with eager-fact consistently outperforming eager-list. 

Overall, the eager and lazy variants based on factorized representation outperform those ba\-sed on listing representation in all but the trivial cases of none or few enumeration requests, where the representation of the query result plays no role. This is as expected, since the enumeration delay and the update time can both remain constant for our queries only if the query result is kept factorized over the views in the view tree.

\begin{figure*}[t]
  \centering   
  \includegraphics[width=0.49\textwidth]{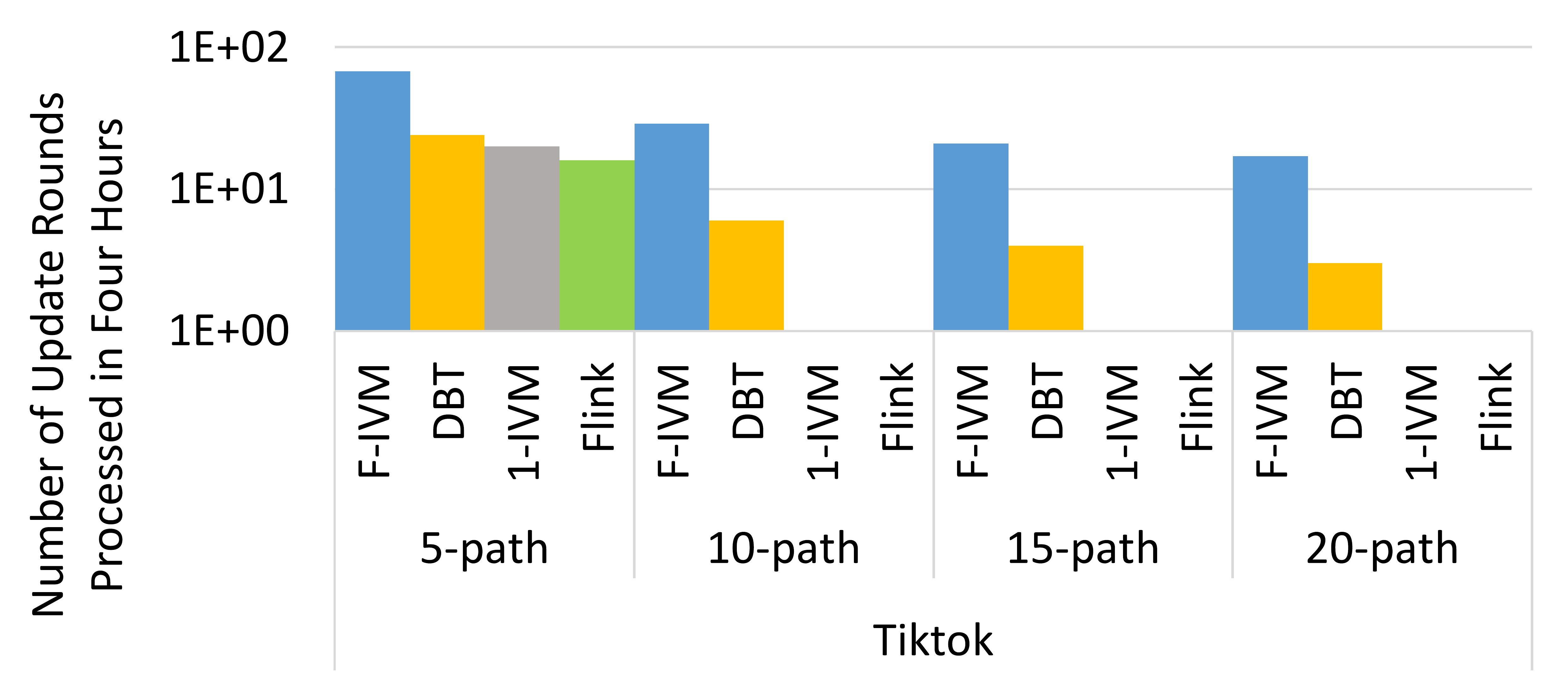}
  \;
  \includegraphics[width=0.49\textwidth]{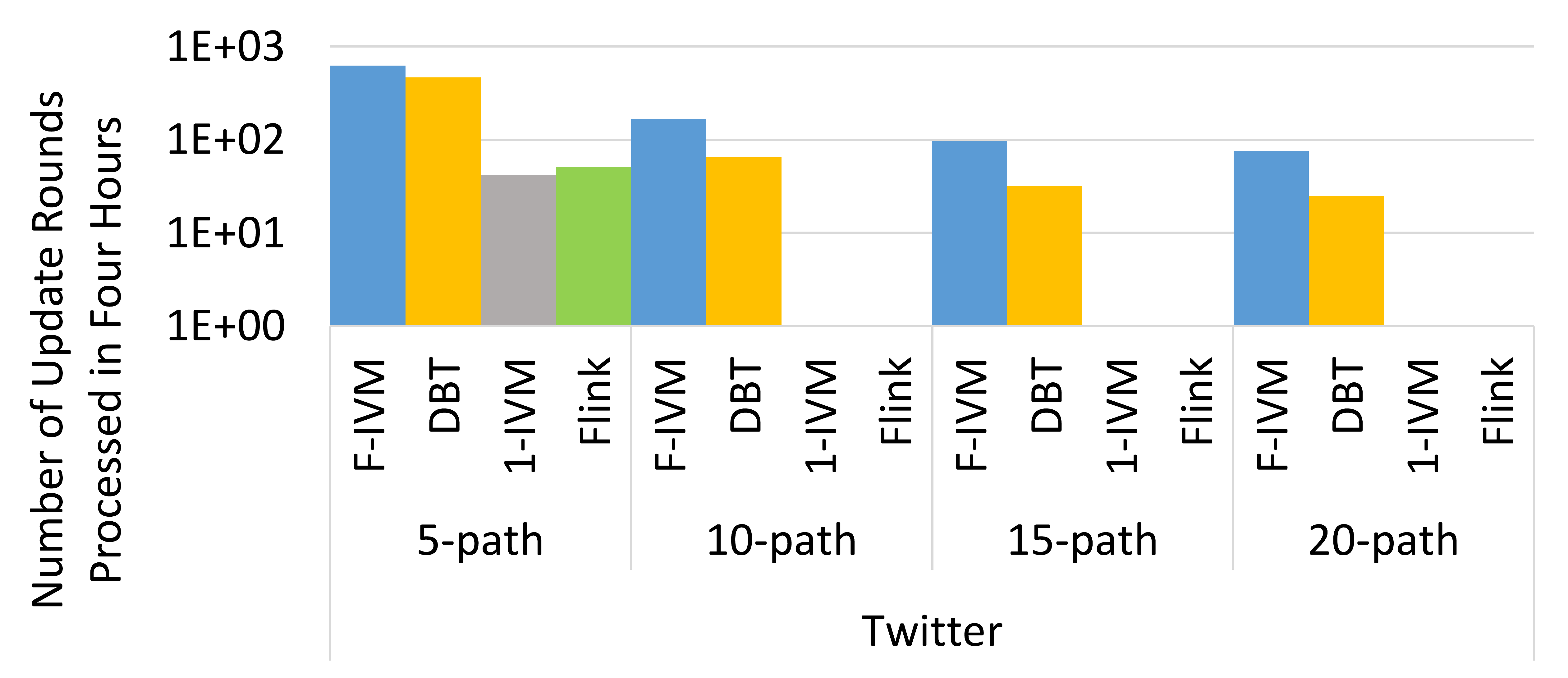}
  \caption{The number of rounds of updates to all relations in the path queries processed within four hours over the {\em TikTok} (left) and {\em Twitter} (right) graph datasets under update batches of size $1,000$. }
  \label{fig:path-all}
\end{figure*}

\subsection{Path Queries}
\label{sec:path-queries}

We investigate the scalability of the maintenance approaches as we  increase the number of joins in the query. We consider the path query with up to 20 self-joins of the edge relation in the {\em TikTok} and {\em Twitter} graphs. The edges are partitioned into batches of 1000 inserts. One round of updates processes one batch of inserts for each copy of the edge relation.
Figure~\ref{fig:path-all} shows the number of rounds of updates processed by each approach within four hours.
\DF outperforms all other approaches on path queries of any length.
All approaches are slower for {\em TikTok}, since it is more skewed than {\em Twitter}.

Flink and \IVM have a similar poor performance and do not scale for long path queries.
Flink maintains the join result via a left-deep binary view tree and computes the aggregates at the root view.
It projects away the join variable after each join. This reduces the number of columns but not the number of rows in the join result.
For a delta to the bottom relation in the view tree, Flink joins it with all other $k-1$ relations in the query. This triggers $\bigO{N^{\lceil\frac{k}{2}\rceil}}$ inserts to the join result, where $N$ is the number of edges.
\IVM computes the delta query by joining the batch of inserts with $k-1$ relations and has the same complexity as Flink.

\DBT and \DF avoid the materialization of the large join result by pushing the aggregates past the joins at each view. Both of them need $\bigO{N^2}$ time to update each view. Like Flink, \DF uses a left-deep view tree. \DBT uses one view tree per delta query, where the delta relation is a child of the top view and the two subqueries to the left and right of the delta relation have left-deep view trees.
\DF constructs fewer views than \DBT: For 20-path, \DF uses 19 views, while \DBT uses 190 views. This explains the better performance of \DF.

\add{

\subsection{Maintenance of Sum Aggregates}
\label{sec:sum_aggregate_maintenance}

We analyze different strategies for maintaining a sum of one variable on top of a natural join. We measure the average throughput of reevaluation and incremental maintenance under updates of size $1,000$ to all the relations of {\em Retailer} and {\em Housing}. For the former dataset, we sum the inventory units for products in {\tt Inventory}; for the latter, we sum over the common join variable. We also benchmark two reevaluation strategies that recompute the results from scratch on every update: \DFRE denotes reevaluation using variable orders and \DBTRE denotes reevaluation using DBToaster. Table~\ref{table:sum_aggregate_cost_comparison} summarizes the results.

\DF achieves the highest average throughput in both cases. For {\em Retailer}, the maintenance cost is dominated by the update on {\tt Inventory}. 
\DBT's recursive delta compilation materializes $13$ views representing connected subqueries: five group-by aggregates over the input relations, {\tt Inv}, {\tt It}, {\tt W}, {\tt L}, and {\tt C}; one group-by aggregate joining {\tt L} and {\tt C}; six views joining {\tt Inv} with subsets of the others, namely \{{\tt It}\}, \{{\tt It}, {\tt W}\}, \{{\tt It}, {\tt W}, {\tt L}\}, \{{\tt W}\}, \{{\tt W}, {\tt L}\}, and \{{\tt W}, {\tt L}, {\tt C}\}; and the final aggregate.
The two views joining {\tt Inv} with \{ {\tt W}, {\tt L} \} and \{ {\tt It}, {\tt W}, {\tt L} \} require linear maintenance for a single-tuple change in {\tt Inventory}.
\IVM recomputes deltas from scratch on each update using only the input relations with no aggregates on top of them. Updates to {\tt Inventory} are efficient due to small sizes of the other relations. 
\DF uses the given variable order to materialize $9$ views, four of them over {\tt Inventory}, \{{\tt Inv}\}, \{{\tt Inv}, {\tt It}\}, \{ {\tt Inv}, {\tt It}, {\tt W} \}, and the final sum, but each with constant maintenance for single-tuple updates to this relation.
In contrast to \IVM, our approach materializes precomputed views in which all nonjoin variables are aggregated away. 
In the {\em Housing} schema, both \DF and \DBT benefit from this preaggregation, and since the query is a star join, both materialize the same views. \DBT computes {\tt SUM(1)} and {\tt SUM(postcode)} for each {\tt postcode} in the delta for {\tt Inventory}, although only the count suffices.
Figure~\ref{table:sum_aggregate_cost_comparison} also shows that the reevaluation strategies significantly underperform the incremental approaches.

\begin{figure}[t]
\begin{center}
{
\renewcommand{\arraystretch}{1.3}
\begin{small}
\begin{tabular}{@{}l@{~}c@{~}c@{~}c@{~}c@{~}c@{}}
\toprule
  & \DF &  \DBT & \IVM & \DFRE & \DBTRE \\
\midrule
Retailer \quad & $2,955,045$ &  $1,250,262$ & $2,925,828$ & $3,785^{*}$ & $3,491^{*}$ \\
Housing  \quad & $22,857,143$ & $17,834,395$ & $2,403,433$ & $79,226$ & $364^{*}$ \\
\bottomrule
\end{tabular}
\end{small}
}
\end{center}
\caption{The average throughput (tuples/sec) of reevaluation and incremental maintenance of a sum aggregate under updates of size $1,000$ to all relations of the {\em Retailer} and {\em Housing} datasets with a one-hour timeout (denoted by the symbol$^{*}$).}
\label{table:sum_aggregate_cost_comparison}
\end{figure}


\subsection{Matrix Chain Multiplication}
\label{sec:matrix_chain_multiplication}

\begin{figure*}[t]
  \centering   
  \includegraphics[width=0.47\textwidth]{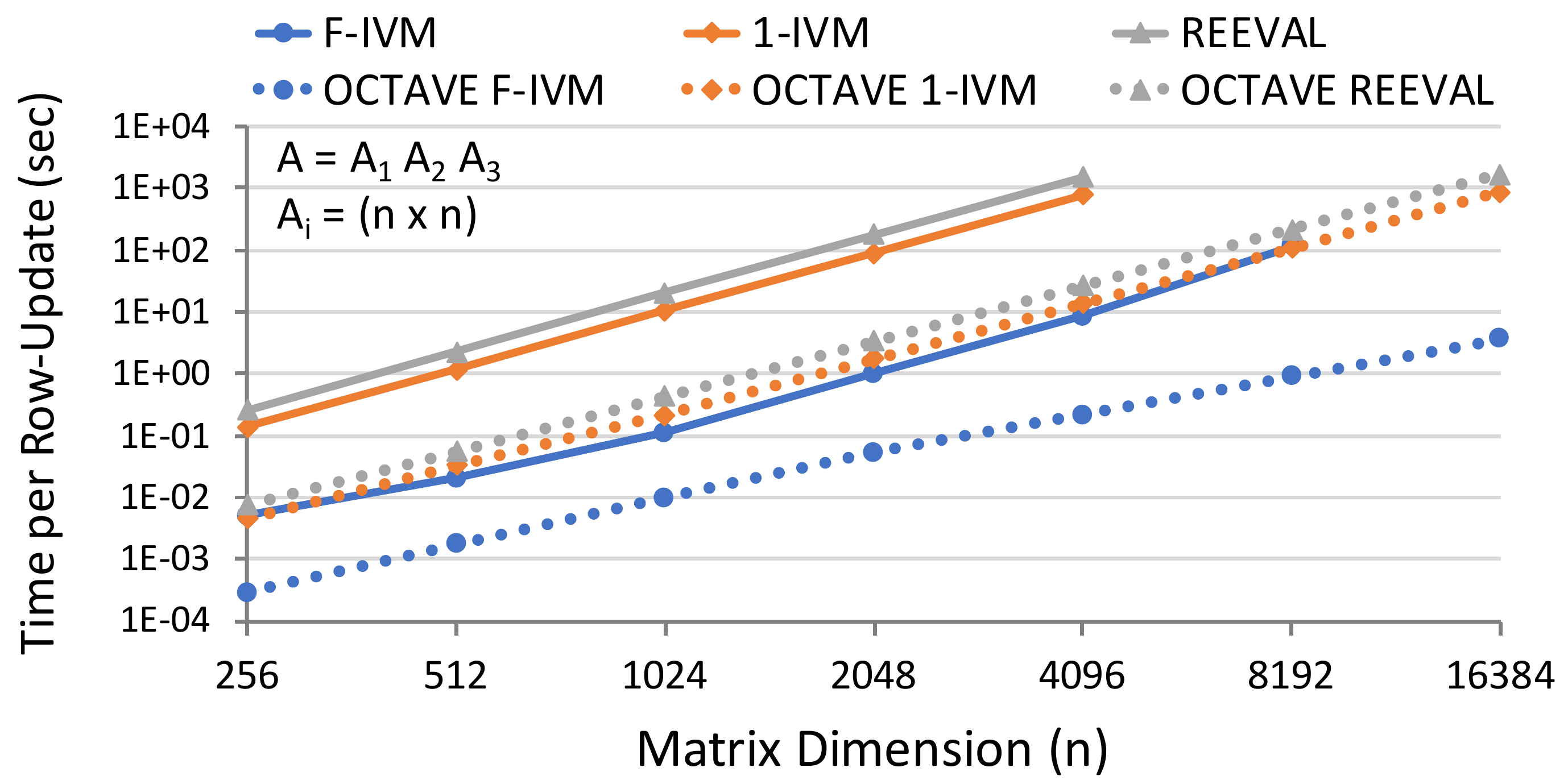}
  \qquad
  \includegraphics[width=0.47\textwidth]{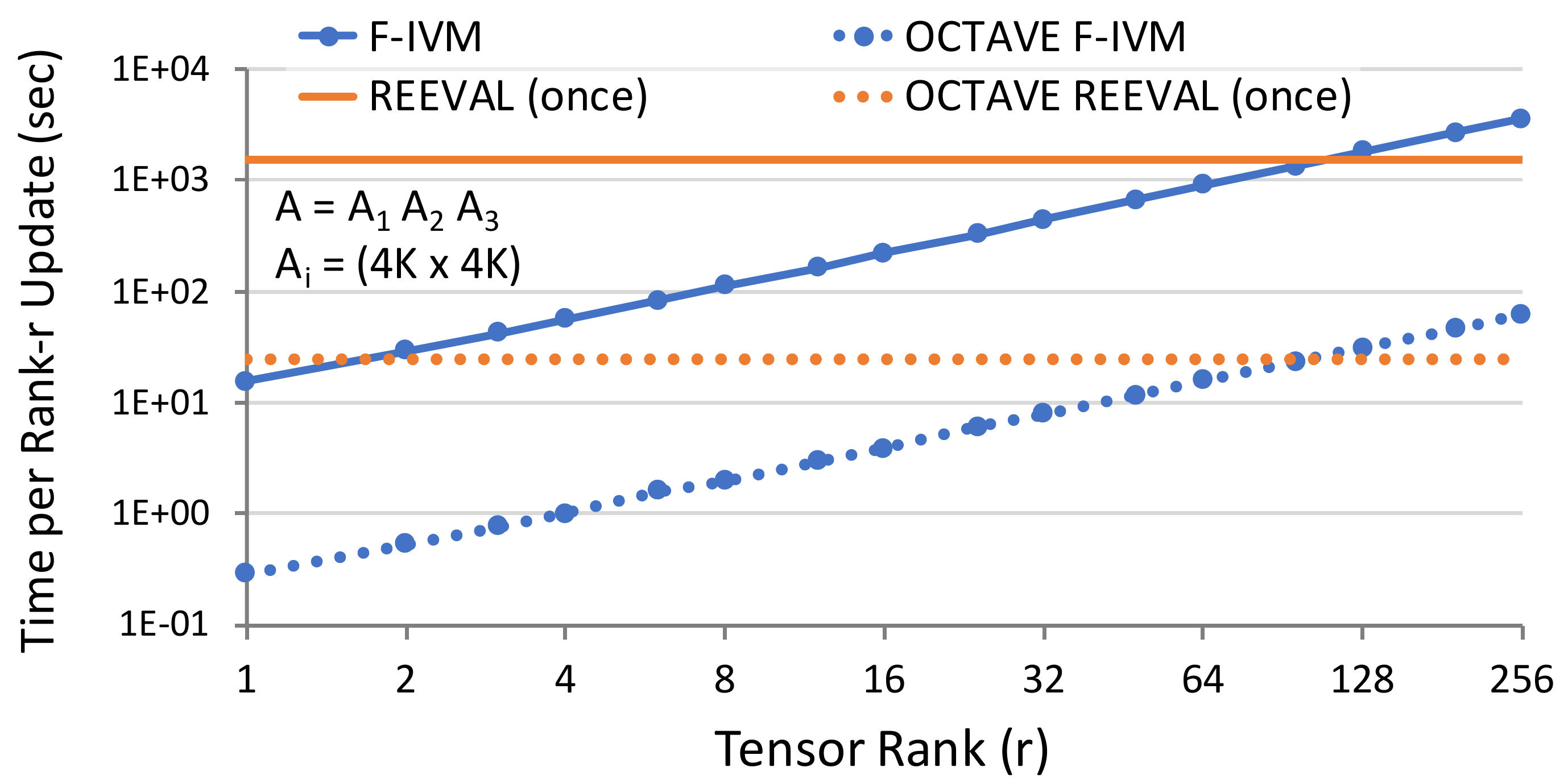}
  \caption{Incremental maintenance and reevaluation of the product of three $(n \times n)$ matrices, $A = A_1 \, A_2 \, A_3$: (left) one-row updates in $A_2$; (right) rank-$r$ updates in $A_2$ for $n=4,096$ using the DBToaster and Octave runtime environments. }
  \label{fig:MCM}
\end{figure*}

We consider the problem of maintaining the multiplication $A = A_1 \, A_2 \, A_3$ of three $(n \times n)$ matrices under changes to $A_2$. We compare \DF with factorized updates, \IVM that recomputes the delta $\delta{A} = A_1 \, \delta{A_2} 
\, A_3$ from scratch, and REEVAL that recomputes the entire product from scratch on every update. 
\DBT becomes \IVM in this particular setting.
We consider two different implementations of these maintenance strategies: The first uses DBToaster's hash maps to store matrices, while the second uses Octave, a numerical tool that stores matrices in dense arrays and offers highly-optimized BLAS routines for matrix multiplication~\cite{Whaley1999}. In both cases, matrix-matrix multiplication takes $\bigO{n^{\alpha}}$ for $\alpha > 2$; for instance, $\alpha = 2.8074$ for 
Strassen's algorithm.

We first consider updates to one row in $A_2$. For \IVM, the delta $\delta{A_{12}} = A_1 \, \delta{A_2}$ might  contain non-zero changes to all $n^2$ matrix entries, thus computing $\delta{A} = \delta{A_{12}} \, A_3$ requires full matrix-matrix multiplication. REEVAL updates $A_2$ first before computing two matrix-matrix multiplications. \DF factorizes $\delta{A_2}$ into a product of two vectors $\delta{A_2} = u \TR{v}$, which are used to compute $\delta{A_{12}} = (A_1 \, u) \, \TR{v} = u_1 \, \TR{v}$ and $\delta{A} = u_1 \, (\TR{v} \, A_3) = u_1 \, v_1$. Both deltas involve only matrix-vector multiplications computed in $\bigO{n^2}$ time. 
Figure~\ref{fig:MCM} (left) shows the average time needed to process an update to one randomly selected row in $A_2$ for different matrix sizes. REEVAL performs two matrix-matrix multiplications, while \IVM performs only one. In the hash-based implementation, the gap between \DF and \IVM grows from $28$x for $n=256$ to $92$x for $n=4,096$; similarly, in the Octave implementation, the same gap grows from $16$x for $n=256$ to $236$x for $n=16,384$. This confirms the difference in the asymptotic complexity of these strategies.

Our next experiment considers rank-$r$ updates to $A_2$, which can be decomposed into a sum of $r$ rank-$1$ tensors, $\delta{A_2} = \sum_{i\in[r]} u_i \TR{v_i}$. \DF processes $\delta{A_2}$ as a sequence of $r$ rank-$1$ updates in $\bigO{rn^2}$ time, while both REEVAL and \IVM take as input one full matrix $\delta{A_2}$ and maintain the product in $\bigO{n^3}$ time per each rank-$r$ update. \IVM has the same performance as REEVAL. Figure~\ref{fig:MCM} (right) shows that the average time \DF takes to process a rank-$r$ update for different $r$ values and the matrix size $4,096$ is linear in the tensor rank $r$. 
Under both implementations in DBToaster and Octave, incremental computation is faster than reevaluation for updates with rank $r\leq 96$.
With larger matrix sizes, the gap between reevaluation and incremental computation increases, which enables incremental maintenance for updates of higher ranks.

\begin{figure*}[t]
  \centering   
  \includegraphics[width=0.48\textwidth]{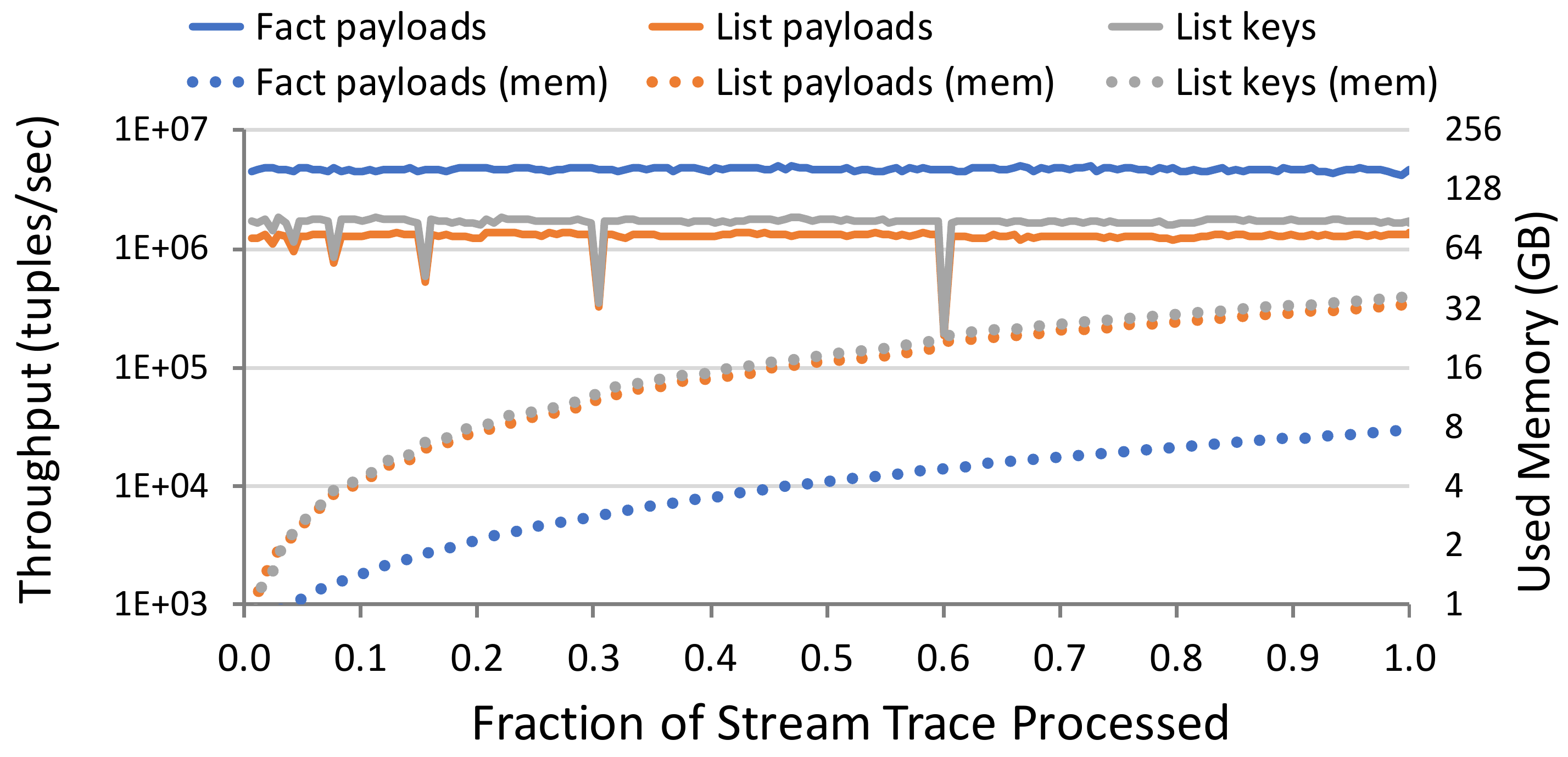}
  \quad
  \includegraphics[width=0.48\textwidth]{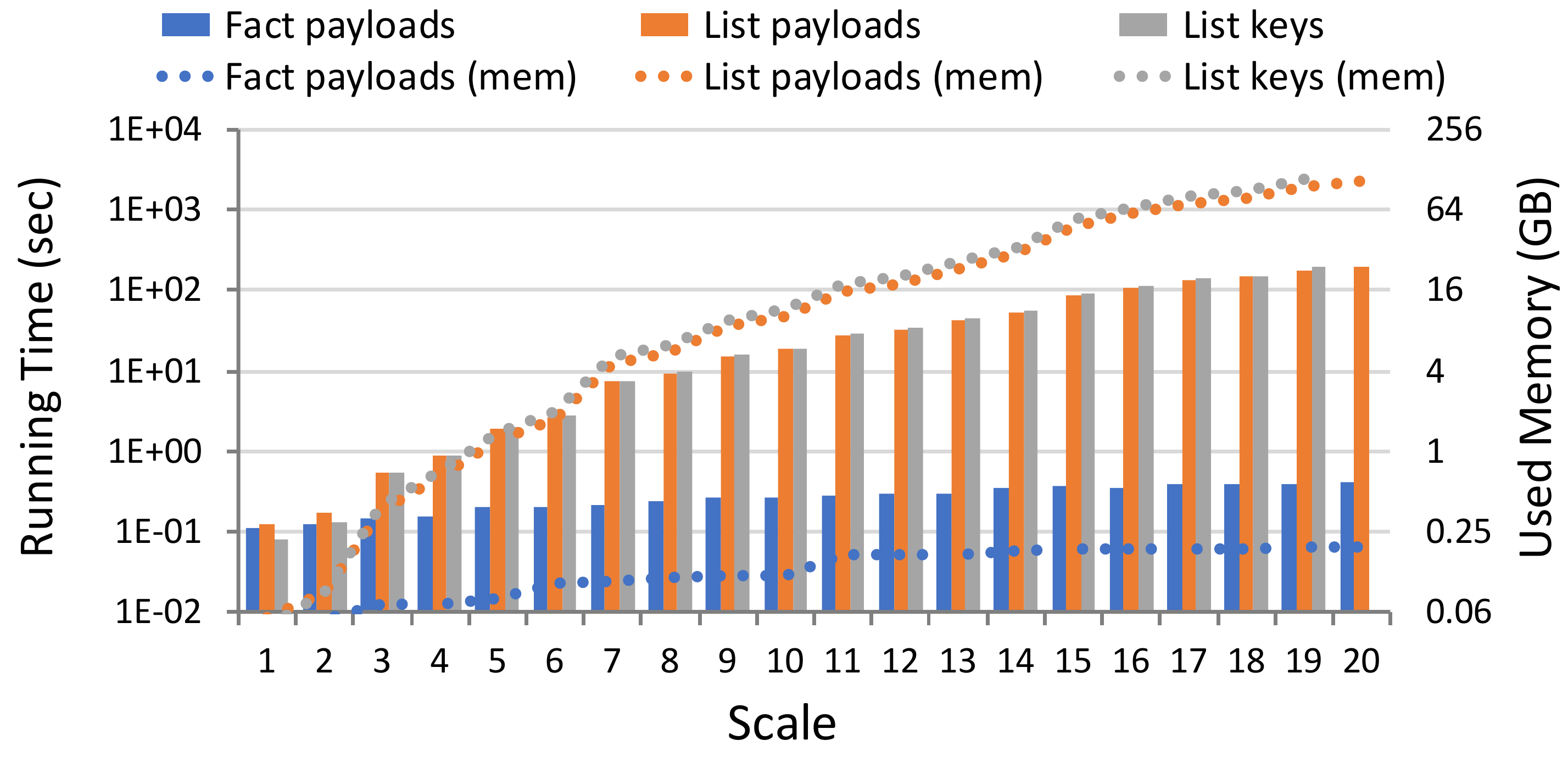}
  \caption{Incremental maintenance using relational and factorized payloads for the natural joins of the {\em Retailer}  (left) and of the {\em Housing} (right) datasets under updates of size $1,000$ to the largest relation ({\em Retailer}) and all input relations ({\em Housing}).}
  \label{fig:FullJoin_Factorized_Relational}
\end{figure*}

\subsection{Factorized Computation of Conjunctive Queries}
\label{sec:factorized_conjunctive}

We analyze \DF on queries whose results are stored as keys with integer multiplicities using listing representation ({\tt List$\;$keys}) and as relational payloads using factorized and listing representations ({\tt Fact$\;$payloads} and {\tt List$\;$payloads}).
Figure~\ref{fig:FullJoin_Factorized_Relational} (left) considers the natural join of {\em Retailer} under updates to the largest relation. The factorized payloads reduce the memory consumption by $4.4$x, from $34$GB to $7.8$GB, improve the average throughput by $2.8$x and $3.7$x (and the overall run time by $3.2$x and $4.2$x) compared to using the two listing encodings.
Figure~\ref{fig:FullJoin_Factorized_Relational} (right) considers the natural join of {\em Housing} under updates to all input relations.  The number of tuples in the dataset varies from $150,000$ (scale 1) to $1,400,000$ (scale 20), while the size of the listing (factorized) representation of natural join grows cubically (linearly) with the scale factor. The two listing encodings blow up the memory consumption and computation time for large scales. Storing tuples in the listing representation using payloads instead of keys avoids the need for hashing wide keys, which makes the joins slightly cheaper. For {\em Housing} and factorized representation, the root view stores $25,000$ values of the join variable regardless of the scale. The root's children map these values to relational payloads for each relation. For the largest scale, {\tt Fact$\;$payloads} is $481$x faster and takes $548$x less memory than {\tt List$\;$payloads} ($410$ms vs. $197$s, $195$MB vs. $104$GB), and {\tt List$\;$keys} exceeds the available memory.

}

%% file: related.tex
\section{Related Work}
\label{sec:related_work}

To the best of our knowledge, ours is the first approach to propose factorized IVM for a range of distinct applications. It extends non-trivially two lines of prior work: higher-order delta-based IVM and factorized computation of in-database analytics. 

Our view language is modelled on functional aggregate queries over semirings~\cite{FAQ:PODS:2016} and generalized multiset relations over rings~\cite{DBT:VLDBJ:2014}; the latter allowed us to adapt DBToaster to factorized IVM.

{\bf IVM.} IVM is a well-studied area spanning more than three deca\-des~\cite{Chirkova:Views:2012:FTD,SalemBCL00,TangSEKF20}. Prior work extensively studied IVM for various query languages and showed that the time complexity of IVM is lower than of recomputation. We go beyond prior work on higher-order IVM for queries with joins and aggregates, as realized in DBToaster~\cite{DBT:VLDBJ:2014}, and propose a unified approach for factorized computation of aggregates over joins~\cite{BKOZ:PVLDB:2013}, factorized incremental computation of linear algebra~\cite{NEK:SIGMOD:2014}, and in-database machine learning over database joins~\cite{SOC:SIGMOD:2016}. DBToaster uses one materialization hierarchy per relation in the query, whereas \DF uses one view tree for all relations. DBToaster can thus have much higher space requirements and update times. As we observed experimentally, it does not consider the maintenance of composite aggregates such as the covariance matrix.
IVM over array data~\cite{Zhao:2017:ArrayIVM} targets scientific workloads but without exploiting data factorization.

\DF over the relational payload ring strict\-ly subsumes prior work on factorized IVM for acyclic joins \cite{DynYannakakis:SIGMOD:2017} as it can support arbitrary joins.
\DF has efficient support for free-connex acyclic~\cite{DynYannakakis:SIGMOD:2017} and $q$-hierarchical queries~\cite{Nicole:PODS:2017}.
Exploiting key attributes to enable succinct delta representations and accelerate maintenance complements our approach~\cite{Katsis:idIVM:2015}.
Our framework generalizes the main idea of the LINVIEW approach~\cite{NEK:SIGMOD:2014} for maintaining matrix computation over arbitrary joins. 
Unlike approaches that exploit the {\em append-only} nature of data streams~\cite{YangGO17}, \DF allows for both data insertions and deletions.
\DF can be used to improve the memory-efficiency of systems that integrate IVM into compilers to speed up the search in abstract syntax trees~\cite{BalakrishnanNKZ21}.
Such systems  suffer from the high storage overhead of systems such as DBToaster that maintain significantly more views than F-IVM.

Commercial DBMSs support IVM for restricted cla\-sses of queries, e.g., Oracle~\cite{Oracle:RestrictionsIVM} and SQLServer \cite{SQLServer:RestrictionsIVM}.
LogicBlox supports higher-order IVM for Datalog meta-programs~\cite{LB:SIGMOD:2015,GOW:PVLDB:2015}. Trill is a streaming engine that supports incremental processing of relational-style que\-ries but no complex aggregates like covariance matrices~\cite{chandramouli2014trill}. Differential Dataflow~\cite{McSherry:DiffDataflow:2013} supports incremental processing for  programs with recursion. There is a distinct line of work on maintenance for recursive Datalog~\cite{Motik:FBF:2019}.

{\bf Static In-DB analytics.} The emerging area of in-data\-base analytics has been overviewed in two tutorials~\cite{Polyzotis:SIGMOD:Tutorial:17,Kumar:SIGMOD:Tutorial:17} and a recent keynote~\cite{Olteanu:VLDBKeynote:2020}.  Several systems support analytics over normalized data via a tight integration of databases and machine learning~\cite{MLlib:JMLR:2016,MADlib:2012,Rusu:2015,Polyzotis:SIGMOD:Tutorial:17,Kumar:SIGMOD:Tutorial:17}. Other systems integrate with R to enable in-situ data processing using domain-specia\-lized routines~\cite{ZCDDMMFSS12,Brown:SciDB:2010:SIGMOD}. The closest in spirit to our approach is work on learning models over factorized joins \cite{Rendle13,SOC:SIGMOD:2016,OS:PVLDB:16,ANNOS:TODS:2020}, pushing ML tasks past joins~\cite{Kumar:InDBMS:2012,LMFAO:SIGMOD:2019} and on in-database linear algebra~\cite{Boehm:VLDB:2016,Arun:VLDB:2017,Figaro:SIGMOD:2022}, yet they do not consider incremental maintenance.

{\bf Learning.} There is a wealth of work in the ML community on incremental or online learning over {\em arbitrary} relations~\cite{OnlineML:2011}. Our approach learns over {\em joins} and crucially exploits the join dependencies in the underlying training dataset to improve the runtime performance.

%% file: conclusion.tex
\section{Conclusion and Future Work}
\label{sec:conclusion}
This article introduces \DF, a system that unifies the task of maintaining a variety of analytics over normalized data under updates. We show its applicability to learning linear regression models, building Chow Liu trees, and query evaluation with listing/factorized result representation. \DF recovers the best known complexities for free-connex acyclic and $q$-hierarchical queries. A prior version of this work~\cite{FIVM:SIGMOD:2018} also discusses the application of \DF to  matrix chain multiplication. These tasks use the same computation paradigm that factorizes the representation and the computation of the keys, the payloads, and the updates. Their differences are confined to the definition of the sum and product operations in a suitable ring. \DF is publicly available and was implemented as an extension of DBToaster~\cite{DBT:VLDBJ:2014}, a state-of-the-art system for incremental maintenance, and shown to outperform competitors by orders of magnitude in both time and space. 

Going forward, we would like to apply this approach to further tasks such as inference in probabilistic graphical models and more complex machine learning tasks.

\DF inherits the limitations of DBToaster, in particular it is single-threaded. A promising avenue of research is to build \DF on top of an open-source parallel and distributed framework such as Apache Flink. Another goal is to extend \DF to support further SQL operators such as theta joins, nested subqueries, and NULLs, which are relevant in practice.

\begin{quote}{\em \hspace*{-2em} If You Liked It, Then You Should Put A Ring On It.
\hspace*{15.5em} -- Beyonc\'e.}
\end{quote}